\documentclass[aps,prc,twoside,twocolumn,nofootinbib,10pt,showpacs,floatfix]{revtex4-1}
\usepackage{amsmath,amssymb}
\usepackage{graphicx,bm}
\usepackage{slashed}
\usepackage{epstopdf}
\usepackage{ulem} 
\usepackage[usenames]{color}
\usepackage{float}
\usepackage{hyperref}
\usepackage{subfigure}
\usepackage{subfigure}
\usepackage{rotating}
\usepackage{color}
\usepackage{multirow}
\usepackage{dcolumn}
\usepackage{overpic}
\usepackage{booktabs}
\usepackage{makecell}
\usepackage{color}
\usepackage{diagbox}

\renewcommand\sout{\bgroup \color{red} \ULdepth=-.5ex \ULset}

\newsavebox{\tablebox}
\usepackage{multirow}
\usepackage{dcolumn}
\usepackage{overpic}
\usepackage{booktabs}
\usepackage{makecell}
\usepackage{slashed}
\usepackage{epstopdf}
\usepackage{diagbox}
\usepackage{array}
\newcolumntype{|}{!{\vline}}
\begin{document}
\title{Fully heavy pentaquark states in constituent quark model}
\author{Hong-Tao An$^{1,2}$}\email{anht14@lzu.edu.cn}
\author{Si-Qiang Luo$^{1,2}$}\email{luosq15@lzu.edu.cn}
\author{Zhan-Wei Liu$^{1,2,3}$}\email{liuzhanwei@lzu.edu.cn}
\author{Xiang Liu$^{1,2,3}$}\email{xiangliu@lzu.edu.cn}
\affiliation{
$^1$School of Physical Science and Technology, Lanzhou University, Lanzhou 730000, China\\
$^2$Research Center for Hadron and CSR Physics, Lanzhou University and Institute of Modern Physics of CAS, Lanzhou 730000, China\\
$^3$Lanzhou Center for Theoretical Physics, Key Laboratory of Theoretical Physics of Gansu Province, and Frontiers Science Center for Rare Isotopes, Lanzhou University, Lanzhou 730000, China}

\begin{abstract}
The LHCb collaboration reported a fully charmed tetraquark state X(6900) in the invariant mass spectrum of $J/\psi$ pairs in 2020.
This discovery inspires us to further study the fully heavy pentaquark system.
In this work, we investigate systematically all possible configurations for ground fully heavy pentaquark system via the variational method in the constituent quark model.
According to our calculations, we further analyze the relative lengths between quarks and the contributions to the pentaquark masses from different terms of the Hamiltonian.
We think no stable states exist in fully heavy pentaquark system.
We hope that our study will be helpful to explore for fully heavy pentaquark states.
\end{abstract}
\maketitle
\section{Introduction}\label{sec1}

After the birth of the quark model for baryons and mesons,
people naturally propose the multiquark states beyond the traditional hadrons \cite{GellMann:1964nj,Zweig:1981pd,Zweig:1964jf}.
Since 2003 many experimental discoveries support the possible existence of multiquark configurations.
For example, a series of charmoniumlike XYZ states have been observed in experiment \cite{Choi:2003ue,LHCb:2016nsl,BESIII:2016bnd,Ablikim:2015vvn,Ablikim:2017oaf,Belle:2011aa,Ablikim:2016qzw,Mizuk:2008me}. $d^{*} (2380)$ was measured  by CELSIUS/WASA \cite{Faldt:2011zv} and WASA-at-COSY Collaborations \cite{Adlarson:2011bh,Adlarson:2012fe}, and it is expected to be a six-quark configuration only composed of u,d quarks.
The LHCb Collaboration has reported $P_{c}$ states which can be the hidden-charm molecular pentaquark states \cite{Aaij:2016phn,Aaij:2015tga,Aaij:2019vzc}.
Recently, a narrow doubly charmed tetraquark state named as the $T_{cc}$ state was observed at LHC \cite{LHCb:2021vvq,LHCb:2021auc}, and it is an explicitly exotic hadron.

Moreover, the LHCb collaboration noticed a narrow structure in $J/\psi$-pair invariant mass of approximately 6.9 GeV with significance greater than 5 $\sigma$ \cite{LHCb:2020bwg}.
This structure is expected to be a $cc\bar{c}\bar{c}$ configuration.
The relevant properties of the fully heavy tetraquark state have been studied, such as the decay behavior \cite{Becchi:2020uvq}, inner configuration \cite{Wan:2020fsk,Guo:2020pvt,Ke:2021iyh}, mass spectra \cite{Jin:2020jfc,Li:2021ygk,Lu:2020cns,Deng:2020iqw,Albuquerque:2020hio,Wang:2020dlo,Zhang:2020xtb,Giron:2020wpx,Faustov:2020qfm,Gordillo:2020sgc,Weng:2020jao}, and the production mechanism \cite{Wang:2020gmd,Wang:2020wrp,Dong:2020nwy,Maciula:2020wri,Karliner:2020dta,Feng:2020riv,Ma:2020kwb,Zhu:2020xni,Zhu:2020xni,Szczurek:2021orw,Wang:2020wrp}.
The discovery of fully heavy tetraquark state naturally makes us speculate that the fully heavy pentaquark state may also exist.

If one replaces the $J/\psi$ meson with the $\Omega_{ccc}$  baryon, we can obtain a fully heavy charmed pentaquark configuration.
Inspired by these, we study systematically all possible fully heavy pentaquark configurations in the constituent quark model.

For the constituent quark model, various versions of nonrelativistic and relativistic models were proposed and widely applied in studying the hadron properties. Almost all of them incorporate both the short-range one-gluon-exchange (OGE) force and the term representing the color confinement in either the coordinate or momentum space. Bhaduri et al. used phenomenological nonrelativistic potentials to fit the low-lying charmonium spectra \cite{Bhaduri:1981pn}. The $qq\bar{Q}\bar{Q}$ states have been investigated via the variational method based on simple Gaussian trial function  \cite{Brink:1998as}, and a good stable candidate, the lowest $I(J^{P})=0(1^{+})$ $ud\bar{b}\bar{b}$ state, was predicated and supported by other works \cite{Karliner:2017qjm,Eichten:2017ffp,Lu:2020rog,Cheng:2020wxa,Luo:2017eub,Bicudo:2017szl,Bicudo:2016ooe,Noh:2021lqs}.
Park et al. improved the potential terms in the constituent model and systemically calculated the $P_{c}$ states, the doubly heavy tetraquark states, and many dibaryons with different configurations \cite{Park:2016mez,Park:2016mez,Park:2013fda,Park:2015nha,Park:2018wjk,Park:2017jbn,Park:2016cmg}.
It is interesting to extend the constituent quark model to fully heavy pentaquark states.

The fully heavy pentaquarks have been studied in various models. In the framework of the modified chromomagnetic interaction (CMI) model, the mass spectra for the ground fully heavy pentaquarks $QQQQ\bar{Q}$ has been systematically investigated \cite{An:2020jix}, and a $J^{P}=3/2^{-}$ $ccbb\bar{b}$ state is considered as a good stable candidate which cannot decay through the strong interaction.
In the framework of the chiral quark model and quark delocalization color screening model, Yan et al. systematically investigate the $cccc\bar{c}$ and $bbbb\bar{b}$ states and obtain three
bound three fully heavy pentaquarks \cite{Yan:2021glh}.
However, it still needs to be further confirmed by solving accurately
the five-body problem for the configurations as pointed out in Ref. \cite{Richard:2021jgp}.
Richard et al. have used a potential model to investigate $QQ\bar{Q}\bar{Q}$ tetraquarks \cite{Richard:2018yrm} and $QQQQQQ$ dibaryons \cite{Richard:2020zxb}. Based on these studies, they infer that a serious solution of the potential model does not lead to a proliferation of stable multiquarks.
On the other hand, they also think that the part of the spectrum above the threshold is also extremely instructive \cite{Richard:2021jgp}.

Moreover, the fully heavy $QQQQ\bar{Q}$ pentaquark states was calculated with the QCD sum rule \cite{Zhang:2020vpz},
and the mass spectrums are predicted to be 7.41 GeV for the $cccc\bar{c}$ state, and 21.6 GeV for the $bbbb\bar{b}$ state, respectively.
If the fully heavy pentaquark states are the diquark-diquark-antiquark type, QCD sum rules give that $\rm M_{cccc\bar{c}}=7.93\pm 0.15$ GeV and $\rm M_{bbbb\bar{b}}=23.91\pm 0.15$ GeV \cite{Wang:2021xao}.


This paper is organized as follows.
Firstly, we introduce the Hamiltonian and construct the wave functions of the constituent quark model in Sec. \ref{sec2}.
Then we show the numerical results and discussion for the masses of the fully heavy pentaquarks obtained from the variational method in Sec. \ref{sec4}.
Finally, we give a short summary in Sec. \ref{sec5}.

\section{Hamiltonian and Wave Functions}\label{sec2}
We choose a nonrelativistic Hamiltonian of the following form \cite{Park:2018wjk}
\begin{eqnarray}\label{Eq1}
H=\sum_{i=1}^{5}\left(m_{i}+\frac{\textbf{p}^{2}_{i}}{2m_{i}}\right)-\frac{3}{4}\sum_{i<j}^{5}\frac{\lambda^{c}_{i}}{2}.\frac{\lambda^{c}_{j}}{2}\left(V^{CON}_{ij}+V^{SS}_{ij}\right)
\end{eqnarray}
where $m_{i}$ is the $i$-th (anti)quark mass, the color operator $\lambda^{c}_{i}/2$ is the Gell-Mann matrix for the $i$-th quark and replaced with $-\lambda^{c*}_{i}$ for antiquark.
The $V^{CON}_{ij}$ is the confinement potential between $i$-th quark and $j$-th quark and composed of the linearizing term and the Coulomb potential term while the $V^{SS}_{ij}$ is the hyperfine potential
\begin{eqnarray}\label{Eq2}
V^{CON}_{ij}&=&-\frac{\kappa}{r_{ij}}+\frac{r_{ij}}{a^{2}_{0}}-D,\nonumber\\
V^{SS}_{ij}&=&\frac{\kappa'}{m_{i}m_{j}c^{4}}\frac{1}{r_{0ij}r_{ij}}e^{-r^{2}_{ij}/r^{2}_{0ij}}\sigma_{i}.\sigma_{j},
\end{eqnarray}
where $m_{i}$ ($m_{j}$) is the mass of the $i$-th ($j$-th) quark, and $r_{ij}$ is distance between $i$-th and $j$-th quark.
For $r_{0ij}$ and $\kappa'$, we have
\begin{eqnarray}\label{Eq3}
r_{0ij}&=&1/\left(\alpha+\beta\frac{m_{i}m_{j}}{m_{i}+m_{j}}\right),\nonumber\\
\kappa'&=&\kappa_{0}\left(1+\gamma\frac{m_{i}m_{j}}{m_{i}+m_{j}}\right).
\end{eqnarray}
The parameters in Eqs. (\ref{Eq2}) and (\ref{Eq3}) are chosen from Ref. \cite{Park:2018wjk} and given in Table \ref{para}. Here, $\kappa$ and $\kappa'$ are the couplings of the Coulomb and hyperfine potentials, respectively, and they are proportional to the running coupling constant $\alpha_{s}(r)$ of QCD. The Coulomb and hyperfine interaction can be deduced from the one-gluon-exchange model. $1/a^{2}_{0}$ represents the strength of linear potential. $r_{0ij}$ is the Gaussian-smearing parameter. Further, we introduce $\kappa_{0}$ and $\gamma$ in $\kappa'$ to provide better descriptions for the interaction between different quark pairs. 

\begin{table}[htp]
\caption{Parameters of the Hamiltonian.}\label{para}
\begin{lrbox}{\tablebox}
\renewcommand\arraystretch{1.5}
\renewcommand\tabcolsep{2.8pt}
\begin{tabular}{c|c|c|c}
\midrule[1.5pt]
\toprule[0.5pt]
Parameter&$\kappa$&$a_{0}$&$D$\\
\toprule[0.5pt]
Value&120.0 MeV fm&0.0318119 $\rm (MeV^{-1}fm)^{1/2}$& 983 MeV\\
\toprule[1.0pt]
Parameter&$\alpha$&$\beta$&$m_{c}$\\
\toprule[0.5pt]
Value&1.0499 $\rm fm^{-1}$& 0.0008314 $\rm (MeV fm)^{-1}$&1918 MeV\\
\toprule[1.0pt]
Parameter&$\kappa_{0}$&$\gamma$&$m_{b}$\\
\toprule[0.5pt]
Value&194.144 MeV& 0.00088 $\rm MeV^{-1}$& 5343 MeV\\
\toprule[0.5pt]
\toprule[1.0pt]
\end{tabular}
\end{lrbox}\scalebox{0.923}{\usebox{\tablebox}}
\end{table}

Now we construct the wave function satisfied with Pauli Principle for fully heavy pentaquark states.
The specific wave functions include the flavor, spatial, and color-spin parts.

\subsection{Flavor part}\label{1}

According to flavor symmetry, we can divide the fully heavy pentaquark system into the following three groups:
(1) the first four quarks are identical: the $cccc\bar{c}$, $cccc\bar{b}$, $bbbb\bar{c}$, and $bbbb\bar{b}$ systems;
(2) the first three quarks are identical: the $cccb\bar{c}$, $cccb\bar{b}$, $bbbc\bar{c}$, and $bbbc\bar{b}$ systems;
(3) the two pairs of quarks are identical: the $ccbb\bar{c}$ and $ccbb\bar{b}$ systems. We use the notation $\{1234\}$ ([1234]) to label that the quarks 1, 2, 3, and 4 are fully antisymmetric (symmetric), and the notations such as $\{34\}$ and [123] are similar.
\subsection{Jacobian coordinates and spatial part}

We construct the wave function for the spatial part in a simple Gaussian form. In the center-of-mass frame of the pentaquark system, the number of
Jacobian coordinates of the system is reduced to 4.
In the case where the constituent quark masses are all different, the Jacobian coordinates are as follows \cite{Park:2017jbn}:

\begin{eqnarray}\label{Eq4}
\textbf{x}_{1}&=&\sqrt{\frac{1}{2}}(\textbf{r}_{1}-\textbf{r}_{2});\nonumber\\
\textbf{x}_{2}&=&\sqrt{\frac{2}{3}}\left[\textbf{r}_{3}-\left(\frac{m_{1}\textbf{r}_{1}+m_{2}\textbf{r}_{2}}{m_{1}+m_{2}}\right)\right];\nonumber\\
\textbf{x}_{3}&=&\sqrt{\frac{1}{2}}(\textbf{r}_{4}-\textbf{r}_{5});\nonumber\\
\textbf{x}_{4}&=&\sqrt{\frac{6}{5}}\left[\left(\frac{m_{1}\textbf{r}_{1}+m_{2}\textbf{r}_{2}+m_{3}\textbf{r}_{3}}{m_{1}+m_{2}+m_{3}})-(\frac{m_{4}\textbf{r}_{4}+m_{5}\textbf{r}_{5}}{m_{4}+m_{5}}\right)\right].\quad
\end{eqnarray}

The Jacobian coordinates in Eq. (\ref{Eq4}) can be used for the $ccbb\bar{c}$ and $ccbb\bar{b}$ states if the masses are arranged as follows:
\begin{eqnarray}\label{Eq5}
m_{1}=m_{2}=m_{c}, m_{3}=m_{\bar{c}}, m_{4}=m_{5}=m_{b}  &~\rm{for}~ & ccbb\bar{c}, \nonumber\\
m_{1}=m_{2}=m_{b}, m_{3}=m_{\bar{b}}, m_{4}=m_{5}=m_{c}  &~\rm{for}~ & ccbb\bar{b}.
\end{eqnarray}
Then a single Gaussian form can
accommodate the required symmetry property:
\begin{eqnarray}\label{Eq6}
R=\exp[-C_{11}\textbf{x}^{2}_{1}-C_{22}\textbf{x}^{2}_{2}-C_{33}\textbf{x}^{2}_{3}-C_{44}\textbf{x}^{2}_{4}],
\end{eqnarray}
where $C_{11}$, $C_{22}$, $C_{33}$, and $C_{44}$ are the variational parameters. In this work we only consider the S-wave pentaquarks. Then spatial function in Eq.(\ref{Eq6}) is symmetric between 1 and 2, and at the same time symmetric between 4 and 5.
We will denote this symmetry property of spatial function by [12]3[45].

For the $cccb\bar{c}$, $cccb\bar{b}$, $bbbc\bar{c}$, and $bbbc\bar{b}$ states, we set the specific masses in the Jacobian coordinates of Eq. (\ref{Eq4}) as
\begin{eqnarray}\label{Eq5}
m_{1}=m_{2}=m_{3}=m_{c}, m_{4}=m_{b}, m_{5}=m_{\bar{c}} &~\rm{for}~ & cccb\bar{c}, \nonumber\\
m_{1}=m_{2}=m_{3}=m_{c}, m_{4}=m_{b}, m_{5}=m_{\bar{b}} &~\rm{for}~ & cccb\bar{b}, \nonumber\\
m_{1}=m_{2}=m_{3}=m_{b}, m_{4}=m_{c}, m_{5}=m_{\bar{c}} &~\rm{for}~ & bbbc\bar{c}, \nonumber\\
m_{1}=m_{2}=m_{3}=m_{b}, m_{4}=m_{c}, m_{5}=m_{\bar{b}} &~\rm{for}~ & bbbc\bar{b}.
\end{eqnarray}
Similarly, we also give a single Gaussian form of $cccb\bar{c}$, $cccb\bar{b}$, $bbbc\bar{c}$, and $bbbc\bar{b}$ states by
\begin{eqnarray}\label{Eq7}
R=\exp[-C_{11}(\textbf{x}^{2}_{1}+\textbf{x}^{2}_{2})-C_{22}\textbf{x}^{2}_{3}-C_{33}\textbf{x}^{2}_{4}],
\end{eqnarray}
where $C_{11}$, $C_{22}$, and $C_{33}$ are the variational parameters.
The spatial function in Eq.(\ref{Eq7}) is symmetric among 1,2, and 3.
We can denote this symmetry property of spatial function by [123]45.

For $cccc\bar{c}$, $cccc\bar{b}$, $bbbb\bar{c}$, and $bbbb\bar{b}$ states, their spatial wave function needs to have the [1234]5 property.
According to discussion in Ref. \cite{Park:2016cmg}, the four Jacobian coordinates can be
\begin{eqnarray}\label{Eq8}
\textbf{x}_{1}&=&\frac{1}{2}(\textbf{r}_{1}-\textbf{r}_{2}+\textbf{r}_{3}-\textbf{r}_{4});\nonumber\\
\textbf{x}_{2}&=&\frac{1}{2}(\textbf{r}_{1}-\textbf{r}_{2}-\textbf{r}_{3}+\textbf{r}_{4});\nonumber\\
\textbf{x}_{3}&=&\frac{1}{2}(\textbf{r}_{1}+\textbf{r}_{2}-\textbf{r}_{3}-\textbf{r}_{4});\nonumber\\
\textbf{x}_{4}&=&\frac{1}{2\sqrt{5}}(\textbf{r}_{1}+\textbf{r}_{2}+\textbf{r}_{3}+\textbf{r}_{4}-4\textbf{r}_{5}).
\end{eqnarray}
and the spatial wave function is
\begin{eqnarray}\label{Eq9}
R=\exp[-(C_{11}(\textbf{x}_{1}^{2}+\textbf{x}_{2}^{2}+\textbf{x}_{3}^{2})-C_{22}(\textbf{x}_{4})^{2})],
\end{eqnarray}
where $C_{11}$ and $C_{22}$ are variational parameters.

At the same time, it is useful to introduce the kinetic term in the center-of-mass frame
\begin{eqnarray}\label{Eq10}
T_{c}=\sum^{5}_{i=1}\frac{\textbf{p}^{2}_{i}}{2m_{i}}= \frac{\textbf{p}^{2}_{x_{1}}}{2m'_{1}}+\frac{\textbf{p}^{2}_{x_{2}}}{2m'_{2}}+\frac{\textbf{p}^{2}_{x_{3}}}{2m'_{3}}
+\frac{\textbf{p}^{2}_{x_{4}}}{2m'_{4}},\nonumber\\
\end{eqnarray}
where different states have different reduced masses $m'_{i}$ and we show them in Table.\ref{m}.

\begin{table}[htp]
\caption{The value of reduced mass $m'_{i}$ in Eq. (\ref{Eq10}) for different states.}\label{m}
\begin{lrbox}{\tablebox}
\renewcommand\arraystretch{1.9}
\renewcommand\tabcolsep{2.85pt}
\begin{tabular}{c|cccc|c|cccc}
\toprule[1.0pt]
\toprule[0.5pt]
States&$m'_{1}$&$m'_{2}$&$m'_{3}$&$m'_{4}$&States&$m'_{1}$&$m'_{2}$&$m'_{3}$&$m'_{4}$\\
\toprule[0.5pt]
$cccc\bar{c}$&$m_{c}$&$m_{c}$&$m_{c}$&$m_{c}$&$cccb\bar{c}$&$m_{c}$&$m_{c}$&$\frac{2m_{c}m_{b}}{m_{c}+m_{b}}$&$\frac{5m_{c}(m_{c}+m_{b})}{2(4m_{c}+m_{b})}$\\
$bbbb\bar{b}$&$m_{b}$&$m_{b}$&$m_{b}$&$m_{b}$&$bbbc\bar{b}$&$m_{b}$&$m_{b}$&$\frac{2m_{c}m_{b}}{m_{c}+m_{b}}$&$\frac{5m_{b}(m_{c}+m_{b})}{2(4m_{b}+m_{c})}$\\
$cccc\bar{b}$&$m_{c}$&$m_{c}$&$m_{c}$&$\frac{5m_{c}m_{b}}{4m_{c}+m_{b}}$&$cccb\bar{b}$&$m_{c}$&$m_{c}$&$m_{b}$&$\frac{5m_{c}m_{b}}{3m_{c}+2m_{b}}$\\
$bbbb\bar{c}$&$m_{b}$&$m_{b}$&$m_{b}$&$\frac{5m_{b}m_{c}}{4m_{b}+m_{c}}$&$bbbc\bar{c}$&$m_{b}$&$m_{b}$&$m_{c}$&$\frac{5m_{c}m_{b}}{3m_{b}+2m_{c}}$\\
$cccb\bar{b}$&$m_{c}$&$m_{c}$&$m_{b}$&$\frac{5m_{c}m_{b}}{3m_{c}+2m_{b}}$&$bbbc\bar{c}$&$m_{b}$&$m_{b}$&$m_{c}$&$\frac{5m_{c}m_{b}}{3m_{b}+2m_{c}}$\\
\toprule[0.5pt]
\toprule[1.0pt]
\end{tabular}
\end{lrbox}\scalebox{0.91}{\usebox{\tablebox}}
\end{table}

\subsection{Color-spin part}

Because the spatial and flavor parts are exchange symmetric, we need to require the color $\otimes$ spin part to be exchange antisymmetric due to Pauli Principle.
Further, according to these three groups of the Sec. \ref{1}, we need to construct the color $\otimes$ spin part, which satisfies \{1234\}5, \{123\}45, and \{12\}\{34\}5 symmetries, respectively.
We consider symmetry properties without the particle 5 because the particle 5 is an antiquark.

The Young tableau, which represents the irreducible bases of the permutation group,
enables us to easily identify the multiquark configuration with certain symmetry properties \cite{Park:2017jbn}.
In this part, we use the Young tableau, Young diagram, and Young-Yamanouchi basis vector to describe the symmetry of a state.
We first start by separately discussing the color and spin wave functions, and then provide the color $\otimes$ spin wave functions.

For the possible color states, we only consider the color singlets because of color confinement.
Here, the color part is based on the SU(3) symmetry.
We can construct three color singlets and use the corresponding Young tableau to represent them:
\begin{align}\label{eq-color1}
C_{1}=\begin{tabular}{|c|c|}
\hline
1 &  2   \\
\cline{1-2}
\multicolumn{1}{|c|}{3} \\
\cline{1-1}
\multicolumn{1}{|c|}{4}  \\
\cline{1-1}
\end{tabular}_{3}
\otimes
(5)_{\bar{3}},
C_{2}=\begin{tabular}{|c|c|}
\hline
1 &  3    \\
\cline{1-2}
\multicolumn{1}{|c|}{2} \\
\cline{1-1}
\multicolumn{1}{|c|}{4}  \\
\cline{1-1}
\end{tabular}_{3}
\otimes
(5)_{\bar{3}},
C_{3}=\begin{tabular}{|c|c|}
\hline
1 &  4   \\
\cline{1-2}
\multicolumn{1}{|c|}{2} \\
\cline{1-1}
\multicolumn{1}{|c|}{3} \\
\cline{1-1}
\end{tabular}_{3}
\otimes
(5)_{\bar{3}}.
\end{align}
According to these three Young tableaux, we can obtain the corresponding Young diagram without particle 5:
\begin{align}\label{color}
\begin{tabular}{|c|c|}
\hline
 $\quad$ &  $\quad$   \\
\cline{1-2}
\multicolumn{1}{|c|}{\quad} \\
\cline{1-1}
\multicolumn{1}{|c|}{\quad}  \\
\cline{1-1}
\end{tabular}.
\end{align}

The spin part is based on the SU(2) symmetry.
For ground $QQQQ\bar{Q}$ system, all possible total spins are $J=$ 5/2, 3/2, and 1/2, respectively.
Here, we show corresponding Young tableaux for different spin states:
\begin{equation}\label{eq-color1}
\begin{split}
J=\frac{5}{2}:&\begin{tabular}{|c|c|c|c|c|}
\hline
1 &  2 &3&4&5   \\
\cline{1-5}
\end{tabular}_{1},\\
J=\frac{3}{2}:&\begin{tabular}{|c|c|c|c|c|}
\hline
1 &  2 &3&4   \\
\cline{1-4}
5\\
\cline{1-1}
\end{tabular}_{1},
\begin{tabular}{|c|c|c|c|c|}
\hline
1 &  2 &3&5   \\
\cline{1-4}
4\\
\cline{1-1}
\end{tabular}_{2},
\begin{tabular}{|c|c|c|c|c|}
\hline
1 &  2 &4&5   \\
\cline{1-4}
3\\
\cline{1-1}
\end{tabular}_{3},
\begin{tabular}{|c|c|c|c|c|}
\hline
1 &  3 &4&5   \\
\cline{1-4}
2\\
\cline{1-1}
\end{tabular}_{4},\\
J=\frac{1}{2}:&\begin{tabular}{|c|c|c|c|c|}
\hline
1 &  2 &3   \\
\cline{1-3}
4&5\\
\cline{1-2}
\end{tabular}_{1},
\begin{tabular}{|c|c|c|c|c|}
\hline
1 &  2 &4   \\
\cline{1-3}
3&5\\
\cline{1-2}
\end{tabular}_{2},
\begin{tabular}{|c|c|c|c|c|}
\hline
1 &  3 &4   \\
\cline{1-3}
2&5\\
\cline{1-2}
\end{tabular}_{3},
\begin{tabular}{|c|c|c|c|c|}
\hline
1 &  3 &5  \\
\cline{1-3}
2&4\\
\cline{1-2}
\end{tabular}_{4},
\begin{tabular}{|c|c|c|c|c|}
\hline
1 & 2 &5  \\
\cline{1-3}
3&4\\
\cline{1-2}
\end{tabular}_{5}.
\end{split}
\end{equation}
According to these ten Young tableaux, we can obtain three Young diagrams without particle 5:


\begin{equation}\label{eq-color2}
\begin{split}
\begin{tabular}{|c|c|c|c|}
\hline
 $\quad$ &  $\quad$& $\quad$& $\quad$  \\
\cline{1-4}
\end{tabular}_{S}
&:J=\frac{5}{2},\frac{3}{2},\\
\begin{tabular}{|c|c|c|}
\hline
 $\quad$ &  $\quad$& $\quad$  \\
\cline{1-3}
\multicolumn{1}{|c|}{\quad} \\
\cline{1-1}
\end{tabular}_{S}
&:J=\frac{3}{2}, \frac{1}{2}, \\
\begin{tabular}{|c|c|c|}
\hline
 $\quad$ &  $\quad$ \\
\cline{1-2}
 $\quad$ &  $\quad$ \\
\cline{1-2}
\end{tabular}_{S}
&:J=\frac{1}{2}.
\end{split}
\end{equation}

By using the Clebsch-Gordan (CG) coefficient of the permutation group $S_{n}$, one obtains the coupling scheme designed to construct the the color $\otimes$ spin wave functions. The detailed procedure can be found in Ref. \cite{An:2020jix}. As an example, we show the two corresponding Young-Yamanouchi basis vectors with \{1234\}5 symmetry
\begin{widetext}
\begin{align}\label{colorspin1}
J=\frac{3}{2}: \quad
&\begin{tabular}{|c|}
\hline
1      \\
\cline{1-1}
2      \\
\cline{1-1}
3      \\
\cline{1-1}
4      \\
\cline{1-1}
\end{tabular}_{CS_1}
=
\frac{1}{\sqrt{3}}
\begin{tabular}{|c|c|}
\hline
1 &2     \\
\cline{1-2}
3      \\
\cline{1-1}
4      \\
\cline{1-1}
\end{tabular}_{C_1}
\otimes
\begin{tabular}{|c|c|c|c}
\cline{1-3}
1 &3&4  &5   \\
\cline{1-3}
2      \\
\cline{1-1}
\end{tabular}_{S_5}
-\frac{1}{\sqrt{3}}
\begin{tabular}{|c|c|}
\hline
1 &3     \\
\cline{1-2}
2      \\
\cline{1-1}
4      \\
\cline{1-1}
\end{tabular}_{C_2}
\otimes
\begin{tabular}{|c|c|c|c}
\cline{1-3}
1 &2&4 &5    \\
\cline{1-3}
3      \\
\cline{1-1}
\end{tabular}_{S_4}
+\frac{1}{\sqrt{3}}
\begin{tabular}{|c|c|}
\hline
1 &4    \\
\cline{1-2}
2      \\
\cline{1-1}
3      \\
\cline{1-1}
\end{tabular}_{C_3}
\otimes
\begin{tabular}{|c|c|c|c}
\cline{1-3}
1 &2&3  &5   \\
\cline{1-3}
4      \\
\cline{1-1}
\end{tabular}_{S_3};
\nonumber
\\
J=\frac{1}{2}: \quad
&\begin{tabular}{|c|}
\hline
1      \\
\cline{1-1}
2      \\
\cline{1-1}
3      \\
\cline{1-1}
4      \\
\cline{1-1}
\end{tabular}_{CS_1}
=
\frac{1}{\sqrt{3}}
\begin{tabular}{|c|c|}
\hline
1 &2     \\
\cline{1-2}
3      \\
\cline{1-1}
4      \\
\cline{1-1}
\end{tabular}_{C_1}
\otimes
\begin{tabular}{|c|c|c|}
\hline
1 &3&4     \\
\cline{1-3}
2  & \multicolumn{1}{c}{5}   \\
\cline{1-1}
\end{tabular}_{S_5}
-\frac{1}{\sqrt{3}}
\begin{tabular}{|c|c|}
\hline
1 &3     \\
\cline{1-2}
2      \\
\cline{1-1}
4      \\
\cline{1-1}
\end{tabular}_{C_2}
\otimes
\begin{tabular}{|c|c|c|}
\hline
1 &2&4     \\
\cline{1-3}
3 &  \multicolumn{1}{c}{5}     \\
\cline{1-1}
\end{tabular}_{S_4}
+\frac{1}{\sqrt{3}}
\begin{tabular}{|c|c|}
\hline
1 &4    \\
\cline{1-2}
2      \\
\cline{1-1}
3      \\
\cline{1-1}
\end{tabular}_{C_3}
\otimes
\begin{tabular}{|c|c|c|}
\hline
1 &2&3     \\
\cline{1-3}
4 & \multicolumn{1}{c}{5}    \\
\cline{1-1}
\end{tabular}_{S_3},
\end{align}
\end{widetext}
with which the color-spin wave functions can be easily written \cite{An:2020jix}.

\section{NUMERICAL RESULTS and discussion}\label{sec4}

In this section, we substitute the wave function and perform variational analysis to determine the masses of ground pentaquark states, corresponding variational parameters, and the relative lengths between quarks in Eqs.(\ref{Eq4}) or (\ref{Eq9}).

Before that we first check the consistence between the experimental masses and the obtained masses of some traditional hadrons using the variational method based on the Hamiltonian in Eq. (\ref{Eq1}) and the parameters in Table \ref{para}. We show the results in Table \ref{tro}. We can see that our values are relatively reliable since the deviations for most states are less than 10 MeV.

\begin{table}[htp]
\caption{Masses of some mesons and baryons obtained from the variational method. The masses and corresponding errors are in units of MeV.
The variational parameters ``a" and ``b" are similar to $C_{ii}$ in Eq. (\ref{Eq6}) and they are in units of $\rm fm^{-2}$.
}\label{tro}
\begin{center}
\begin{lrbox}{\tablebox}
\renewcommand\arraystretch{1.6}
\renewcommand\tabcolsep{2.8pt}
\begin{tabular}{c|c|c|c|c|c|c}
\midrule[1.5pt]
\toprule[0.5pt]
Meson&$J/\psi$&$\eta_{c}$&$\Upsilon$&$\eta_{b}$&$B_{c}$&$B^{*}_{c}$\\
Theoretical&3092.2&2998.5&9468.9&9389.0&6287.9&6350.5\\
Parameters&a=12.5&a=15.0&a=49.7&a=57.4&a=22.9&a=20.2\\
Experimental&3096.9&2983.9&9460.3&9399.0&6274.9&\\
Error&-4.7&14.6&8.6&10.0&13.0&\\
\toprule[0.5pt]
Baryon&$\Omega_{ccc}$&$\Omega_{bbb}$&$\Omega^{*}_{ccb}$&$\Omega_{ccb}$&$\Omega^{*}_{bbc}$&$\Omega_{bbc}$\\
Theoretical&4801.4&14421.6&8063.8&8029.5&11273.2&11234.2\\
\multirow{2}*{Parameters}&\multirow{2}*{a=9.3}&\multirow{2}*{a=32.5}&a=10.4&a=10.8&a=26.0&a=26.8\\
&&&b=15.1&b=16.1&b=14.2&b=15.2\\
\toprule[0.5pt]
Baryon&$\Xi_{cc}$&$\Sigma_{c}$&$\Sigma^{*}_{c}$&$\Lambda_{c}$&$\Lambda$&$\Sigma$\\
Theoretical&3612&2445&2518&2283&1110&1188\\
\multirow{2}*{Parameters}&a=8.0&a=2.1&a=2.0&a=2.8&a=2.7&a=2.1\\
&b=3.2&b=3.7&b=3.4&b=3.7&b=2.7&b=3.1\\
Experimental&3621&2454&2518&2287&1116&1189\\
Error&-9.0&-9.0&0.0&-4.0&-6.0&-1.0\\
\toprule[0.5pt]
\toprule[1.0pt]
\end{tabular}
\end{lrbox}\scalebox{0.94}{\usebox{\tablebox}}
\end{center}
\end{table}

Here, we define the binding energy according to Ref. \cite{Park:2018wjk}:
\begin{eqnarray}\label{Eq21}
B_{T}=M_{pentaquark}-M_{baryon}-M_{meson},
\end{eqnarray}
where $M_{pentaquark}$ is the mass of ground pentaquark states;
$M_{baryon}$ and $M_{meson}$ are the masses of corresponding baryons and mesons in the lowest threshold with the same quantum numbers as the pentaquark, respectively.
For the $J^{P}=1/2^{-}$ pentaquark states, they are octet baryons $+$ pseudoscalar mesons or decuplet baryons $+$ vector mesons.
While for the $J^{P}=3/2^{-}$ pentaquark states, they are decuplet baryons $+$ pseudoscalar mesons or octet baryons $+$ vector mesons.

According to the obtained variational parameters, we have the wave functions and thus can further calculate the internal contributions to the ground pentaquark states and their lowest meson-baryon thresholds, including quark masses part, kinetic energy part, confinement potential part, and hyperfine potential part.
Moreover, in order to understand the composition of the total energy of the pentaquark states in comparison to the lowest meson-baryon threshold, it is important to understand the relative lengths between quarks for the pentaquark states.
These determine the magnitude of the various kinetic energies and the potential energies between quarks \cite{Park:2018wjk}.

Here, we also define the $V^{C}$: the sum of Coulomb potential and linear potential.
In most of the multiquark configurations, the contributions to the bound state from the parts of $V^{C}$ and
kinetic energy are repulsive, and therefore the contribution from the color spin interaction becomes important in these circumstances \cite{Park:2018wjk}.
However, the hyperfine potential is far smaller compared to other contributions in the fully heavy pentaquark system from corresponding tables because the hyperfine potential part is inversely proportional to the quark masses.

Based on these internal contributions, we compare the compositions of the masses from the constituent quark model and from the Chromomagnetic Interaction (CMI) model for fully heavy pentaquark states.
In the CMI model, as discussed in Ref. \cite{An:2020jix}, the mass of a hadron is typically composed
of the sum of the effective quark mass term (including the color interaction term) and the color-spin interaction term.
We want to identify the origin of the effective quark mass term and the color-spin interaction term used in the CMI model from our calculation.
Then, we investigate whether it is sensible to extrapolate these concepts to higher multiquark configurations.

In the following subsections, we discuss systematically the configurations of fully heavy pentaquark states group by group.

\subsection{$cccc\bar{c}$, $bbbb\bar{b}$, $cccc\bar{b}$, and $bbbb\bar{c}$ systems}

\begin{table*}
\caption{ The masses, variational parameters, the internal contribution, and the relative lengths between quarks for $cccc\bar{c}$ system and their lowest baryon-meson thresholds.
Here, $(i,j)$ denotes the contribution from the $i$-th and $j$-th quarks.
The number is given as i=1,2,3,4 for the quarks, and 5 for the antiquark.
The masses and corresponding contributions are in units of MeV, and the relative lengths (variational parameters) are in units of fm ($\rm fm^{-2}$).
}\label{nr1}
\begin{lrbox}{\tablebox}
\renewcommand\arraystretch{1.55}
\renewcommand\tabcolsep{2.3pt}
\begin{tabular}{c|c|ccc|ccc|c|cc|ccc}
\midrule[1.5pt]
\toprule[0.5pt]
\multicolumn{1}{c}{$cccc\bar{c}$}&\multicolumn{4}{r}{The contribution from each term}&\multicolumn{3}{c|}{$V^{C}$}&\multirow{2}*{Overall}&\multicolumn{2}{c}{Present Work}&\multicolumn{2}{c}{CMI Model}\\
\Xcline{1-8}{0.5pt}\Xcline{10-13}{0.5pt}
\multicolumn{1}{r}{$J^{P}=\frac32^{-}$}&&Value&\multicolumn{1}{r}{$\Omega_{ccc}\eta_{c}$}&\multicolumn{1}{r|}{Difference}&$(i,j)$&\multicolumn{1}{r}{$\frac32^{-}$}
&\multicolumn{1}{c|}{$\Omega_{ccc}\eta_{c}$}&&Contribution&Value&Contribution&Value\\ \Xcline{1-13}{0.5pt}
\multicolumn{2}{c}{Mass}&\multicolumn{1}{r}{8144.6}&7800.0&344.6&(1,2)&8.1&\multicolumn{1}{c|}{-22.7($\Omega_{ccc}$)}&
\multirow{10}*{$c$-quark}&4$m_{c}$&7672.0&$\frac{3}{2}$$m_{cc}$&4757.3\\
\Xcline{1-5}{0.5pt}
\multirow{2}*{\makecell[c]{Variational\\ Parameters\\ (fm$^{-2}$)}}&$C_{11}$&\multicolumn{1}{r}{8.6}&9.3&&(1.3)&8.1&\multicolumn{1}{c|}{-22.7($\Omega_{ccc}$)}&
&$\frac{\textbf{p}^{2}_{x_{1}}}{2m'_{1}}+\frac{\textbf{p}^{2}_{x_{2}}}{2m'_{2}}+\frac{\textbf{p}^{2}_{x_{3}}}{2m'_{3}}$&781.1&$\frac12m_{c\bar{c}}$&1534.3\\
&$C_{22}$&\multicolumn{1}{r}{5.3}&15.0&&(2,3)&8.1&\multicolumn{1}{c|}{-22.7($\Omega_{ccc}$)}&&
$\frac{m_{\bar{c}}}{4m_{c}+m_{\bar{c}}}\frac{\textbf{p}^{2}_{x_{1}}}{2m'_{4}}$&32.3\\
\Xcline{1-5}{0.5pt}
\multicolumn{2}{c}{Quark Mass}&\multicolumn{1}{r}{9590.0}&9590.0&0.0&(1,4)&8.1&&&\multirow{2}*{\makecell[l]{$V^{C}(12)+V^{C}(13)+$ \\$V^{C}(23)+V^{C}(14)+$ \\ $V^{C}(24)+V^{C}(34)$}}&\multirow{2}*{48.6}&\\
\multicolumn{2}{c}{\multirow{1}{*}{Confinement Potential}}&\multirow{1}{*}{-2423.8}&\multirow{1}{*}{-2762.7}&\multirow{1}{*}{338.9}&(2,4)&8.1&&&&\\
\Xcline{1-5}{0.5pt}
\multicolumn{2}{c}{\multirow{5}{*}{$V^{C}$ Subtotal}}&\multicolumn{1}{r}{\multirow{5}{*}{33.7}}&\multirow{5}{*}{-305.2}&\multicolumn{1}{c}{\multirow{5}{*}{338.9}}&(3,4)&8.1&&&\multirow{2}*{\makecell[l]{$\frac{1}{2}[V^{C}(12)+V^{C}(13)$ \\$+V^{C}(23)+V^{C}(14)]$ }}&\multirow{2}*{-7.4}&\\
\multicolumn{5}{c}{}&(1,5)&-3.7&&&&\\
\multicolumn{5}{c}{}&(2,5)&-3.7&&&-2D&-1966&\\
\Xcline{10-13}{0.5pt}
\multicolumn{5}{c}{}&(3,5)&-3.7&&&Subtotal&6560.6&&6291.6\\
\Xcline{9-13}{0.5pt}
\multicolumn{5}{c}{}&(4,5)&-3.7&-237.2($\eta_{c}$)&\multirow{6}*{$\bar{c}$-quark}&$m_{\bar{c}}$&1918.0&$\frac12m_{c\bar{c}}$&1534.3\\ \Xcline{1-8}{0.5pt}
\multicolumn{2}{c}{\multirow{2}*{Kinetic Energy}}&\multicolumn{1}{r}{\multirow{2}*{942.4}}&\multirow{2}*{1021.1}&\multirow{2}*{-78.7}&\multicolumn{3}{c|}{Relative Lengths (fm)}&&$\frac{4m_{c}}{4m_{c}+m_{\bar{c}}}\frac{\textbf{p}^{2}_{x_{1}}}{2m'_{4}}$&129.0\\
\Xcline{6-8}{0.5pt}
\multicolumn{5}{c|}{}&(1,2)&0.409&$0.370(\Omega_{ccc})$&&\multirow{2}*{\makecell[l]{$\frac{1}{2}[V^{C}(12)+V^{C}(13)$ \\$+V^{C}(23)+V^{C}(14)]$ }}&\multirow{2}*{-7.4}\\
\multicolumn{2}{c}{\multirow{2}*{CS Interaction}}&\multicolumn{1}{r}{\multirow{2}*{36.0}}&\multirow{2}*{-48.5}&\multirow{2}*{84.5}&(1,3)&0.409&$0.370(\Omega_{ccc})$&&&\\
\multicolumn{5}{c|}{}&(2,3)&0.409&$0.370(\Omega_{ccc})$&&$\frac12D$&-491.5&\\
\Xcline{1-5}{0.5pt}\Xcline{10-13}{0.5pt}
\multicolumn{2}{c}{\multirow{3}*{Total Contribution}}&\multicolumn{1}{r}{\multirow{3}*{1012.1}}&\multirow{3}*{667.4}&\multirow{3}*{344.7}&(1,4)&0.409&&&Subtotal&1548.1&&1534.3\\
\Xcline{9-13}{0.5pt}
\multicolumn{5}{c|}{}&(2,4)&0.409&&\multirow{5}*{\makecell[c]{CS \\ Interaction}}&\multirow{2}*{\makecell[l]{$\frac{7}{12}[V^{S}(12)+V^{S}(13)$\\ $+V^{S}(23)+V^{S}(14)$\\
$+V^{S}(24)+V^{S}(34)]$}}&\multirow{2}*{52.5}&\multirow{2}*{$\frac{7}{6}v_{cc}$}&\multirow{2}*{66.2}\\
\multicolumn{5}{c|}{}&(3,4)&0.409&&&&\\
\multicolumn{5}{c|}{}&(1,5)&0.385&&&\multirow{2}*{\makecell[l]{$-\frac{1}{4}[V^{S}(15)+V^{S}(25)$ \\ $+V^{S}(35)+V^{S}(45)]$}}&\multirow{2}*{-16.5}&\multirow{2}*{$-\frac{1}{3}v_{c\bar{c}}$}&\multirow{2}*{-28.4}\\
\multicolumn{5}{c|}{}&(2,5)&0.385&&&&\\
\Xcline{10-13}{0.5pt}
\multicolumn{5}{c|}{}&(3,5)&0.385&&&Subtotal&35.9&&37.8\\
\Xcline{9-13}{0.5pt}
\multicolumn{5}{c|}{}&(4,5)&0.385&$0.290(\eta_{c})$&Total&&8144.6&&7863.6\\
\toprule[1pt]
\multicolumn{1}{r}{$J^{P}=\frac12^{-}$}&&Value&\multicolumn{1}{r}{$\Omega_{ccc}J/\psi$}&\multicolumn{1}{r|}{Difference}&$(i,j)$&\multicolumn{1}{r}{$\frac32^{-}$}
&\multicolumn{1}{c|}{$\Omega_{ccc}J/\psi$}&&Contribution&Value&Contribution&Value\\ \Xcline{1-13}{0.5pt}
\multicolumn{2}{c}{Mass}&\multicolumn{1}{r}{8193.2}&7893.6&299.6&(1,2)&11.6&\multicolumn{1}{c|}{-22.7($\Omega_{ccc}$)}&
\multirow{10}*{$c$-quark}&4$m_{c}$&7672.0&$\frac{3}{2}$$m_{cc}$&4757.3\\
\Xcline{1-5}{0.5pt}
\multirow{2}*{\makecell[c]{Variational\\ Parameters\\ (fm$^{-2}$)}}&$C_{11}$&\multicolumn{1}{r}{8.2}&9.3&&(1.3)&11.6&\multicolumn{1}{c|}{-22.7($\Omega_{ccc}$)}&
&$\frac{\textbf{p}^{2}_{x_{1}}}{2m'_{1}}+\frac{\textbf{p}^{2}_{x_{2}}}{2m'_{2}}+\frac{\textbf{p}^{2}_{x_{3}}}{2m'_{3}}$&746.9&$\frac12m_{c\bar{c}}$&1534.3\\
&$C_{22}$&\multicolumn{1}{r}{5.2}&12.5&&(2,3)&11.6&\multicolumn{1}{c|}{-22.7($\Omega_{ccc}$)}&&
$\frac{m_{\bar{c}}}{4m_{c}+m_{\bar{c}}}\frac{\textbf{p}^{2}_{x_{1}}}{2m'_{4}}$&31.7\\
\Xcline{1-5}{0.5pt}
\multicolumn{2}{c}{Quark Mass}&\multicolumn{1}{r}{9590.0}&9590.0&0.0&(1,4)&11.6&&&\multirow{2}*{\makecell[l]{$V^{C}(12)+V^{C}(13)+$ \\$V^{C}(23)+V^{C}(14)+$ \\ $V^{C}(24)+V^{C}(34)$}}&\multirow{2}*{69.6}&\\
\multicolumn{2}{c}{\multirow{1}{*}{Confinement Potential}}&\multirow{1}{*}{-2385.0}&\multirow{1}{*}{-2689.7}&\multirow{1}{*}{304.7}&(2,4)&11.6&&&&\\
\Xcline{1-5}{0.5pt}
\multicolumn{2}{c}{\multirow{5}{*}{$V^{C}$ Subtotal}}&\multicolumn{1}{r}{\multirow{5}{*}{72.5}}&\multirow{5}{*}{-232.2}&\multicolumn{1}{c}{\multirow{5}{*}{304.7}}&(3,4)&11.6&&&\multirow{2}*{\makecell[l]{$\frac{1}{2}[V^{C}(12)+V^{C}(13)$ \\$+V^{C}(23)+V^{C}(14)]$ }}&\multirow{2}*{1.4}&\\
\multicolumn{5}{c}{}&(1,5)&0.7&&&&\\
\multicolumn{5}{c}{}&(2,5)&0.7&&&-2D&-1966&\\
\Xcline{10-13}{0.5pt}
\multicolumn{5}{c}{}&(3,5)&0.7&&&Subtotal&6555.6&&6291.6\\
\Xcline{9-13}{0.5pt}
\multicolumn{5}{c}{}&(4,5)&0.7&-164.2($J/\psi$)&\multirow{6}*{$\bar{c}$-quark}&$m_{\bar{c}}$&1918.0&$\frac12m_{c\bar{c}}$&1534.3\\ \Xcline{1-8}{0.5pt}
\multicolumn{2}{c}{\multirow{2}*{Kinetic Energy}}&\multicolumn{1}{r}{\multirow{2}*{905.4}}&\multirow{2}*{945.0}&\multirow{2}*{-39.6}&\multicolumn{3}{c|}{Relative Lengths (fm)}&&$\frac{4m_{c}}{4m_{c}+m_{\bar{c}}}\frac{\textbf{p}^{2}_{x_{1}}}{2m'_{4}}$&126.8\\
\Xcline{6-8}{0.5pt}
\multicolumn{5}{c|}{}&(1,2)&0.416&$0.370(\Omega_{ccc})$&&\multirow{2}*{\makecell[l]{$\frac{1}{2}[V^{C}(12)+V^{C}(13)$ \\$+V^{C}(23)+V^{C}(14)]$ }}&\multirow{2}*{1.4}\\
\multicolumn{2}{c}{\multirow{2}*{CS Interaction}}&\multicolumn{1}{r}{\multirow{2}*{82.8}}&\multirow{2}*{48.3}&\multirow{2}*{34.5}&(1,3)&0.416&$0.370(\Omega_{ccc})$&&&\\
\multicolumn{5}{c|}{}&(2,3)&0.416&$0.370(\Omega_{ccc})$&&$\frac12D$&-491.5&\\
\Xcline{1-5}{0.5pt}\Xcline{10-13}{0.5pt}
\multicolumn{2}{c}{\multirow{3}*{Total Contribution}}&\multicolumn{1}{r}{\multirow{3}*{1060.7}}&\multirow{3}*{761.1}&\multirow{3}*{299.6}&(1,4)&0.416&&&Subtotal&1554.7&&1534.3\\
\Xcline{9-13}{0.5pt}
\multicolumn{5}{c|}{}&(2,4)&0.416&&\multirow{5}*{\makecell[c]{CS \\ Interaction}}&\multirow{2}*{\makecell[l]{$\frac{7}{12}[V^{S}(12)+V^{S}(13)$\\ $+V^{S}(23)+V^{S}(14)$\\
$+V^{S}(24)+V^{S}(34)]$}}&\multirow{2}*{50.9}&\multirow{2}*{$\frac{7}{6}v_{cc}$}&\multirow{2}*{66.2}\\
\multicolumn{5}{c|}{}&(3,4)&0.416&&&&\\
\multicolumn{5}{c|}{}&(1,5)&0.393&&&\multirow{2}*{\makecell[l]{$\frac{1}{2}[V^{S}(15)+V^{S}(25)$ \\ $+V^{S}(35)+V^{S}(45)]$}}&\multirow{2}*{31.8}&\multirow{2}*{$\frac{2}{3}v_{c\bar{c}}$}&\multirow{2}*{56.7}\\
\multicolumn{5}{c|}{}&(2,5)&0.393&&&&\\
\Xcline{10-13}{0.5pt}
\multicolumn{5}{c|}{}&(3,5)&0.393&&&Subtotal&82.8&&123.0\\
\Xcline{9-13}{0.5pt}
\multicolumn{5}{c|}{}&(4,5)&0.393&$0.318(J/\psi)$&Total&&8193.2&&7948.8\\
\toprule[0.5pt]
\toprule[1.0pt]
\end{tabular}
\end{lrbox}\scalebox{0.905}{\usebox{\tablebox}}
\end{table*}

\begin{table*}
\caption{ The masses, variational parameters, the internal contribution, and the relative lengths between quarks for $cccc\bar{b}$ system and their lowest baryon-meson threshold.
The notations are same as those of Table. \ref{nr1}.
}\label{nr2}
\begin{lrbox}{\tablebox}
\renewcommand\arraystretch{1.55}
\renewcommand\tabcolsep{2.32pt}
\begin{tabular}{c|c|ccc|ccc|c|cc|ccc}
\midrule[1.5pt]
\toprule[0.5pt]
\multicolumn{1}{c}{$cccc\bar{b}$}&\multicolumn{4}{r}{The contribution from each term}&\multicolumn{3}{c|}{$V^{C}$}&\multirow{2}*{Overall}&\multicolumn{2}{c}{Present Work}&\multicolumn{2}{c}{CMI Model}\\
\Xcline{1-8}{0.5pt}\Xcline{10-13}{0.5pt}
\multicolumn{1}{r}{$J^{P}=\frac32^{-}$}&&Value&\multicolumn{1}{r}{$\Omega_{ccc}B_{c}$}&\multicolumn{1}{r|}{Difference}&$(i,j)$&\multicolumn{1}{r}{$\frac32^{-}$}
&\multicolumn{1}{c|}{$\Omega_{ccc}B_{c}$}&&Contribution&Value&Contribution&Value\\ \Xcline{1-13}{0.5pt}
\multicolumn{2}{c}{Mass}&\multicolumn{1}{r}{11477.8}&\multicolumn{1}{r}{11089.3}&388.5&(1,2)&-5.6&\multicolumn{1}{c|}{-22.7($\Omega_{ccc}$)}&
\multirow{10}*{$c$-quark}&4$m_{c}$&7672.0&$\frac{3}{2}$$m_{cc}$&4757.3\\
\Xcline{1-5}{0.5pt}
\multirow{2}*{\makecell[c]{Variational\\ Parameters\\ (fm$^{-2}$)}}&$C_{11}$&\multicolumn{1}{r}{8.9}&\multicolumn{1}{r}{9.3}&&(1.3)&-5.6&\multicolumn{1}{c|}{-22.7($\Omega_{ccc}$)}&
&$\frac{\textbf{p}^{2}_{x_{1}}}{2m'_{1}}+\frac{\textbf{p}^{2}_{x_{2}}}{2m'_{2}}+\frac{\textbf{p}^{2}_{x_{3}}}{2m'_{3}}$&812.6&$\frac{m_{c}}{m_{c}+m_{\bar{b}}}m_{c\bar{b}}$&1577.5\\
&$C_{22}$&\multicolumn{1}{r}{8.1}&\multicolumn{1}{r}{22.9}&&(2,3)&-5.6&\multicolumn{1}{c|}{-22.7($\Omega_{ccc}$)}&&
$\frac{m_{\bar{b}}}{4m_{c}+m_{\bar{b}}}\frac{\textbf{p}^{2}_{x_{1}}}{2m'_{4}}$&49.4\\
\Xcline{1-5}{0.5pt}
\multicolumn{2}{c}{Quark Mass}&\multicolumn{1}{r}{13015.0}&\multicolumn{1}{r}{13015.0}&0&(1,4)&-5.6&&&\multirow{2}*{\makecell[l]{$V^{C}(12)+V^{C}(13)+$ \\$V^{C}(23)+V^{C}(14)+$ \\ $V^{C}(24)+V^{C}(34)$}}&\multirow{2}*{-33.3}&\\
\multicolumn{2}{c}{\multirow{1}{*}{Confinement Potential}}&\multicolumn{1}{r}{-2520.8}&\multicolumn{1}{r}{-2940.3}&\multirow{1}{*}{419.5}&(2,4)&-5.6&&&&\\
\Xcline{1-5}{0.5pt}
\multicolumn{2}{c}{\multirow{5}{*}{$V^{C}$ Subtotal}}&\multicolumn{1}{r}{\multirow{5}{*}{-63.3}}&\multicolumn{1}{r}{\multirow{5}{*}{-482.8}}&\multicolumn{1}{c}{\multirow{5}{*}{419.5}}&(3,4)&-5.6&&&\multirow{2}*{\makecell[l]{$\frac{1}{2}[V^{C}(15)+V^{C}(25)$ \\$+V^{C}(35)+V^{C}(45)]$ }}&\multirow{2}*{-15.0}&\\
\multicolumn{5}{c}{}&(1,5)&-7.5&&&&\\
\multicolumn{5}{c}{}&(2,5)&-7.5&&&-2D&-1966.0&\\
\Xcline{10-13}{0.5pt}
\multicolumn{5}{c}{}&(3,5)&-7.5&&&Subtotal&6519.7&&6334.8\\
\Xcline{9-13}{0.5pt}
\multicolumn{5}{c}{}&(4,5)&-7.5&-414.8($B_{c}$)&\multirow{6}*{$\bar{c}$-quark}&$m_{\bar{c}}$&5343.0&$\frac{m_{\bar{b}}}{m_{c}+m_{\bar{b}}}m_{c\bar{b}}$&4744.8\\ \Xcline{1-8}{0.5pt}
\multicolumn{2}{c}{\multirow{2}*{Kinetic Energy}}&\multicolumn{1}{r}{\multirow{2}*{932.9}}&\multicolumn{1}{r}{\multirow{2}*{1037.2}}&\multirow{2}*{-104.3}&\multicolumn{3}{c|}{Relative Lengths (fm)}&&$\frac{4m_{c}}{4m_{c}+m_{\bar{b}}}\frac{\textbf{p}^{2}_{x_{1}}}{2m'_{4}}$&70.9\\
\Xcline{6-8}{0.5pt}
\multicolumn{5}{c|}{}&(1,2)&0.381&$0.370(\Omega_{ccc})$&&\multirow{2}*{\makecell[l]{$\frac{1}{2}[V^{C}(12)+V^{C}(13)$ \\$+V^{C}(23)+V^{C}(14)]$ }}&\multirow{2}*{-15.0}\\
\multicolumn{2}{c}{\multirow{2}*{CS Interaction}}&\multicolumn{1}{r}{\multirow{2}*{50.7}}&\multicolumn{1}{r}{\multirow{2}*{-22.6}}&\multirow{2}*{73.3}&(1,3)&0.381&$0.370(\Omega_{ccc})$&&&\\
\multicolumn{5}{c|}{}&(2,3)&0.381&$0.370(\Omega_{ccc})$&&$\frac12D$&-491.5&\\
\Xcline{1-5}{0.5pt}\Xcline{10-13}{0.5pt}
\multicolumn{2}{c}{\multirow{3}*{Total Contribution}}&\multicolumn{1}{r}{\multirow{3}*{920.3}}&\multicolumn{1}{r}{\multirow{3}*{531.8}}&\multirow{3}*{388.5}&(1,4)&0.381&&&Subtotal&4907.4&&4744.8\\
\Xcline{9-13}{0.5pt}
\multicolumn{5}{c|}{}&(2,4)&0.381&&\multirow{5}*{\makecell[c]{CS \\ Interaction}}&\multirow{2}*{\makecell[l]{$\frac{7}{12}[V^{S}(12)+V^{S}(13)$\\ $+V^{S}(23)+V^{S}(14)$\\
$+V^{S}(24)+V^{S}(34)]$}}&\multirow{2}*{58.7}&\multirow{2}*{$\frac{7}{6}v_{cc}$}&\multirow{2}*{66.2}\\
\multicolumn{5}{c|}{}&(3,4)&0.381&&&&\\
\multicolumn{5}{c|}{}&(1,5)&0.377&&&\multirow{2}*{\makecell[l]{$-\frac{1}{4}[V^{S}(15)+V^{S}(25)$ \\ $+V^{S}(35)+V^{S}(45)]$}}&\multirow{2}*{-8.0}&\multirow{2}*{$-\frac{1}{3}v_{c\bar{b}}$}&\multirow{2}*{-15.7}\\
\multicolumn{5}{c|}{}&(2,5)&0.377&&&&\\
\Xcline{10-13}{0.5pt}
\multicolumn{5}{c|}{}&(3,5)&0.377&&&Subtotal&50.7&&50.5\\
\Xcline{9-13}{0.5pt}
\multicolumn{5}{c|}{}&(4,5)&0.377&$0.234(B_{c})$&Total&&11477.8&&11130.1\\
\toprule[1pt]
\multicolumn{1}{r}{$J^{P}=\frac12^{-}$}&&Value&\multicolumn{1}{r}{$\Omega_{ccc}B^{*}_{c}$}&\multicolumn{1}{r|}{Difference}&$(i,j)$&\multicolumn{1}{r}{$\frac32^{-}$}
&\multicolumn{1}{c|}{$\Omega_{ccc}B^{*}_{c}$}&&Contribution&Value&Contribution&Value\\ \Xcline{1-13}{0.5pt}
\multicolumn{2}{c}{Mass}&\multicolumn{1}{r}{11501.5}&\multicolumn{1}{r}{11151.9}&350.0&(1,2)&-3.6&\multicolumn{1}{c|}{-22.7($\Omega_{ccc}$)}&
\multirow{10}*{$c$-quark}&4$m_{c}$&7672.0&$\frac{3}{2}$$m_{cc}$&4757.3\\
\Xcline{1-5}{0.5pt}
\multirow{2}*{\makecell[c]{Variational\\ Parameters\\ (fm$^{-2}$)}}&$C_{11}$&\multicolumn{1}{r}{8.7}&9.3&&(1.3)&-3.6&\multicolumn{1}{c|}{-22.7($\Omega_{ccc}$)}&
&$\frac{\textbf{p}^{2}_{x_{1}}}{2m'_{1}}+\frac{\textbf{p}^{2}_{x_{2}}}{2m'_{2}}+\frac{\textbf{p}^{2}_{x_{3}}}{2m'_{3}}$&793.7&$\frac{m_{c}}{m_{c}+m_{\bar{b}}}m_{c\bar{b}}$&1577.5\\
&$C_{22}$&\multicolumn{1}{r}{8.0}&20.2&&(2,3)&-3.6&\multicolumn{1}{c|}{-22.7($\Omega_{ccc}$)}&&
$\frac{m_{\bar{b}}}{4m_{c}+m_{\bar{b}}}\frac{\textbf{p}^{2}_{x_{1}}}{2m'_{4}}$&48.9\\
\Xcline{1-5}{0.5pt}
\multicolumn{2}{c}{Quark Mass}&\multicolumn{1}{r}{13015.0}&\multicolumn{1}{r}{13015.0}&0&(1,4)&-3.6&&&\multirow{2}*{\makecell[l]{$V^{C}(12)+V^{C}(13)+$ \\$V^{C}(23)+V^{C}(14)+$ \\ $V^{C}(24)+V^{C}(34)$}}&\multirow{2}*{-21.6}&\\
\multicolumn{2}{c}{\multirow{1}{*}{Confinement Potential}}&\multicolumn{1}{r}{-2499.7}&\multicolumn{1}{r}{-2885.9}&\multirow{1}{*}{386.2}&(2,4)&-3.6&&&&\\
\Xcline{1-5}{0.5pt}
\multicolumn{2}{c}{\multirow{5}{*}{$V^{C}$ Subtotal}}&\multicolumn{1}{r}{\multirow{5}{*}{-42.2}}&\multirow{5}{*}{-428.4}&\multicolumn{1}{c}{\multirow{5}{*}{386.2}}&(3,4)&-3.6&&&\multirow{2}*{\makecell[l]{$\frac{1}{2}[V^{C}(15)+V^{C}(25)$ \\$+V^{C}(35)+V^{C}(45)]$ }}&\multirow{2}*{-10.4}&\\
\multicolumn{5}{c}{}&(1,5)&-5.2&&&&\\
\multicolumn{5}{c}{}&(2,5)&-5.2&&&-2D&-1966.0&\\
\Xcline{10-13}{0.5pt}
\multicolumn{5}{c}{}&(3,5)&-5.2&&&Subtotal&6516.6&&6334.8\\
\Xcline{9-13}{0.5pt}
\multicolumn{5}{c}{}&(4,5)&-5.2&-360.4($B^{*}_{c}$)&\multirow{6}*{$\bar{b}$-quark}&$m_{\bar{c}}$&5343.0&$\frac{m_{\bar{b}}}{m_{c}+m_{\bar{b}}}m_{c\bar{b}}$&4744.8\\ \Xcline{1-8}{0.5pt}
\multicolumn{2}{c}{\multirow{2}*{Kinetic Energy}}&\multicolumn{1}{r}{\multirow{2}*{912.8}}&\multicolumn{1}{r}{\multirow{2}*{981.5}}&\multirow{2}*{-68.7}&\multicolumn{3}{c|}{Relative Lengths (fm)}&&$\frac{4m_{c}}{4m_{c}+m_{\bar{b}}}\frac{\textbf{p}^{2}_{x_{1}}}{2m'_{4}}$&70.2\\
\Xcline{6-8}{0.5pt}
\multicolumn{5}{c|}{}&(1,2)&0.385&$0.370(\Omega_{ccc})$&&\multirow{2}*{\makecell[l]{$\frac{1}{2}[V^{C}(15)+V^{C}(25)$ \\$+V^{C}(35)+V^{C}(45)]$ }}&\multirow{2}*{-10.4}\\
\multicolumn{2}{c}{\multirow{2}*{CS Interaction}}&\multicolumn{1}{r}{\multirow{2}*{73.3}}&\multicolumn{1}{r}{\multirow{2}*{41.3}}&\multirow{2}*{32.0}&(1,3)&0.385&$0.370(\Omega_{ccc})$&&&\\
\multicolumn{5}{c|}{}&(2,3)&0.385&$0.370(\Omega_{ccc})$&&$\frac12D$&-491.5&\\
\Xcline{1-5}{0.5pt}\Xcline{10-13}{0.5pt}
\multicolumn{2}{c}{\multirow{3}*{Total Contribution}}&\multicolumn{1}{r}{\multirow{3}*{943.9}}&\multicolumn{1}{r}{\multirow{3}*{594.4}}&\multirow{3}*{349.7}&(1,4)&0.385&&&Subtotal&4911.3&&4744.8\\
\Xcline{9-13}{0.5pt}
\multicolumn{5}{c|}{}&(2,4)&0.385&&\multirow{5}*{\makecell[c]{CS \\ Interaction}}&\multirow{2}*{\makecell[l]{$\frac{7}{12}[V^{S}(12)+V^{S}(13)$\\ $+V^{S}(23)+V^{S}(14)$\\
$+V^{S}(24)+V^{S}(34)]$}}&\multirow{2}*{57.7}&\multirow{2}*{$\frac{7}{6}v_{cc}$}&\multirow{2}*{66.2}\\
\multicolumn{5}{c|}{}&(3,4)&0.385&&&&\\
\multicolumn{5}{c|}{}&(1,5)&0.382&&&\multirow{2}*{\makecell[l]{$\frac{1}{2}[V^{S}(15)+V^{S}(25)$ \\ $+V^{S}(35)+V^{S}(45)]$}}&\multirow{2}*{15.6}&\multirow{2}*{$\frac{2}{3}v_{c\bar{b}}$}&\multirow{2}*{31.5}\\
\multicolumn{5}{c|}{}&(2,5)&0.382&&&&\\
\Xcline{10-13}{0.5pt}
\multicolumn{5}{c|}{}&(3,5)&0.382&&&Subtotal&73.3&&97.7\\
\Xcline{9-13}{0.5pt}
\multicolumn{5}{c|}{}&(4,5)&0.382&$0.250(B_{c})$&Total&&11501.3&&11177.3\\
\toprule[0.5pt]
\toprule[1.0pt]
\end{tabular}
\end{lrbox}\scalebox{0.903}{\usebox{\tablebox}}
\end{table*}

\begin{table*}
\caption{ The masses, variational parameters, the internal contribution, and the relative lengths between quarks for $bbbb\bar{c}$ system and their lowest baryon-meson threshold.
The notations are same as those of Table. \ref{nr1}.
}\label{nr3}
\begin{lrbox}{\tablebox}
\renewcommand\arraystretch{1.55}
\renewcommand\tabcolsep{2.32pt}
\begin{tabular}{c|c|ccc|ccc|c|cc|ccc}
\midrule[1.5pt]
\toprule[0.5pt]
\multicolumn{1}{c}{$bbbb\bar{c}$}&\multicolumn{4}{r}{The contribution from each term}&\multicolumn{3}{c|}{$V^{C}$}&\multirow{2}*{Overall}&\multicolumn{2}{c}{Present Work}&\multicolumn{2}{c}{CMI Model}\\
\Xcline{1-8}{0.5pt}\Xcline{10-13}{0.5pt}
\multicolumn{1}{r}{$J^{P}=\frac32^{-}$}&&Value&\multicolumn{1}{r}{$\Omega_{bbb}B_{c}$}&\multicolumn{1}{r|}{Difference}&$(i,j)$&\multicolumn{1}{r}{Vaule}
&\multicolumn{1}{c|}{$\Omega_{bbb}B_{c}$}&&Contribution&Value&Contribution&Value\\ \Xcline{1-13}{0.5pt}
\multicolumn{2}{c}{Mass}&\multicolumn{1}{r}{20974.5}&20709.5&265.0&(1,2)&-80.8&\multicolumn{1}{c|}{-287.9($\Omega_{bbb}$)}&
\multirow{10}*{$b$-quark}&4$m_{b}$&21372.0&$\frac{3}{2}$$m_{bb}$&14309.4\\
\Xcline{1-5}{0.5pt}
\multirow{2}*{\makecell[c]{Variational\\ Parameters\\ (fm$^{-2}$)}}&$C_{11}$&\multicolumn{1}{r}{24.9}&32.5&&(1.3)&-80.8&\multicolumn{1}{c|}{-287.9($\Omega_{bbb}$)}&
&$\frac{\textbf{p}^{2}_{x_{1}}}{2m'_{1}}+\frac{\textbf{p}^{2}_{x_{2}}}{2m'_{2}}+\frac{\textbf{p}^{2}_{x_{3}}}{2m'_{3}}$&816.6&$\frac{m_{b}}{m_{b}+m_{\bar{c}}}m_{b\bar{c}}$&4783.6\\
&$C_{22}$&\multicolumn{1}{r}{9.4}&22.9&&(2,3)&-80.8&\multicolumn{1}{c|}{-287.9($\Omega_{bbb}$)}&&
$\frac{m_{\bar{b}}}{4m_{b}+m_{\bar{b}}}\frac{\textbf{p}^{2}_{x_{1}}}{2m'_{4}}$&20.5\\
\Xcline{1-5}{0.5pt}
\multicolumn{2}{c}{Quark Mass}&\multicolumn{1}{r}{23290.0}&23290.0&0&(1,4)&-80.8&&&\multirow{2}*{\makecell[l]{$V^{C}(12)+V^{C}(13)+$ \\$V^{C}(23)+V^{C}(14)+$ \\ $V^{C}(24)+V^{C}(34)$}}&\multirow{2}*{-484.8}&\\
\multicolumn{2}{c}{\multirow{1}{*}{Confinement Potential}}&\multirow{1}{*}{-3394.7}&\multirow{1}{*}{-3735.8}&\multirow{1}{*}{341.1}&(2,4)&-80.8&&&&\\
\Xcline{1-5}{0.5pt}
\multicolumn{2}{c}{\multirow{5}{*}{$V^{C}$ Subtotal}}&\multicolumn{1}{r}{\multirow{5}{*}{-937.1}}&\multirow{5}{*}{-1278.3}&\multicolumn{1}{c}{\multirow{5}{*}{341.2}}&(3,4)&-80.8&&&\multirow{2}*{\makecell[l]{$\frac{1}{2}[V^{C}(12)+V^{C}(13)$ \\$+V^{C}(23)+V^{C}(14)]$ }}&\multirow{2}*{-226.2}&\\
\multicolumn{5}{c}{}&(1,5)&-113.1&&&&\\
\multicolumn{5}{c}{}&(2,5)&-113.1&&&-2D&-1966.0&\\
\Xcline{10-13}{0.5pt}
\multicolumn{5}{c}{}&(3,5)&-113.1&&&Subtotal&19532.1&&19093.0\\
\Xcline{9-13}{0.5pt}
\multicolumn{5}{c}{}&(4,5)&-113.1&-414.8($B_{c}$)&\multirow{6}*{$\bar{c}$-quark}&$m_{\bar{c}}$&1918.0&$\frac{m_{c}}{m_{b}+m_{\bar{c}}}m_{b\bar{c}}$&1538.6\\ \Xcline{1-8}{0.5pt}
\multicolumn{2}{c}{\multirow{2}*{Kinetic Energy}}&\multicolumn{1}{r}{\multirow{2}*{1065.3}}&\multirow{2}*{1183.2}&\multirow{2}*{-117.9}&\multicolumn{3}{c|}{Relative Lengths (fm)}&&$\frac{4m_{b}}{4m_{b}+m_{\bar{b}}}\frac{\textbf{p}^{2}_{x_{1}}}{2m'_{4}}$&228.2\\
\Xcline{6-8}{0.5pt}
\multicolumn{5}{c|}{}&(1,2)&0.261&$0.197(\Omega_{bbb})$&&\multirow{2}*{\makecell[l]{$\frac{1}{2}[V^{C}(12)+V^{C}(13)$ \\$+V^{C}(23)+V^{C}(14)]$ }}&\multirow{2}*{-226.2}\\
\multicolumn{2}{c}{\multirow{2}*{CS Interaction}}&\multicolumn{1}{r}{\multirow{2}*{14.0}}&\multirow{2}*{-27.9}&\multirow{2}*{41.9}&(1,3)&0.261&$0.197(\Omega_{bbb})$&&&\\
\multicolumn{5}{c|}{}&(2,3)&0.261&$0.197(\Omega_{bbb})$&&$\frac12D$&-491.5&\\
\Xcline{1-5}{0.5pt}\Xcline{10-13}{0.5pt}
\multicolumn{2}{c}{\multirow{3}*{Total Contribution}}&\multicolumn{1}{r}{\multirow{3}*{142.2}}&\multirow{3}*{-123.0}&\multirow{3}*{265.1}&(1,4)&0.261&&&Subtotal&1428.5&&1538.6\\
\Xcline{9-13}{0.5pt}
\multicolumn{5}{c|}{}&(2,4)&0.261&&\multirow{5}*{\makecell[c]{CS \\ Interaction}}&\multirow{2}*{\makecell[l]{$\frac{7}{12}[V^{S}(12)+V^{S}(13)$\\ $+V^{S}(23)+V^{S}(14)$\\
$+V^{S}(24)+V^{S}(34)]$}}&\multirow{2}*{31.2}&\multirow{2}*{$\frac{7}{6}v_{bb}$}&\multirow{2}*{35.8}\\
\multicolumn{5}{c|}{}&(3,4)&0.261&&&&\\
\multicolumn{5}{c|}{}&(1,5)&0.225&&&\multirow{2}*{\makecell[l]{$-\frac{1}{4}[V^{S}(15)+V^{S}(25)$ \\ $+V^{S}(35)+V^{S}(45)]$}}&\multirow{2}*{-17.3}&\multirow{2}*{$-\frac{1}{3}v_{b\bar{c}}$}&\multirow{2}*{-15.7}\\
\multicolumn{5}{c|}{}&(2,5)&0.254&&&&\\
\Xcline{10-13}{0.5pt}
\multicolumn{5}{c|}{}&(3,5)&0.254&&&Subtotal&14.0&&20.1\\
\Xcline{9-13}{0.5pt}
\multicolumn{5}{c|}{}&(4,5)&0.254&$0.234(B_{c})$&Total&&20974.6&&20651.7\\
\toprule[1pt]
\multicolumn{1}{r}{$J^{P}=\frac12^{-}$}&&Value&\multicolumn{1}{r}{$\Omega_{bbb}B^{*}_{c}$}&\multicolumn{1}{r|}{Difference}&$(i,j)$&\multicolumn{1}{r}{Value}
&\multicolumn{1}{c|}{$\Omega_{bbb}B^{*}_{c}$}&&Contribution&Value&Contribution&Value\\ \Xcline{1-13}{0.5pt}
\multicolumn{2}{c}{Mass}&\multicolumn{1}{r}{21025.6}&20772.1&253.4&(1,2)&-77.0&\multicolumn{1}{c|}{-287.8($\Omega_{bbb}$)}&
\multirow{10}*{$b$-quark}&4$m_{b}$&21372.0&$\frac{3}{2}$$m_{bb}$&14309.4\\
\Xcline{1-5}{0.5pt}
\multirow{2}*{\makecell[c]{Variational\\ Parameters\\ (fm$^{-2}$)}}&$C_{11}$&\multicolumn{1}{r}{23.7}&\multicolumn{1}{r}{32.5}&&(1.3)&-77.0&\multicolumn{1}{c|}{-287.8($\Omega_{bbb}$)}&
&$\frac{\textbf{p}^{2}_{x_{1}}}{2m'_{1}}+\frac{\textbf{p}^{2}_{x_{2}}}{2m'_{2}}+\frac{\textbf{p}^{2}_{x_{3}}}{2m'_{3}}$&778.9&$\frac{m_{b}}{m_{b}+m_{\bar{c}}}m_{b\bar{c}}$&4783.6\\
&$C_{22}$&\multicolumn{1}{r}{9.2}&\multicolumn{1}{r}{20.2}&&(2,3)&-77.0&\multicolumn{1}{c|}{-287.8($\Omega_{bbb}$)}&&
$\frac{m_{\bar{b}}}{4m_{b}+m_{\bar{b}}}\frac{\textbf{p}^{2}_{x_{1}}}{2m'_{4}}$&20.0\\
\Xcline{1-5}{0.5pt}
\multicolumn{2}{c}{Quark Mass}&\multicolumn{1}{r}{23290.0}&\multicolumn{1}{r}{23290.0}&0.0&(1,4)&-77.0&&&\multirow{2}*{\makecell[l]{$V^{C}(12)+V^{C}(13)+$ \\$V^{C}(23)+V^{C}(14)+$ \\ $V^{C}(24)+V^{C}(34)$}}&\multirow{2}*{-462.0}&\\
\multicolumn{2}{c}{\multirow{1}{*}{Confinement Potential}}&\multirow{1}{*}{-3350.5}&\multirow{1}{*}{-3681.4}&\multirow{1}{*}{330.9}&(2,4)&-77.0&&&&\\
\Xcline{1-5}{0.5pt}
\multicolumn{2}{c}{\multirow{5}{*}{$V^{C}$ Subtotal}}&\multicolumn{1}{r}{\multirow{5}{*}{-893.0}}&\multirow{5}{*}{-1223.9}&\multicolumn{1}{c}{\multirow{5}{*}{330.9}}&(3,4)&-77.0&&&\multirow{2}*{\makecell[l]{$\frac{1}{2}[V^{C}(12)+V^{C}(13)$ \\$+V^{C}(23)+V^{C}(14)]$ }}&\multirow{2}*{-215.6}&\\
\multicolumn{5}{c}{}&(1,5)&-107.8&&&&\\
\multicolumn{5}{c}{}&(2,5)&-107.8&&&-2D&-1966.0&\\
\Xcline{10-13}{0.5pt}
\multicolumn{5}{c}{}&(3,5)&-107.8&&&Subtotal&19527.3&&19093.0\\
\Xcline{9-13}{0.5pt}
\multicolumn{5}{c}{}&(4,5)&-107.8&-360.4($B^{*}_{c}$)&\multirow{6}*{$\bar{c}$-quark}&$m_{\bar{c}}$&1918.0&$\frac{m_{c}}{m_{b}+m_{\bar{c}}}m_{b\bar{c}}$&1538.6\\ \Xcline{1-8}{0.5pt}
\multicolumn{2}{c}{\multirow{2}*{Kinetic Energy}}&\multicolumn{1}{r}{\multirow{2}*{1022.2}}&\multirow{2}*{1127.5}&\multirow{2}*{-105.5}&\multicolumn{3}{c|}{Relative Lengths (fm)}&&$\frac{4m_{b}}{4m_{b}+m_{\bar{b}}}\frac{\textbf{p}^{2}_{x_{1}}}{2m'_{4}}$&223.3\\
\Xcline{6-8}{0.5pt}
\multicolumn{5}{c|}{}&(1,2)&0.266&$0.197(\Omega_{bbb})$&&\multirow{2}*{\makecell[l]{$\frac{1}{2}[V^{C}(12)+V^{C}(13)$ \\$+V^{C}(23)+V^{C}(14)]$ }}&\multirow{2}*{-215.6}\\
\multicolumn{2}{c}{\multirow{2}*{CS Interaction}}&\multicolumn{1}{r}{\multirow{2}*{63.7}}&\multirow{2}*{36.0}&\multirow{2}*{27.7}&(1,3)&0.266&$0.197(\Omega_{bbb})$&&&\\
\multicolumn{5}{c|}{}&(2,3)&0.266&$0.197(\Omega_{bbb})$&&$\frac12D$&-491.5&\\
\Xcline{1-5}{0.5pt}\Xcline{10-13}{0.5pt}
\multicolumn{2}{c}{\multirow{3}*{Total Contribution}}&\multicolumn{1}{r}{\multirow{3}*{192.9}}&\multirow{3}*{-60.4}&\multirow{3}*{253.3}&(1,4)&0.266&&&Subtotal&1434.2&&1538.6\\
\Xcline{9-13}{0.5pt}
\multicolumn{5}{c|}{}&(2,4)&0.266&&\multirow{5}*{\makecell[c]{CS \\ Interaction}}&\multirow{2}*{\makecell[l]{$\frac{7}{12}[V^{S}(12)+V^{S}(13)$\\ $+V^{S}(23)+V^{S}(14)$\\
$+V^{S}(24)+V^{S}(34)]$}}&\multirow{2}*{30.3}&\multirow{2}*{$\frac{7}{6}v_{bb}$}&\multirow{2}*{35.8}\\
\multicolumn{5}{c|}{}&(3,4)&0.266&&&&\\
\multicolumn{5}{c|}{}&(1,5)&0.231&&&\multirow{2}*{\makecell[l]{$-\frac{1}{4}[V^{S}(15)+V^{S}(25)$ \\ $+V^{S}(35)+V^{S}(45)]$}}&\multirow{2}*{33.5}&\multirow{2}*{$\frac{2}{3}v_{b\bar{b}}$}&\multirow{2}*{31.5}\\
\multicolumn{5}{c|}{}&(2,5)&0.231&&&&\\
\Xcline{10-13}{0.5pt}
\multicolumn{5}{c|}{}&(3,5)&0.231&&&Subtotal&63.7&&67.2\\
\Xcline{9-13}{0.5pt}
\multicolumn{5}{c|}{}&(4,5)&0.231&$0.250(B^{*}_{c})$&Total&&21025.3&&20698.8\\
\toprule[0.5pt]
\toprule[1.0pt]
\end{tabular}
\end{lrbox}\scalebox{0.903}{\usebox{\tablebox}}
\end{table*}

\begin{table*}
\caption{ The masses, variational parameters, the internal contributions, and the relative lengths between quarks for $bbbb\bar{b}$ system and their lowest baryon-meson threshold.
The notations are same as those of Table. \ref{nr1}.
}\label{nr4}
\begin{lrbox}{\tablebox}
\renewcommand\arraystretch{1.55}
\renewcommand\tabcolsep{2.32pt}
\begin{tabular}{c|c|ccc|ccc|c|cc|ccc}
\midrule[1.5pt]
\toprule[0.5pt]
\multicolumn{1}{c}{$bbbb\bar{b}$}&\multicolumn{4}{r}{The contribution from each term}&\multicolumn{3}{c|}{$V^{C}$}&\multirow{2}*{Overall}&\multicolumn{2}{c}{Present Work}&\multicolumn{2}{c}{CMI Model}\\
\Xcline{1-8}{0.5pt}\Xcline{10-13}{0.5pt}
\multicolumn{1}{r}{$J^{P}=\frac32^{-}$}&&Value&\multicolumn{1}{r}{$\Omega_{ccc}\eta_{b}$}&\multicolumn{1}{r|}{Difference}&$(i,j)$&\multicolumn{1}{r}{Vaule}
&\multicolumn{1}{c|}{$\Omega_{bbb}\eta_{b}$}&&Contribution&Value&Contribution&Value\\ \Xcline{1-13}{0.5pt}
\multicolumn{2}{c}{Mass}&\multicolumn{1}{r}{24210.7}&23810.6&400.1&(1,2)&-113.7&\multicolumn{1}{c|}{-287.9($\Omega_{bbb}$)}&
\multirow{10}*{$b$-quark}&4$m_{b}$&21372.0&$\frac{3}{2}$$m_{bb}$&14309.4\\
\Xcline{1-5}{0.5pt}
\multirow{2}*{\makecell[c]{Variational\\ Parameters\\ (fm$^{-2}$)}}&$C_{11}$&\multicolumn{1}{r}{28.2}&32.5&&(1.3)&-113.7&\multicolumn{1}{c|}{-287.9($\Omega_{bbb}$)}&
&$\frac{\textbf{p}^{2}_{x_{1}}}{2m'_{1}}+\frac{\textbf{p}^{2}_{x_{2}}}{2m'_{2}}+\frac{\textbf{p}^{2}_{x_{3}}}{2m'_{3}}$&926.1&$\frac12m_{b\bar{b}}$&4722.5\\
&$C_{22}$&\multicolumn{1}{r}{17.5}&57.4&&(2,3)&-113.7&\multicolumn{1}{c|}{-287.9($\Omega_{bbb}$)}&&
$\frac{m_{\bar{b}}}{4m_{b}+m_{\bar{b}}}\frac{\textbf{p}^{2}_{x_{1}}}{2m'_{4}}$&38.3\\
\Xcline{1-5}{0.5pt}
\multicolumn{2}{c}{Quark Mass}&\multicolumn{1}{r}{26715.0}&26715.0&0&(1,4)&-113.7&&&\multirow{2}*{\makecell[l]{$V^{C}(12)+V^{C}(13)+$ \\$V^{C}(23)+V^{C}(14)+$ \\ $V^{C}(24)+V^{C}(34)$}}&\multirow{2}*{-682.2}&\\
\multicolumn{2}{c}{\multirow{1}{*}{Confinement Potential}}&\multirow{1}{*}{-3649.6}&\multirow{1}{*}{-4200.2}&\multirow{1}{*}{550.6}&(2,4)&-113.7&&&&\\
\Xcline{1-5}{0.5pt}
\multicolumn{2}{c}{\multirow{5}{*}{$V^{C}$ Subtotal}}&\multicolumn{1}{r}{\multirow{5}{*}{-1192.1}}&\multirow{5}{*}{-1742.7}&\multicolumn{1}{c}{\multirow{5}{*}{550.6}}&(3,4)&-113.7&&&\multirow{2}*{\makecell[l]{$\frac{1}{2}[V^{C}(12)+V^{C}(13)$ \\$+V^{C}(23)+V^{C}(14)]$ }}&\multirow{2}*{-254.8}&\\
\multicolumn{5}{c}{}&(1,5)&-127.4&&&&\\
\multicolumn{5}{c}{}&(2,5)&-127.4&&&-2D&-1966.0&\\
\Xcline{10-13}{0.5pt}
\multicolumn{5}{c}{}&(3,5)&-127.4&&&Subtotal&19433.4&&19031.8\\
\Xcline{9-13}{0.5pt}
\multicolumn{5}{c}{}&(4,5)&-127.4&-879.1($\eta_{b}$)&\multirow{6}*{$\bar{b}$-quark}&$m_{\bar{b}}$&5343.0&$\frac12m_{b\bar{b}}$&4722.5\\ \Xcline{1-8}{0.5pt}
\multicolumn{2}{c}{\multirow{2}*{Kinetic Energy}}&\multicolumn{1}{r}{\multirow{2}*{1117.7}}&\multirow{2}*{1338.5}&\multirow{2}*{-220.8}&\multicolumn{3}{c|}{Relative Lengths (fm)}&&$\frac{4m_{b}}{4m_{b}+m_{\bar{b}}}\frac{\textbf{p}^{2}_{x_{1}}}{2m'_{4}}$&153.2\\
\Xcline{6-8}{0.5pt}
\multicolumn{5}{c|}{}&(1,2)&0.225&$0.197(\Omega_{bbb})$&&\multirow{2}*{\makecell[l]{$\frac{1}{2}[V^{C}(12)+V^{C}(13)$ \\$+V^{C}(23)+V^{C}(14)]$ }}&\multirow{2}*{-254.8}\\
\multicolumn{2}{c}{\multirow{2}*{CS Interaction}}&\multicolumn{1}{r}{\multirow{2}*{27.6}}&\multirow{2}*{-41.7}&\multirow{2}*{69.3}&(1,3)&0.225&$0.197(\Omega_{bbb})$&&&\\
\multicolumn{5}{c|}{}&(2,3)&0.409&$0.197(\Omega_{bbb})$&&$\frac12D$&-491.5&\\
\Xcline{1-5}{0.5pt}\Xcline{10-13}{0.5pt}
\multicolumn{2}{c}{\multirow{3}*{Total Contribution}}&\multicolumn{1}{r}{\multirow{3}*{-46.8}}&\multirow{3}*{-446.9}&\multirow{3}*{400.1}&(1,4)&0.225&&&Subtotal&4749.9&&4722.5\\
\Xcline{9-13}{0.5pt}
\multicolumn{5}{c|}{}&(2,4)&0.225&&\multirow{5}*{\makecell[c]{CS \\ Interaction}}&\multirow{2}*{\makecell[l]{$\frac{7}{12}[V^{S}(12)+V^{S}(13)$\\ $+V^{S}(23)+V^{S}(14)$\\
$+V^{S}(24)+V^{S}(34)]$}}&\multirow{2}*{40.2}&\multirow{2}*{$\frac{7}{6}v_{bb}$}&\multirow{2}*{35.8}\\
\multicolumn{5}{c|}{}&(3,4)&0.225&&&&\\
\multicolumn{5}{c|}{}&(1,5)&0.212&&&\multirow{2}*{\makecell[l]{$-\frac{1}{4}[V^{S}(15)+V^{S}(25)$ \\ $+V^{S}(35)+V^{S}(45)]$}}&\multirow{2}*{-12.6}&\multirow{2}*{$-\frac{1}{3}v_{b\bar{b}}$}&\multirow{2}*{-15.3}\\
\multicolumn{5}{c|}{}&(2,5)&0.212&&&&\\
\Xcline{10-13}{0.5pt}
\multicolumn{5}{c|}{}&(3,5)&0.212&&&Subtotal&27.6&&20.5\\
\Xcline{9-13}{0.5pt}
\multicolumn{5}{c|}{}&(4,5)&0.212&$0.148(\eta_{b})$&Total&&24210.9&&23774.8\\
\toprule[1pt]
\multicolumn{1}{r}{$J^{P}=\frac12^{-}$}&&Value&\multicolumn{1}{r}{$\Omega_{bbb}\Upsilon$}&\multicolumn{1}{r|}{Difference}&$(i,j)$&\multicolumn{1}{r}{Value}
&\multicolumn{1}{c|}{$\Omega_{bbb}\Upsilon$}&&Contribution&Value&Contribution&Value\\ \Xcline{1-13}{0.5pt}
\multicolumn{2}{c}{Mass}&\multicolumn{1}{r}{24248.0}&23890.5&357.5&(1,2)&-110.0&\multicolumn{1}{c|}{-287.8($\Omega_{bbb}$)}&
\multirow{10}*{$b$-quark}&4$m_{b}$&21372.0&$\frac{3}{2}$$m_{bb}$&14309.4\\
\Xcline{1-5}{0.5pt}
\multirow{2}*{\makecell[c]{Variational\\ Parameters\\ (fm$^{-2}$)}}&$C_{11}$&\multicolumn{1}{r}{27.2}&32.5&&(1.3)&-110.0&\multicolumn{1}{c|}{-287.8($\Omega_{bbb}$)}&
&$\frac{\textbf{p}^{2}_{x_{1}}}{2m'_{1}}+\frac{\textbf{p}^{2}_{x_{2}}}{2m'_{2}}+\frac{\textbf{p}^{2}_{x_{3}}}{2m'_{3}}$&891.3&$\frac12m_{b\bar{b}}$&4722.5\\
&$C_{22}$&\multicolumn{1}{r}{17.2}&49.7&&(2,3)&-110.0&\multicolumn{1}{c|}{-287.8($\Omega_{bbb}$)}&&
$\frac{m_{\bar{b}}}{4m_{b}+m_{\bar{b}}}\frac{\textbf{p}^{2}_{x_{1}}}{2m'_{4}}$&37.5\\
\Xcline{1-5}{0.5pt}
\multicolumn{2}{c}{Quark Mass}&\multicolumn{1}{r}{26715.0}&26715.0&0&(1,4)&-110.0&&&\multirow{2}*{\makecell[l]{$V^{C}(12)+V^{C}(13)+$ \\$V^{C}(23)+V^{C}(14)+$ \\ $V^{C}(24)+V^{C}(34)$}}&\multirow{2}*{-660.0}&\\
\multicolumn{2}{c}{\multirow{1}{*}{Confinement Potential}}&\multirow{1}{*}{-3609.6}&\multirow{1}{*}{-4117.8}&\multirow{1}{*}{508.2}&(2,4)&-110.0&&&&\\
\Xcline{1-5}{0.5pt}
\multicolumn{2}{c}{\multirow{5}{*}{$V^{C}$ Subtotal}}&\multicolumn{1}{r}{\multirow{5}{*}{-1152.1}}&\multirow{5}{*}{-1660.0}&\multicolumn{1}{c}{\multirow{5}{*}{508.0}}&(3,4)&-110.0&&&\multirow{2}*{\makecell[l]{$\frac{1}{2}[V^{C}(12)+V^{C}(13)$ \\$+V^{C}(23)+V^{C}(14)]$ }}&\multirow{2}*{-246.0}&\\
\multicolumn{5}{c}{}&(1,5)&-123.0&&&&\\
\multicolumn{5}{c}{}&(2,5)&-123.0&&&-2D&-1966.0&\\
\Xcline{10-13}{0.5pt}
\multicolumn{5}{c}{}&(3,5)&-123.0&&&Subtotal&19428.8&&19031.8\\
\Xcline{9-13}{0.5pt}
\multicolumn{5}{c}{}&(4,5)&-123.0&-796.7($\Upsilon$)&\multirow{6}*{$\bar{b}$-quark}&$m_{\bar{b}}$&5343.0&$\frac12m_{b\bar{b}}$&4722.5\\ \Xcline{1-8}{0.5pt}
\multicolumn{2}{c}{\multirow{2}*{Kinetic Energy}}&\multicolumn{1}{r}{\multirow{2}*{1078.9}}&\multirow{2}*{1253.2}&\multirow{2}*{-174.3}&\multicolumn{3}{c|}{Relative Lengths (fm)}&&$\frac{4m_{b}}{4m_{b}+m_{\bar{b}}}\frac{\textbf{p}^{2}_{x_{1}}}{2m'_{4}}$&150.1\\
\Xcline{6-8}{0.5pt}
\multicolumn{5}{c|}{}&(1,2)&0.228&$0.197(\Omega_{bbb})$&&\multirow{2}*{\makecell[l]{$\frac{1}{2}[V^{C}(12)+V^{C}(13)$ \\$+V^{C}(23)+V^{C}(14)]$ }}&\multirow{2}*{-246.0}\\
\multicolumn{2}{c}{\multirow{2}*{CS Interaction}}&\multicolumn{1}{r}{\multirow{2}*{63.6}}&\multirow{2}*{40.0}&\multirow{2}*{23.6}&(1,3)&0.228&$0.197(\Omega_{bbb})$&&&\\
\multicolumn{5}{c|}{}&(2,3)&0.228&$0.197(\Omega_{bbb})$&&$\frac12D$&-491.5&\\
\Xcline{1-5}{0.5pt}\Xcline{10-13}{0.5pt}
\multicolumn{2}{c}{\multirow{3}*{Total Contribution}}&\multicolumn{1}{r}{\multirow{3}*{-9.6}}&\multirow{3}*{-366.8}&\multirow{3}*{357.3}&(1,4)&0.228&&&Subtotal&4755.6&&4722.5\\
\Xcline{9-13}{0.5pt}
\multicolumn{5}{c|}{}&(2,4)&0.228&&\multirow{5}*{\makecell[c]{CS \\ Interaction}}&\multirow{2}*{\makecell[l]{$\frac{7}{12}[V^{S}(12)+V^{S}(13)$\\ $+V^{S}(23)+V^{S}(14)$\\
$+V^{S}(24)+V^{S}(34)]$}}&\multirow{2}*{39.1}&\multirow{2}*{$\frac{7}{6}v_{bb}$}&\multirow{2}*{35.8}\\
\multicolumn{5}{c|}{}&(3,4)&0.228&&&&\\
\multicolumn{5}{c|}{}&(1,5)&0.216&&&\multirow{2}*{\makecell[l]{$-\frac{1}{4}[V^{S}(15)+V^{S}(25)$ \\ $+V^{S}(35)+V^{S}(45)]$}}&\multirow{2}*{24.5}&\multirow{2}*{$\frac{2}{3}v_{b\bar{b}}$}&\multirow{2}*{30.7}\\
\multicolumn{5}{c|}{}&(2,5)&0.216&&&&\\
\Xcline{10-13}{0.5pt}
\multicolumn{5}{c|}{}&(3,5)&0.216&&&Subtotal&63.6&&66.5\\
\Xcline{9-13}{0.5pt}
\multicolumn{5}{c|}{}&(4,5)&0.216&$0.160(\Upsilon)$&Total&&24248.0&&23820.8\\
\toprule[0.5pt]
\toprule[1.0pt]
\end{tabular}
\end{lrbox}\scalebox{0.903}{\usebox{\tablebox}}
\end{table*}

Firstly we investigate the $cccc\bar{c}$, $bbbb\bar{b}$, $cccc\bar{b}$, and $bbbb\bar{c}$ systems.
These four systems need to satisfy the \{1234\}5 symmetry.
There are only $J^{P}=3/2^{-}$ and a $J^{P}=1/2^{-}$ states in every system.
We show their masses, variational parameters, the internal contribution, the relative lengths between quarks, and their lowest baryon-meson threshold in Tables \ref{nr1}-\ref{nr4}, respectively.

Among four systems, it is $J^{P}=1/2^{-}$ $bbbb\bar{c}$ state that is most likely to be stable against the strong decay according to the Tables \ref{nr1}-\ref{nr4}.
However, even this state is still much above the corresponding lowest baryon-meson threshold, and its binding energy $B_{T}=+253.4$ MeV.
Thus there are no bound states in these four systems, and they all are unstable states which can decay into a baryon and a meson through the strong interaction.

\subsubsection{Internal contributions}

Here, we take the $cccc\bar{c}$ system as an example.
According to Table \ref{nr1},
the masse of $J^{P}=3/2^{-}$ and $J^{P}=1/2^{-}$ $cccc\bar{c}$ states are 8144.6 MeV and 8193.2 MeV, respectively.
Meanwhile, their binding energy $B_{T}$ are $+344.6$ MeV and $+299.6$ MeV, respectively.
Thus, they are both obviously higher than the corresponding rearrangement baryon-meson thresholds.

For internal contributions, the quark contents of the pentaquark state are the same as the corresponding rearrangement decay threshold.
Thus, the quark mass term need not be considered.
Moreover, the contribution from the hyperfine potential term is negligible relative to the contributions from other terms.
As for the kinetic energy, the $J^{P}=3/2^{-}$ ($J^{P}=1/2^{-}$) pentaquark state has 942.4 (905.4) MeV, which
can be understood as the sum of three internal kinetic energies:
kinetic energies of the three internal $c-c$ between $(ccc)$, the $c-\bar{c}$, and the $(ccc)-(c\bar{c})$ pairs.
Accordingly, the sum of the internal kinetic energies of $\Omega_{ccc}\eta_{c}$ or $\Omega_{ccc}J/\psi$ states
comes from the three internal $c-c$ between $(ccc)$ and $c-\bar{c}$.
Therefore, $cccc\bar{c}$ system has an additional kinetic energy needed to bring the $\Omega_{ccc}\eta_{c}$ or $\Omega_{ccc}J/\psi$ into a compact configuration.
The actual kinetic energies of the $c-c$ of $(ccc)$ and $c-\bar{c}$ in the pentaquark state are smaller than those inside the $\Omega_{ccc}\eta_{c}$ and $\Omega_{ccc}J/\psi$ system.
Meanwhile, even if considering the additional kinetic energy between the $(ccc)-(c\bar{c})$ pairs, the total kinetic energies in the $cccc\bar{c}$ states are still smaller than that of the lowest baryon-meson threshold. The relative length between the pair $c-c$ ($c-\bar{c}$) are longer in the pentaquark.
Thus the contributions from $V^{C}$ are thought to be attractive but much smaller than the contribution from the meson-baryon threshold,
which is the main reason why these pentaquark states all have positive binding energies $B_{T}$. The $cccc\bar{c}$ states cannot bind into a compact configuration.

Compared to the $cccc\bar{c}$ states, the $V^{C}$ between the heavy quarks seems to be more attractive in the $bbbb\bar{b}$ states, which is also consistent with Refs. \cite{Park:2018wjk,Karliner:2017qjm}.
The reason comes from the smaller relative length between the pair $b-b$ ($b-\bar{b}$)
which is 0.55 times that of $c-c$ ($c-\bar{c}$) in pentaquark states.
These show that the relative heavy quark pairs ($bb$ and $b\bar{b}$) are much more compact than the relative light quark pairs ($cc$ and $c\bar{c}$).
However, the quark-antiquark distances are still longer than those of $\Upsilon$ and $\eta_{b}$,
and this leads to a smaller attraction in the $bbbb\bar{b}$ states.
Thus, the $bbbb\bar{b}$ states still have a large positive binding energy.

\subsubsection{The comparison with CMI model}
Let us compare the masses of three states calculated from the constituent quark model and from the CMI model \cite{An:2020jix}.
Here, we take the $bbbb\bar{c}$ system as an example.
In the CMI model, the masses of $J^{P}=3/2^{-}$ and $1/2^{-}$ states are constructed as the following formulas:

\begin{eqnarray}\label{Eq41}
M_{J^{P}=3/2^{-}}&=&\frac{3}{2}m_{bb}+m_{b\bar{c}}+\frac{7}{6}v_{bb}-\frac{1}{3}v_{b\bar{c}}, \nonumber\\
M_{J^{P}=1/2^{-}}&=&\frac{3}{2}m_{bb}+m_{b\bar{c}}+\frac{7}{6}v_{bb}+\frac{2}{3}v_{b\bar{c}},
\end{eqnarray}
where $m_{bb}$ and $m_{b\bar{c}}$ are the parameters which combined effective quark masses and color interaction between two quarks,
and $v_{bb}$ and $v_{b\bar{c}}$ are the parameters for the color-spin interaction.
These parameters are determined from the traditional hadron masses.
We can divide the internal contributions from constituent quark model and CMI model into the $b$ effective quark mass, $\bar{c}$ effective quark mass, and the color-spin interaction term, and then compare with them in Table \ref{nr3}.

In our opinion, the effective quark mass term (including the color interaction term) of the CMI model absorbs the quark mass term, confinement potential term, and kinetic term of the constituent quark model.
Here, we give some explanations of the division of the effective quark mass.
For the $-D$ term and the $V^{C}$ term,
it is divided into each quark by multiplying a factor of 1/2 \cite{Park:2018wjk}.
For the kinetic term,
it is divided according to their relative contribution depending on the mass of the single quark.
Similarly, the division of the $m_{b\bar{c}}$ in the CMI model is also dependedt on the mass of the single quark.

Now, we compare the values from the constituent quark model and the CMI model in Tables \ref{nr1}-\ref{nr4}.
Note that these parameters of both different models are determined from the traditional hadron masses and can describe the traditional hadron mass spectrum well.
These two different models used in the fully heavy pentaquark system have some existing differences.

On the whole,
the effective quark masses from the constituent quark model are systematically larger than those from the CMI model according to Tables \ref{nr1}-\ref{nr4}.
For example, the $b$ effective quark masses are 19433.4 MeV in the consistent quark model and about 400 MeV larger than that of the CMI model in the $bbbb\bar b$ state with $J^P=3/2^-$.
Meanwhile, we have noticed similar situations for the $qq\bar{Q}\bar{Q}$ system according to the CMI model \cite{Weng:2021hje} and the constituent quark model \cite{Park:2018wjk}.
It seems that the effective quark masses should have different values in the meson, the baryon, the tetraquark states, and the pentaquark states.

On the contrary, the color-spin terms from two different models have much similarity.
Thus the mass gaps of the CMI model are relatively reliable.

\subsection{$cccb\bar{c}$, $bbbc\bar{b}$, $cccb\bar{b}$, and $bbbc\bar{c}$ systems}

\begin{table*}
\caption{ The masses, variational parameters, the contribution from each term in the Hamiltonian, and the relative lengths between quarks for $cccb\bar{c}$ system and their baryon-meson thresholds.
The notations are same as those of Table. \ref{nr1}.
}\label{nr5}
\begin{lrbox}{\tablebox}
\renewcommand\arraystretch{1.55}
\renewcommand\tabcolsep{2.185pt}
\begin{tabular}{c|c|ccc|ccc|c|cr|ccc}
\midrule[1.5pt]
\toprule[0.5pt]
\multicolumn{1}{c}{$cccb\bar{c}$}&\multicolumn{5}{c}{The contribution from each term}&\multicolumn{2}{c|}{$V^{C}$}&\multirow{2}*{Overall}&\multicolumn{2}{c}{Present Work}&\multicolumn{2}{c}{CMI Model}\\
\Xcline{1-8}{0.5pt}\Xcline{10-13}{0.5pt}
\multicolumn{1}{r}{$J^{P}=\frac32^{-}$}&&Value&\multicolumn{1}{r}{$\Omega^{*}_{ccb}\eta_{c}$}&\multicolumn{1}{r|}{Difference}&$(i,j)$&\multicolumn{1}{r}{Vaule}
&\multicolumn{1}{c|}{$\Omega^{*}_{ccb}\eta_{c}$}&&Contribution&Value&Contribution&Value\\ \Xcline{1-13}{0.5pt}
\multicolumn{2}{c}{Mass}&\multicolumn{1}{r}{11443.7}&\multicolumn{1}{r}{11062.3}&381.4&(1,2)&-4.6&\multicolumn{1}{c|}{-45.4($\Omega^{*}_{ccb}$)}&\multirow{6}*{$c$-quark}&3$m_{c}$&5754.0&$\frac38$ $m_{cc}$&1189.3\\
\Xcline{1-5}{0.5pt}
\multirow{3}*{\makecell[c]{Variational\\ Parameters\\ (fm$^{-2}$)}}&$C_{11}$&\multicolumn{1}{r}{8.6}&\multicolumn{1}{r}{10.4}&&(1,3)&-4.6&
&&$[\frac{\textbf{p}^{2}_{x_{1}}}{2m'_{1}}+\frac{\textbf{p}^{2}_{x_{2}}}{2m'_{2}}]$+
$[\frac{m_{c}+m_{b}}{4m_{c}+m_{b}}\frac{\textbf{p}^{2}_{x_{4}}}{2m'_{4}}]$&\multirow{1}*{\makecell[r]{$523.1$\\ $+115.6$}}&$\frac{m_{c}}{m_{b}+m_{c}}$$\frac98$ $m_{cb}$&1802.4\\
&\multicolumn{1}{c|}{$C_{22}$}&\multicolumn{1}{r}{10.1}&\multicolumn{1}{r}{15.1}&&(2,3)&-4.6&&
&\multirow{3}*{\makecell[c]{$V^{C}(12)+V^{C}(13)+V^{C}(23)$\\$\frac{1}{2}[V^{C}(14)+V^{C}(14)+V^{C}(34)]$\\ $\frac{1}{2}[V^{C}(15)+V^{C}(23)+V^{C}(14)]$\\ $-\frac32D$ }}&\multirow{3}*{\makecell[r]{-13.9\\-20.4\\-20.8\\-1474.5}}&$\frac{1}{2}$$\frac98$ $m_{c\bar{c}}$&1726.1\\
&\multicolumn{1}{c|}{$C_{33}$}&\multicolumn{1}{r}{9.5}&\multicolumn{1}{r}{15.0}&&(1,4)&-13.6&\multicolumn{1}{c|}{-109.4($\Omega^{*}_{ccb}$)}&&&\\
\Xcline{1-5}{0.5pt}
\multicolumn{2}{c}{Quark Mass}&\multicolumn{1}{r}{13015.0}&\multicolumn{1}{r}{13015.0}&0.0&(2,4)&-13.6&\multicolumn{1}{c|}{-109.4($\Omega^{*}_{ccb}$)}&&&\\
\Xcline{10-13}{0.5pt}
\multicolumn{2}{c}{\multirow{2}{*}{\makecell[c]{Confinement\\ Potential}}}&\multicolumn{1}{r}{\multirow{2}{*}{-2548.3}}&\multicolumn{1}{r}{\multirow{2}{*}{-2936.8}}&\multicolumn{1}{c|}{\multirow{2}{*}{388.5}}&(3,4)&-13.6&&&Subtotal&4863.1&&4717.8\\
\Xcline{9-13}{0.5pt}
\multicolumn{2}{c}{\multirow{5}{*}{\makecell[c]{$V^{C}$ \\ Subtotal}}}&\multicolumn{1}{r}{\multirow{5}{*}{-90.8}}&\multicolumn{1}{r}{\multirow{5}{*}{-479.3}}&\multicolumn{1}{c|}{\multirow{5}{*}{388.5}}&(1,5)&-13.8&
&\multirow{5}*{$b$-quark}&\multirow{4}*{\makecell[c]{$m_{b}$\\ $\frac{m_{c}}{m_{c}+m_{b}}\frac{\textbf{p}^{2}_{x_{3}}}{2m'_{3}}$+$\frac{m_{c}}{m_{c}+m_{b}}\frac{3m_{c}}{4m_{c}+m_{b}}\frac{\textbf{p}^{2}_{x_{4}}}{2m'_{4}}$\\ $\frac{1}{2}[V^{C}(14)+V^{C}(24)+V^{C}(34)]$\\ $\frac{1}{2}V^{C}(45)$\\ $-\frac{1}{2}D$}}&\multirow{4}*{\makecell[r]{5343.0\\ 55.1\\+24.2\\-20.4\\2.7\\-491.5}}&$\frac{m_{b}}{m_{b}+m_{c}}$$\frac98$ $m_{cb}$&5415.7\\
\Xcline{1-5}{0.5pt}
\multicolumn{5}{c}{}&(2,5)&-13.8&&&&&$-\frac{m_{b}}{m_{c}+m_{b}}\frac{1}{8}m_{b\bar{c}}$&-592.9\\
\multicolumn{5}{c}{}&(3,5)&-13.8&-237.2($\eta_{c}$)&&&&\\
\multicolumn{5}{c}{}&(4,5)&5.4&&&&&\\
\Xcline{6-8}{0.5pt}\Xcline{10-13}{0.5pt}
\multicolumn{5}{c|}{}&\multicolumn{3}{c|}{Relative Lengths (fm)}&&Subtotal&4913.1&&4822.8\\
\Xcline{1-13}{0.5pt}
\multicolumn{2}{c}{\multirow{3}*{\makecell[c]{Kinetic \\ Energy}}}&\multicolumn{1}{r}{\multirow{3}{*}{937.8}}&\multicolumn{1}{r}{\multirow{3}{*}{1038.2}}&\multicolumn{1}{c|}{\multirow{3}{*}{-100.4}}&(1,2)&0.384&$0.349(\Omega^{*}_{ccb})$
&\multirow{5}*{$\bar{c}$-quark}&\multirow{4}*{\makecell[c]{$m_{c}$\\ $\frac{m_{b}}{m_{c}+m_{b}}\frac{\textbf{p}^{2}_{x_{3}}}{2m'_{3}}$+$\frac{m_{b}}{m_{c}+m_{b}}\frac{3m_{\bar{c}}}{4m_{c}+m_{\bar{b}}}\frac{\textbf{p}^{2}_{x_{4}}}{2m'_{4}}$\\ $\frac{1}{2}[V^{C}(15)+V^{C}(25)+V^{C}(35)]$\\ $\frac{1}{2}V^{C}(45)$\\ $-\frac{1}{2}D$}}&\multirow{4}*{\makecell[r]{1918.0\\ 153.4\\+67.4\\-20.8\\2.7\\-491.5}}&$\frac{1}{2}$$\frac98$ $m_{c\bar{c}}$&1726.1\\
\multicolumn{5}{c|}{}&(1,3)&0.384&&&&&$-\frac{m_{c}}{m_{c}+m_{b}}\frac{1}{8}m_{b\bar{c}}$&-197.3\\
\multicolumn{5}{c|}{}&(2,3)&0.384&&&&&\\
\multicolumn{5}{c|}{}&(1,4)&0.373&$0.305(\Omega^{*}_{ccb})$&&&\\
\Xcline{10-13}{0.5pt}
\multicolumn{5}{c|}{}&(2,4)&0.373&$0.305(\Omega^{*}_{ccb})$&&Subtotal&1629.2&&1528.8\\
\Xcline{9-13}{0.5pt}
\multicolumn{2}{c}{\multirow{2}*{\makecell[c]{CS \\ Interaction}}}&\multicolumn{1}{r}{\multirow{2}{*}{25.6}}&\multicolumn{1}{r}{\multirow{2}{*}{-54.0}}&\multicolumn{1}{c|}{\multirow{2}{*}{79.6}}&(3,4)&0.328&&\multirow{4}*{\makecell[c]{CS\\  Interaction}}&\multirow{3}*{\makecell[c]{$\frac{5}{8}[V^{S}(12)+V^{S}(13)+V^{S}(23)]$\\$-\frac{5}{24}[V^{S}(14)+V^{S}(24)+V^{S}(34)]$
\\$\frac{5}{24}[V^{S}(15)+V^{S}(25)+V^{S}(35)]$\\$-\frac{1}{8}V^{S}(45)$}}&\multirow{3}*{\makecell[r]{32.3\\5.0\\-10.8\\-1.0}}
&\multirow{3}*{\makecell[c]{$\frac{5}{8}v_{cc}$\\$\frac{5}{24}v_{cb}$\\$-\frac{5}{24}v_{c\bar{c}}$\\$-\frac{1}{24}v_{b\bar{c}}$}}
&\multirow{3}*{\makecell[c]{35.5\\6.6\\-17.7\\-2.0}}\\
\multicolumn{5}{c|}{}&(1,5)&0.369&&&&\\
\multicolumn{2}{c}{\multirow{3}*{\makecell[c]{Total \\ Contribution}}}&\multicolumn{1}{c}{\multirow{3}*{\makecell[c]{872.6}}}&\multicolumn{1}{r}{\multirow{3}*{\makecell[c]{504.9}}}&\multicolumn{1}{c|}{\multirow{3}*{\makecell[c]{367.7}}}&(2,5)&0.369&&&&\\
\Xcline{10-13}{0.5pt}
\multicolumn{5}{c|}{}&(3,5)&0.369&$0.290(\eta_{c})$&&Subtotal&25.6&&22.3\\
\Xcline{9-13}{0.5pt}
\multicolumn{5}{c|}{}&(4,5)&0.354&&Total&&11431.0&&11091.5\\
\toprule[1.2pt]
\multicolumn{1}{r}{$J^{P}=\frac12^{-}$}&&Value&\multicolumn{1}{r}{$\Omega_{ccb}\eta_{c}$}&\multicolumn{1}{r|}{Difference}&$(i,j)$&\multicolumn{1}{r}{Vaule}
&\multicolumn{1}{c|}{$\Omega_{ccb}\eta_{c}$}&&Contribution&Value&Contribution&Value\\ \Xcline{1-13}{0.5pt}
\multicolumn{2}{c}{Mass}&\multicolumn{1}{r}{11438.2}&11028.0&410.2&(1,2)&-0.8&\multicolumn{1}{c|}{-52.8($\Omega_{ccb}$)}&\multirow{6}*{$c$-quark}&3$m_{c}$&5754.0&$\frac38$ $m_{cc}$&1189.3\\
\Xcline{1-5}{0.5pt}
\multirow{3}*{\makecell[c]{Variational\\ Parameters\\ (fm$^{-2}$)}}&$C_{11}$&\multicolumn{1}{r}{8.4}&\multicolumn{1}{r}{10.8}&&(1,3)&-0.8&
&&$[\frac{\textbf{p}^{2}_{x_{1}}}{2m'_{1}}+\frac{\textbf{p}^{2}_{x_{2}}}{2m'_{2}}]$+$[\frac{m_{c}+m_{b}}{4m_{c}+m_{b}}\frac{\textbf{p}^{2}_{x_{4}}}{2m'_{4}}]$
&\multirow{1}*{\makecell[r]{$509.4$\\ $+145.1$}}&$\frac{m_{c}}{m_{b}+m_{c}}$$\frac98$ $m_{cb}$&1802.4\\
&\multicolumn{1}{c|}{$C_{22}$}&\multicolumn{1}{r}{7.4}&\multicolumn{1}{r}{16.1}&&(2,3)&-0.8&&
&\multirow{3}*{\makecell[c]{$V^{C}(12)+V^{C}(13)+V^{C}(23)$\\$\frac{1}{2}[V^{C}(14)+V^{C}(14)+V^{C}(34)]$\\ $\frac{1}{2}[V^{C}(15)+V^{C}(23)+V^{C}(14)]$\\ $-\frac32D$ }}&\multirow{3}*{\makecell[r]{-2.3\\-25.1\\-24.1\\-1474.5}}&$\frac{1}{2}$$\frac98$ $m_{c\bar{b}}$&1726.1\\
&\multicolumn{1}{c|}{$C_{33}$}&\multicolumn{1}{r}{11.9}&\multicolumn{1}{r}{15.0}&&(1,4)&-16.8&&&&\\
\Xcline{1-5}{0.5pt}
\multicolumn{2}{c}{Quark Mass}&\multicolumn{1}{r}{13015.0}&\multicolumn{1}{r}{13015.0}&0.0&(2,4)&-16.8&\multicolumn{1}{c|}{-109.4($\Omega_{ccb}$)}&&&\\
\Xcline{10-13}{0.5pt}
\multicolumn{2}{c}{\multirow{2}{*}{\makecell[c]{Confinement\\ Potential}}}&\multicolumn{1}{r}{\multirow{2}{*}{-2563.7}}&\multicolumn{1}{r}{\multirow{2}{*}{-2966.3}}&\multicolumn{1}{c|}{\multirow{2}{*}{402.6}}&(3,4)&-16.8&
\multicolumn{1}{c|}{-109.4($\Omega_{ccb}$)}&&Subtotal&4882.5&&4717.8\\
\Xcline{9-13}{0.5pt}
\multicolumn{2}{c}{\multirow{5}{*}{\makecell[c]{$V^{C}$ \\ Subtotal}}}&\multicolumn{1}{r}{\multirow{5}{*}{-106.2}}&\multicolumn{1}{r}{\multirow{5}{*}{-508.8}}&\multicolumn{1}{c|}{\multirow{5}{*}{402.6}}&(1,5)&-16.1&&\multirow{5}*{$b$-quark}&\multirow{4}*{\makecell[c]{$m_{b}$\\ $\frac{m_{c}}{m_{c}+m_{b}}\frac{\textbf{p}^{2}_{x_{3}}}{2m'_{3}}$+$\frac{m_{c}}{m_{c}+m_{b}}\frac{3m_{c}}{4m_{c}+m_{b}}\frac{\textbf{p}^{2}_{x_{4}}}{2m'_{4}}$\\ $\frac{1}{2}[V^{C}(14)+V^{C}(24)+V^{C}(34)]$\\ $\frac{1}{2}V^{C}(45)$\\ $-\frac{1}{2}D$}}&\multirow{4}*{\makecell[r]{5343.0\\ 40.3\\+30.4\\-25.1\\-2.7\\-491.5}}&$\frac{m_{b}}{m_{b}+m_{c}}$$\frac98$ $m_{cb}$&5415.7\\
\Xcline{1-5}{0.5pt}
\multicolumn{5}{c}{}&(2,5)&-16.1&&&&&$-\frac{m_{b}}{m_{c}+m_{b}}\frac{1}{8}m_{b\bar{c}}$&-592.9\\
\multicolumn{5}{c}{}&(3,5)&-16.1&\multicolumn{1}{c|}{-237.2($\eta_{c}$)}&&&&\\
\multicolumn{5}{c}{}&(4,5)&-5.4&&&&&\\
\Xcline{6-8}{0.5pt}\Xcline{10-13}{0.5pt}
\multicolumn{5}{c|}{}&\multicolumn{3}{c|}{Relative Lengths (fm)}&&Subtotal&4897.4&&4822.8\\
\Xcline{1-13}{0.5pt}
\multicolumn{2}{c}{\multirow{3}*{\makecell[c]{Kinetic \\ Energy}}}&\multicolumn{1}{r}{\multirow{3}{*}{922.0}}&\multicolumn{1}{r}{\multirow{3}{*}{1067.7}}&\multicolumn{1}{c|}{\multirow{3}{*}{-145.7}}&(1,2)&0.389&$0.342(\Omega_{ccb})$&\multirow{5}*{$\bar{c}$-quark}&\multirow{4}*{\makecell[c]{$m_{b}$\\ $\frac{m_{b}}{m_{c}+m_{b}}\frac{\textbf{p}^{2}_{x_{3}}}{2m'_{3}}$+$\frac{m_{b}}{m_{c}+m_{b}}\frac{3m_{\bar{c}}}{4m_{c}+m_{\bar{b}}}\frac{\textbf{p}^{2}_{x_{4}}}{2m'_{4}}$\\ $\frac{1}{2}[V^{C}(15)+V^{C}(25)+V^{C}(35)]$\\ $\frac{1}{2}V^{C}(45)$\\ $-\frac{1}{2}D$}}&\multirow{4}*{\makecell[r]{1918.0\\ 112.2\\+84.6\\-24.1\\-2.7\\-491.5}}&$\frac{1}{2}$$\frac98$ $m_{c\bar{c}}$&1726.1\\
\multicolumn{5}{c|}{}&(1,3)&0.389&&&&&$-\frac{m_{c}}{m_{c}+m_{b}}\frac{1}{8}m_{b\bar{c}}$&-197.3\\
\multicolumn{5}{c|}{}&(2,3)&0.389&&&&&\\
\multicolumn{5}{c|}{}&(1,4)&0.370&$0.297(\Omega_{ccb})$&&&\\
\Xcline{10-13}{0.5pt}
\multicolumn{5}{c|}{}&(2,4)&0.370&$0.297(\Omega_{ccb})$&&Subtotal&1596.5&&1528.8\\
\Xcline{9-13}{0.5pt}
\multicolumn{2}{c}{\multirow{2}*{\makecell[c]{CS \\ Interaction}}}&\multicolumn{1}{r}{\multirow{2}{*}{32.9}}&\multicolumn{1}{r}{\multirow{2}{*}{-88.4}}&\multicolumn{1}{c|}{\multirow{2}{*}{121.3}}&(3,4)&0.370&&\multirow{4}*{\makecell[c]{CS\\  Interaction}}&\multirow{3}*{\makecell[c]{$\frac{5}{8}[V^{S}(12)+V^{S}(13)+V^{S}(23)]$
\\\\$+\frac{3}{8}V^{S}(45)$}}&\multirow{3}*{\makecell[c]{30.4\\\\2.5}}
&\multirow{3}*{\makecell[c]{$\frac{5}{8}v_{cc}$\\\\$\frac{1}{8}v_{b\bar{c}}$}}
&\multirow{3}*{\makecell[c]{35.5\\\\5.9}}\\
\multicolumn{5}{c|}{}&(1,5)&0.371&&&&\\
\multicolumn{2}{c}{\multirow{3}*{\makecell[c]{Total \\ Contribution}}}&\multicolumn{1}{r}{\multirow{3}*{\makecell[c]{848.7}}}&\multicolumn{1}{r}{\multirow{3}*{\makecell[c]{470.5}}}&\multicolumn{1}{c|}{\multirow{3}*{\makecell[c]{378.2}}}&(2,5)&0.371&&&&\\
\Xcline{10-13}{0.5pt}
\multicolumn{5}{c|}{}&(3,5)&0.371&0.290($\eta_{c}$)&&Subtotal&32.9&&41.4\\
\Xcline{9-13}{0.5pt}
\multicolumn{5}{c|}{}&(4,5)&0.414&&Total&&11409.3&&11110.8\\
\toprule[0.5pt]
\toprule[1.0pt]
\end{tabular}
\end{lrbox}\scalebox{0.90}{\usebox{\tablebox}}
\end{table*}

\begin{table*}
\caption{ The masses, variational parameters, the contribution from each term in the Hamiltonian, and the relative lengths between quarks for $bbbc\bar{b}$ system and their baryon-meson thresholds.
The notations are same as those of Table. \ref{nr1}.
}\label{nr6}
\centering
\begin{lrbox}{\tablebox}
\renewcommand\arraystretch{1.55}
\renewcommand\tabcolsep{2.15pt}
\begin{tabular}{c|c|ccc|ccc|c|cr|ccc}
\midrule[1.5pt]
\toprule[0.5pt]
\multicolumn{1}{c}{$bbbc\bar{b}$}&\multicolumn{5}{c}{The contribution from each term}&\multicolumn{2}{c|}{$V^{C}$}&\multirow{2}*{Overall}&\multicolumn{2}{c}{Present Work}&\multicolumn{2}{c}{CMI Model}\\
\Xcline{1-8}{0.5pt}\Xcline{10-13}{0.5pt}
\multicolumn{1}{r}{$J^{P}=\frac32^{-}$}&&Value&\multicolumn{1}{r}{$\Omega^{*}_{bbc}\eta_{b}$}&\multicolumn{1}{r|}{Difference}&$(i,j)$&\multicolumn{1}{r}{Vaule}
&\multicolumn{1}{c|}{$\Omega^{*}_{bbc}\eta_{b}$}&&Contribution&Value&Contribution&Value\\ \Xcline{1-13}{0.5pt}
\multicolumn{2}{c}{Mass}&\multicolumn{1}{r}{21091.6}&\multicolumn{1}{r}{20662.2}&429.4&(1,2)&-46.2&\multicolumn{1}{c|}{-235.9($\Omega^{*}_{bbc}$)}&\multirow{6}*{$b$-quark}
&3$m_{b}$&16029.0&$\frac38$ $m_{bb}$&3573.6\\
\Xcline{1-5}{0.5pt}
\multirow{3}*{\makecell[c]{Variational\\ Parameters\\ (fm$^{-2}$)}}&$C_{11}$&\multicolumn{1}{r}{20.6}&\multicolumn{1}{r}{26.0}&&(1,3)&-46.2&
&&$[\frac{\textbf{p}^{2}_{x_{1}}}{2m'_{1}}+\frac{\textbf{p}^{2}_{x_{2}}}{2m'_{2}}]$+
$[\frac{m_{c}+m_{b}}{4m_{b}+m_{c}}\frac{\textbf{p}^{2}_{x_{4}}}{2m'_{4}}]$&\multirow{1}*{\makecell[r]{$450.9$\\ $+80.0$}}&$\frac{m_{b}}{m_{b}+m_{c}}$$\frac98$ $m_{cb}$&5416.5\\
&\multicolumn{1}{c|}{$C_{22}$}&\multicolumn{1}{r}{15.5}&\multicolumn{1}{r}{14.2}&&(2,3)&-46.2&&
&\multirow{3}*{\makecell[c]{$V^{C}(12)+V^{C}(13)+V^{C}(23)$\\$\frac{1}{2}[V^{C}(14)+V^{C}(14)+V^{C}(34)]$\\ $\frac{1}{2}[V^{C}(15)+V^{C}(23)+V^{C}(14)]$\\ $-\frac32D$ }}&\multirow{3}*{\makecell[r]{-138.5\\-168.4\\-177.3\\-1474.5}}&$\frac{1}{2}$$\frac98$ $m_{b\bar{b}}$&5312.8\\
&\multicolumn{1}{c|}{$C_{33}$}&\multicolumn{1}{r}{18.3}&\multicolumn{1}{r}{57.4}&&(1,4)&-112.2&\multicolumn{1}{c|}{-131.2($\Omega^{*}_{bbc}$)}&&&\\
\Xcline{1-5}{0.5pt}
\multicolumn{2}{c}{Quark Mass}&\multicolumn{1}{r}{23290.0}&\multicolumn{1}{r}{23290.0}&0.0&(2,4)&-112.2&\multicolumn{1}{c|}{-131.2($\Omega^{*}_{bbc}$)}&&&\\
\Xcline{10-13}{0.5pt}
\multicolumn{2}{c}{\multirow{2}{*}{\makecell[c]{Confinement\\ Potential}}}&\multicolumn{1}{r}{\multirow{2}{*}{-3256.0}}&\multicolumn{1}{r}{\multirow{2}{*}{-3834.9}}&\multicolumn{1}{c|}{\multirow{2}{*}{578.9}}&(3,4)&-112.2&
&&Subtotal&14601.2&&14302.9\\
\Xcline{9-13}{0.5pt}
\multicolumn{2}{c}{\multirow{5}{*}{\makecell[c]{$V^{C}$ \\ Subtotal}}}&\multicolumn{1}{r}{\multirow{5}{*}{-798.5}}&\multicolumn{1}{r}{\multirow{5}{*}{-1377.4}}&\multicolumn{1}{c|}{\multirow{5}{*}{578.9}}&(1,5)&-118.2&
&\multirow{5}*{$c$-quark}&\multirow{4}*{\makecell[c]{$m_{c}$\\ $\frac{m_{b}}{m_{c}+m_{b}}\frac{\textbf{p}^{2}_{x_{3}}}{2m'_{3}}$+$\frac{m_{b}}{m_{c}+m_{b}}\frac{3m_{b}}{4m_{b}+m_{c}}\frac{\textbf{p}^{2}_{x_{4}}}{2m'_{4}}$\\ $\frac{1}{2}[V^{C}(14)+V^{C}(24)+V^{C}(34)]$\\ $\frac{1}{2}V^{C}(45)$\\ $-\frac{1}{2}D$}}&\multirow{4}*{\makecell[r]{1918.0\\ 236.1\\+129.8\\-168.4\\15.6\\-491.5}}&$\frac{m_{c}}{m_{b}+m_{c}}$$\frac98$ $m_{cb}$&1802.7\\
\Xcline{1-5}{0.5pt}
\multicolumn{5}{c}{}&(2,5)&-118.2&&&&&$\frac{-m_{c}}{m_{c}+m_{b}}\frac{1}{8}m_{c\bar{b}}$&-197.4\\
\multicolumn{5}{c}{}&(3,5)&-118.2&-879.1($\eta_{b}$)&&&&\\
\multicolumn{5}{c}{}&(4,5)&31.3&&&&&\\
\Xcline{6-8}{0.5pt}\Xcline{10-13}{0.5pt}
\multicolumn{5}{c|}{}&\multicolumn{3}{c|}{Relative Lengths (fm)}&&Subtotal&1639.6&&1605.3\\
\Xcline{1-13}{0.5pt}
\multicolumn{2}{c}{\multirow{3}*{\makecell[c]{Kinetic \\ Energy}}}&\multicolumn{1}{r}{\multirow{3}{*}{1028.0}}&\multicolumn{1}{r}{\multirow{3}{*}{1251.3}}&\multicolumn{1}{c|}{\multirow{3}{*}{-223.3}}&(1,2)&0.248&$0.221(\Omega^{*}_{bbc})$
&\multirow{5}*{$\bar{b}$-quark}&\multirow{4}*{\makecell[c]{$m_{b}$\\ $\frac{m_{c}}{m_{c}+m_{b}}\frac{\textbf{p}^{2}_{x_{3}}}{2m'_{3}}$+$\frac{m_{c}}{m_{c}+m_{b}}\frac{3m_{b}}{4m_{b}+m_{c}}\frac{\textbf{p}^{2}_{x_{4}}}{2m'_{4}}$\\ $\frac{1}{2}[V^{C}(15)+V^{C}(25)+V^{C}(35)]$\\ $\frac{1}{2}V^{C}(45)$\\ $-\frac{1}{2}D$}}&\multirow{4}*{\makecell[r]{5343.0\\84.8\\+46.6\\-177.3\\15.6\\-491.5}}&$\frac{1}{2}$$\frac98$ $m_{b\bar{b}}$&5312.8\\
\multicolumn{5}{c|}{}&(1,3)&0.248&&&&&$\frac{-m_{b}}{m_{c}+m_{b}}\frac{1}{8}m_{c\bar{b}}$&-593.0\\
\multicolumn{5}{c|}{}&(2,3)&0.248&&&&&\\
\multicolumn{5}{c|}{}&(1,4)&0.269&$0.281(\Omega^{*}_{bbc})$&&&\\
\Xcline{10-13}{0.5pt}
\multicolumn{5}{c|}{}&(2,4)&0.269&$0.281(\Omega^{*}_{bbc})$&&Subtotal&4821.2&&4719.8\\
\Xcline{9-13}{0.5pt}
\multicolumn{2}{c}{\multirow{2}*{\makecell[c]{CS \\ Interaction}}}&\multicolumn{1}{r}{\multirow{2}{*}{19.8}}&\multicolumn{1}{r}{\multirow{2}{*}{-44.2}}&\multicolumn{1}{c|}{\multirow{2}{*}{64.0}}&(3,4)&0.269&&\multirow{4}*{\makecell[c]{CS\\  Interaction}}&\multirow{3}*{\makecell[c]{$\frac{5}{8}[V^{S}(12)+V^{S}(13)+V^{S}(23)]$\\$-\frac{5}{24}[V^{S}(14)+V^{S}(24)+V^{S}(34)]$
\\$\frac{5}{24}[V^{S}(15)+V^{S}(25)+V^{S}(35)]$\\$-\frac{1}{8}V^{S}(45)$}}&\multirow{3}*{\makecell[r]{18.4\\8.5\\-5.5\\-1.5}}
&\multirow{3}*{\makecell[c]{$\frac{5}{8}v_{bb}$\\$\frac{5}{24}v_{cb}$\\$-\frac{5}{24}v_{b\bar{b}}$\\$-\frac{1}{24}v_{c\bar{b}}$}}
&\multirow{3}*{\makecell[c]{19.2\\6.6\\-9.6\\-2.0}}\\
\multicolumn{5}{c|}{}&(1,5)&0.264&&&&\\
\multicolumn{2}{c}{\multirow{3}*{\makecell[c]{Total \\ Contribution}}}&\multicolumn{1}{r}{\multirow{3}*{\makecell[c]{249.5}}}&\multicolumn{1}{r}{\multirow{3}*{\makecell[c]{-170.3}}}&\multicolumn{1}{c|}{\multirow{3}*{\makecell[c]{419.6}}}
&(2,5)&0.264&&&&\\
\Xcline{10-13}{0.5pt}
\multicolumn{5}{c|}{}&(3,5)&0.264&$0.148(\eta_{b})$&&Subtotal&19.8&&14.2\\
\Xcline{9-13}{0.5pt}
\multicolumn{5}{c|}{}&(4,5)&0.286&&Total&&21081.8&&20641.1\\
\toprule[1.2pt]
\multicolumn{1}{r}{$J^{P}=\frac12^{-}$}&&Value&\multicolumn{1}{r}{$\Omega_{bbc}\eta_{b}$}&\multicolumn{1}{r|}{Difference}&$(i,j)$&\multicolumn{1}{r}{Vaule}
&\multicolumn{1}{c|}{$\Omega_{bbc}\eta_{b}$}&&Contribution&Value&Contribution&Value\\ \Xcline{1-13}{0.5pt}
\multicolumn{2}{c}{Mass}&\multicolumn{1}{r}{21079.0}&\multicolumn{1}{r}{20623.2}&455.8&(1,2)&-38.3&\multicolumn{1}{c|}{-243.2($\Omega_{bbc}$)}&\multirow{6}*{$b$-quark}
&3$m_{b}$&16029.0&$\frac38$ $m_{bb}$&3573.6\\
\Xcline{1-5}{0.5pt}
\multirow{3}*{\makecell[c]{Variational\\ Parameters\\ (fm$^{-2}$)}}&$C_{11}$&\multicolumn{1}{r}{17.8}&\multicolumn{1}{r}{26.8}&&(1,3)&-38.3&
&&$[\frac{\textbf{p}^{2}_{x_{1}}}{2m'_{1}}+\frac{\textbf{p}^{2}_{x_{2}}}{2m'_{2}}]$+
$[\frac{m_{c}+m_{b}}{4m_{b}+m_{c}}\frac{\textbf{p}^{2}_{x_{4}}}{2m'_{4}}]$&\multirow{1}*{\makecell[r]{$388.9$\\ $+101.1$}}&$\frac{m_{b}}{m_{b}+m_{c}}$$\frac98$ $m_{cb}$&5416.5\\
&\multicolumn{1}{c|}{$C_{22}$}&\multicolumn{1}{r}{14.8}&\multicolumn{1}{r}{15.2}&&(2,3)&-38.3&&
&\multirow{3}*{\makecell[c]{$V^{C}(12)+V^{C}(13)+V^{C}(23)$\\$\frac{1}{2}[V^{C}(14)+V^{C}(14)+V^{C}(34)]$\\ $\frac{1}{2}[V^{C}(15)+V^{C}(23)+V^{C}(14)]$\\ $-\frac32D$ }}&\multirow{3}*{\makecell[r]{-115.0\\-180.0\\-183.5\\-1474.5}}&$\frac{1}{2}$$\frac98$ $m_{b\bar{b}}$&5312.8\\
&\multicolumn{1}{c|}{$C_{33}$}&\multicolumn{1}{r}{23.1}&\multicolumn{1}{r}{57.4}&&(1,4)&-120.0&\multicolumn{1}{c|}{-145.0($\Omega_{bbc}$)}&&&\\
\Xcline{1-5}{0.5pt}
\multicolumn{2}{c}{Quark Mass}&\multicolumn{1}{r}{23290.0}&\multicolumn{1}{r}{23290.0}&0.0&(2,4)&-120.0&\multicolumn{1}{c|}{-145.0($\Omega_{bbc}$)}&&&\\
\Xcline{10-13}{0.5pt}
\multicolumn{2}{c}{\multirow{2}{*}{\makecell[c]{Confinement\\ Potential}}}&\multicolumn{1}{r}{\multirow{2}{*}{-3270.6}}&\multicolumn{1}{r}{\multirow{2}{*}{-3869.8}}&\multicolumn{1}{c|}{\multirow{2}{*}{599.2}}&(3,4)&-120.0&
&&Subtotal&14601.2&&14302.9\\
\Xcline{9-13}{0.5pt}
\multicolumn{2}{c}{\multirow{5}{*}{\makecell[c]{$V^{C}$ \\ Subtotal}}}&\multicolumn{1}{r}{\multirow{5}{*}{-813.0}}&\multicolumn{1}{r}{\multirow{5}{*}{-1412.3}}&\multicolumn{1}{c|}{\multirow{5}{*}{599.2}}&(1,5)&-122.3&
&\multirow{5}*{$c$-quark}&\multirow{4}*{\makecell[c]{$m_{c}$\\ $\frac{m_{b}}{m_{c}+m_{b}}\frac{\textbf{p}^{2}_{x_{3}}}{2m'_{3}}$+$\frac{m_{b}}{m_{c}+m_{b}}\frac{3m_{b}}{4m_{b}+m_{c}}\frac{\textbf{p}^{2}_{x_{4}}}{2m'_{4}}$\\ $\frac{1}{2}[V^{C}(14)+V^{C}(24)+V^{C}(34)]$\\ $\frac{1}{2}V^{C}(45)$\\ $-\frac{1}{2}D$}}&\multirow{4}*{\makecell[r]{1918.0\\ 225.4\\+164.2\\-183.5\\14.4\\-491.5}}&$\frac{m_{c}}{m_{b}+m_{c}}$$\frac98$ $m_{cb}$&1802.7\\
\Xcline{1-5}{0.5pt}
\multicolumn{5}{c}{}&(2,5)&-122.3&&&&&$\frac{-m_{c}}{m_{c}+m_{b}}\frac{1}{8}m_{c\bar{b}}$&-197.4\\
\multicolumn{5}{c}{}&(3,5)&-122.3&-879.1($\eta_{b}$)&&&&\\
\multicolumn{5}{c}{}&(4,5)&28.9&&&&&\\
\Xcline{6-8}{0.5pt}\Xcline{10-13}{0.5pt}
\multicolumn{5}{c|}{}&\multicolumn{3}{c|}{Relative Lengths (fm)}&&Subtotal&1639.6&&1605.3\\
\Xcline{1-13}{0.5pt}
\multicolumn{2}{c}{\multirow{3}*{\makecell[c]{Kinetic \\ Energy}}}&\multicolumn{1}{r}{\multirow{3}{*}{1019.4}}&\multicolumn{1}{r}{\multirow{3}{*}{1286.4}}&\multicolumn{1}{c|}{\multirow{3}{*}{-267.0}}&(1,2)&0.267&$0.217(\Omega_{bbc})$
&\multirow{5}*{$\bar{b}$-quark}&\multirow{4}*{\makecell[c]{$m_{b}$\\ $\frac{m_{c}}{m_{c}+m_{b}}\frac{\textbf{p}^{2}_{x_{3}}}{2m'_{3}}$+$\frac{m_{c}}{m_{c}+m_{b}}\frac{3m_{b}}{4m_{b}+m_{c}}\frac{\textbf{p}^{2}_{x_{4}}}{2m'_{4}}$\\ $\frac{1}{2}[V^{C}(15)+V^{C}(25)+V^{C}(35)]$\\ $\frac{1}{2}V^{C}(45)$\\ $-\frac{1}{2}D$}}&\multirow{4}*{\makecell[r]{5343.0\\80.9\\+58.9\\-183.5\\14.4\\-491.5}}&$\frac{1}{2}$$\frac98$ $m_{b\bar{b}}$&5312.8\\
\multicolumn{5}{c|}{}&(1,3)&0.267&&&&&$\frac{-m_{b}}{m_{c}+m_{b}}\frac{1}{8}m_{c\bar{b}}$&-593.0\\
\multicolumn{5}{c|}{}&(2,3)&0.267&&&&&\\
\multicolumn{5}{c|}{}&(1,4)&0.263&$0.272(\Omega_{bbc})$&&&\\
\Xcline{10-13}{0.5pt}
\multicolumn{5}{c|}{}&(2,4)&0.263&$0.272(\Omega_{bbc})$&&Subtotal&4821.2&&4719.8\\
\Xcline{9-13}{0.5pt}
\multicolumn{2}{c}{\multirow{2}*{\makecell[c]{CS \\ Interaction}}}&\multicolumn{1}{r}{\multirow{2}{*}{20.7}}&\multicolumn{1}{r}{\multirow{2}{*}{-83.3}}&\multicolumn{1}{c|}{\multirow{2}{*}{104.0}}&(3,4)&0.263&&\multirow{4}*{\makecell[c]{CS\\  Interaction}}&\multirow{3}*{\makecell[c]{$\frac{5}{8}[V^{S}(12)+V^{S}(13)+V^{S}(23)]$
\\\\$+\frac{3}{8}V^{S}(45)$}}&\multirow{3}*{\makecell[c]{16.2\\\\4.5}}
&\multirow{3}*{\makecell[c]{$\frac{5}{8}v_{bb}$\\\\$\frac{1}{8}v_{c\bar{b}}$}}
&\multirow{3}*{\makecell[c]{19.2\\\\5.9}}\\
\multicolumn{5}{c|}{}&(1,5)&0.261&&&&\\
\multicolumn{2}{c}{\multirow{3}*{\makecell[c]{Total \\ Contribution}}}&\multicolumn{1}{r}{\multirow{3}*{\makecell[c]{227.1}}}&\multicolumn{1}{r}{\multirow{3}*{\makecell[c]{-209.3}}}&\multicolumn{1}{c|}{\multirow{3}*{\makecell[c]{436.4}}}
&(2,5)&0.261&&&&\\
\Xcline{10-13}{0.5pt}
\multicolumn{5}{c|}{}&(3,5)&0.261&$0.148(\eta_{b})$&&Subtotal&20.7&&25.1\\
\Xcline{9-13}{0.5pt}
\multicolumn{5}{c|}{}&(4,5)&0.292&&Total&&21082.7&&20653.1\\
\toprule[0.5pt]
\toprule[1.0pt]
\end{tabular}
\end{lrbox}\scalebox{0.898}{\usebox{\tablebox}}
\end{table*}

\begin{table*}
\caption{ The masses, variational parameters, the contribution from each term in the Hamiltonian, and the relative lengths between quarks for $cccb\bar{b}$ system and their baryon-meson thresholds.
The notations are same as those of Table. \ref{nr1}.
}\label{nr7}
\begin{lrbox}{\tablebox}
\renewcommand\arraystretch{1.55}
\renewcommand\tabcolsep{2.185pt}
\begin{tabular}{c|c|ccc|ccc|c|cr|ccc}
\midrule[1.5pt]
\toprule[0.5pt]
\multicolumn{1}{c}{$cccb\bar{b}$}&\multicolumn{5}{c}{The contribution from each term}&\multicolumn{2}{c|}{$V^{C}$}&\multirow{2}*{Overall}&\multicolumn{2}{c}{Present Work}&\multicolumn{2}{c}{CMI Model}\\
\Xcline{1-8}{0.5pt}\Xcline{10-13}{0.5pt}
\multicolumn{1}{r}{$J^{P}=\frac32^{-}$}&&Value&\multicolumn{1}{r}{$\Omega_{ccc}\eta_{b}$}&\multicolumn{1}{r|}{Difference}&$(i,j)$&\multicolumn{1}{r}{Vaule}
&\multicolumn{1}{c|}{$\Omega_{ccc}\eta_{b}$}&&Contribution&Value&Contribution&Value\\ \Xcline{1-13}{0.5pt}
\multicolumn{2}{c}{Mass}&\multicolumn{1}{r}{14687.2}&14190.4&496.8&(1,2)&-2.4&\multicolumn{1}{c|}{-22.7($\Omega_{ccc}$)}&\multirow{6}*{$c$-quark}&3$m_{c}$&5754.0&$\frac38$ $m_{cc}$&1189.3\\
\Xcline{1-5}{0.5pt}
\multirow{3}*{\makecell[c]{Variational\\ Parameters\\ (fm$^{-2}$)}}&$C_{11}$&\multicolumn{1}{r}{8.7}&\multicolumn{1}{r}{9.3}&&(1,3)&-2.4&
\multicolumn{1}{c|}{-22.7($\Omega_{ccc}$)}&&$[\frac{\textbf{p}^{2}_{x_{1}}}{2m'_{1}}+\frac{\textbf{p}^{2}_{x_{2}}}{2m'_{2}}]$+
$[\frac{2m_{\bar{b}}}{3m_{c}+2m_{\bar{b}}}\frac{\textbf{p}^{2}_{x_{4}}}{2m'_{4}}]$&\multirow{1}*{\makecell[r]{$526.9$\\ $+165.7$}}&$\frac{m_{c}}{m_{b}+m_{c}}$$\frac98$ $m_{cb}$&1802.4\\
&\multicolumn{1}{c|}{$C_{22}$}&\multicolumn{1}{r}{15.8}&\multicolumn{1}{r}{57.4}&&(2,3)&-2.4&\multicolumn{1}{c|}{-22.7($\Omega_{ccc}$)}&
&\multirow{3}*{\makecell[c]{$V^{C}(12)+V^{C}(13)+V^{C}(23)$\\$\frac{1}{2}[V^{C}(14)+V^{C}(14)+V^{C}(34)]$\\ $\frac{1}{2}[V^{C}(15)+V^{C}(23)+V^{C}(14)]$\\ $-\frac32D$ }}&\multirow{3}*{\makecell[r]{-7.2\\-78.6\\-78.6\\-1474.5}}&$\frac{m_{c}}{m_{b}+m_{c}}$$\frac98$ $m_{c\bar{b}}$&1788.4\\
&\multicolumn{1}{c|}{$C_{33}$}&\multicolumn{1}{r}{13.6}&&&(1,4)&-52.4&&&&\\
\Xcline{1-5}{0.5pt}
\multicolumn{2}{c}{Quark Mass}&\multicolumn{1}{r}{16440.0}&\multicolumn{1}{r}{16440.0}&0.0&(2,4)&-52.4&&&&\\
\Xcline{10-13}{0.5pt}
\multicolumn{2}{c}{\multirow{2}{*}{\makecell[c]{Confinement\\ Potential}}}&\multicolumn{1}{r}{\multirow{2}{*}{-2746.5}}&\multicolumn{1}{r}{\multirow{2}{*}{3404.6}}&\multicolumn{1}{c|}{\multirow{2}{*}{658.1}}&(3,4)&-52.4&&&Subtotal&4807.7&&4780.1\\
\Xcline{9-13}{0.5pt}
\multicolumn{2}{c}{\multirow{5}{*}{\makecell[c]{$V^{C}$ \\ Subtotal}}}&\multicolumn{1}{r}{\multirow{5}{*}{-289.0}}&\multicolumn{1}{r}{\multirow{5}{*}{-947.1}}&\multicolumn{1}{c|}{\multirow{5}{*}{658.1}}&(1,5)&-52.4&&\multirow{5}*{$b$-quark}&\multirow{4}*{\makecell[c]{$m_{b}$\\ $\frac{1}{2}\frac{\textbf{p}^{2}_{x_{3}}}{2m'_{3}}$+$\frac{1}{2}\frac{3m_{\bar{c}}}{3m_{c}+2m_{\bar{b}}}\frac{\textbf{p}^{2}_{x_{4}}}{2m'_{4}}$\\ $\frac{1}{2}[V^{C}(14)+V^{C}(24)+V^{C}(34)]$\\ $\frac{1}{2}V^{C}(45)$\\ $-\frac{1}{2}D$}}&\multirow{4}*{\makecell[r]{5343.0\\ 86.6\\+44.6\\-78.6\\16.2\\-491.5}}&$\frac{m_{b}}{m_{b}+m_{c}}$$\frac98$ $m_{cb}$&5415.7\\
\Xcline{1-5}{0.5pt}
\multicolumn{5}{c}{}&(2,5)&-52.4&&&&&$-\frac{1}{2}\frac{1}{8}m_{b\bar{b}}$&-590.3\\
\multicolumn{5}{c}{}&(3,5)&-52.4&&&&&\\
\multicolumn{5}{c}{}&(4,5)&32.4&-897.1($\eta_{b}$)&&&&\\
\Xcline{6-8}{0.5pt}\Xcline{10-13}{0.5pt}
\multicolumn{5}{c|}{}&\multicolumn{3}{c|}{Relative Lengths (fm)}&&Subtotal&4920.3&&4825.4\\
\Xcline{1-13}{0.5pt}
\multicolumn{2}{c}{\multirow{3}*{\makecell[c]{Kinetic \\ Energy}}}&\multicolumn{1}{r}{\multirow{3}{*}{955.0}}&\multicolumn{1}{r}{\multirow{3}{*}{1191.5}}&\multicolumn{1}{c|}{\multirow{3}{*}{-236.5}}&(1,2)&0.382&$0.370(\Omega_{ccc})$&\multirow{5}*{$\bar{b}$-quark}&\multirow{4}*{\makecell[c]{$m_{b}$\\ $\frac{1}{2}\frac{\textbf{p}^{2}_{x_{3}}}{2m'_{3}}$+$\frac{1}{2}\frac{3m_{\bar{c}}}{3m_{c}+2m_{\bar{b}}}\frac{\textbf{p}^{2}_{x_{4}}}{2m'_{4}}$\\ $\frac{1}{2}[V^{C}(15)+V^{C}(25)+V^{C}(35)]$\\ $\frac{1}{2}V^{C}(45)$\\ $-\frac{1}{2}D$}}&\multirow{4}*{\makecell[r]{5343.0\\ 86.6\\+44.6\\-78.6\\16.2\\-491.5}}&$\frac{m_{b}}{m_{b}+m_{c}}$$\frac98$ $m_{c\bar{b}}$&5326.0\\
\multicolumn{5}{c|}{}&(1,3)&0.382&$0.370(\Omega_{ccc})$&&&&$-\frac{1}{2}\frac{1}{8}m_{b\bar{b}}$&-590.3\\
\multicolumn{5}{c|}{}&(2,3)&0.382&$0.370(\Omega_{ccc})$&&&&\\
\multicolumn{5}{c|}{}&(1,4)&0.328&&&&\\
\Xcline{10-13}{0.5pt}
\multicolumn{5}{c|}{}&(2,4)&0.328&&&Subtotal&4920.3&&4735.7\\
\Xcline{9-13}{0.5pt}
\multicolumn{2}{c}{\multirow{2}*{\makecell[c]{CS \\ Interaction}}}&\multicolumn{1}{r}{\multirow{2}{*}{38.5}}&\multicolumn{1}{r}{\multirow{2}{*}{-36.4}}&\multicolumn{1}{c|}{\multirow{2}{*}{74.9}}&(3,4)&0.328&&\multirow{4}*{\makecell[c]{CS\\  Interaction}}&\multirow{3}*{\makecell[c]{$\frac{5}{8}[V^{S}(12)+V^{S}(13)+V^{S}(23)]$\\$-\frac{5}{24}[V^{S}(14)+V^{S}(24)+V^{S}(34)]$
\\$\frac{5}{24}[V^{S}(15)+V^{S}(25)+V^{S}(35)]$\\$-\frac{1}{8}V^{S}(45)$}}&\multirow{3}*{\makecell[c]{31.2\\6.3\\-6.3\\-1.0}}
&\multirow{3}*{\makecell[c]{$\frac{5}{8}v_{cc}$\\$\frac{5}{24}v_{cb}$\\$-\frac{5}{24}v_{c\bar{b}}$\\$-\frac{1}{24}v_{b\bar{b}}$}}
&\multirow{3}*{\makecell[c]{35.5\\6.6\\-9.8\\-1.9}}\\
\multicolumn{5}{c|}{}&(1,5)&0.328&&&&\\
\multicolumn{2}{c}{\multirow{3}*{\makecell[c]{Total \\ Contribution}}}&\multicolumn{1}{c}{\multirow{3}*{\makecell[c]{704.5}}}&\multicolumn{1}{c}{\multirow{3}*{\makecell[c]{208.0}}}&\multicolumn{1}{c|}{\multirow{3}*{\makecell[c]{496.5}}}&(2,5)&0.328&&&&\\
\Xcline{10-13}{0.5pt}
\multicolumn{5}{c|}{}&(3,5)&0.328&&&Subtotal&30.3&&30.4\\
\Xcline{9-13}{0.5pt}
\multicolumn{5}{c|}{}&(4,5)&0.283&$0.148(\eta_{b})$&Total&&14678.6&&14371.6\\
\toprule[1.2pt]
\multicolumn{1}{r}{$J^{P}=\frac12^{-}$}&&Value&\multicolumn{1}{r}{$\Omega_{ccc}\Upsilon$}&\multicolumn{1}{r|}{Difference}&$(i,j)$&\multicolumn{1}{r}{Vaule}
&\multicolumn{1}{c|}{$\Omega_{ccc}\Upsilon$}&&Contribution&Value&Contribution&Value\\ \Xcline{1-13}{0.5pt}
\multicolumn{2}{c}{Mass}&\multicolumn{1}{r}{14676.3}&14270.3&406.0&(1,2)&-3.0&\multicolumn{1}{c|}{-22.7($\Omega_{ccc}$)}&\multirow{6}*{$c$-quark}&3$m_{c}$&5754.0&$\frac38$ $m_{cc}$&1189.3\\
\Xcline{1-5}{0.5pt}
\multirow{3}*{\makecell[c]{Variational\\ Parameters\\ (fm$^{-2}$)}}&$C_{11}$&\multicolumn{1}{r}{8.8}&\multicolumn{1}{r}{9.3}&&(1,3)&-3.0&
\multicolumn{1}{c|}{-22.7($\Omega_{ccc}$)}&&$[\frac{\textbf{p}^{2}_{x_{1}}}{2m'_{1}}+\frac{\textbf{p}^{2}_{x_{2}}}{2m'_{2}}]$+
$[\frac{2m_{\bar{b}}}{3m_{c}+2m_{\bar{b}}}\frac{\textbf{p}^{2}_{x_{4}}}{2m'_{4}}]$&\multirow{1}*{\makecell[r]{$533.3$\\ $+160.2$}}&$\frac{m_{c}}{m_{b}+m_{c}}$$\frac98$ $m_{cb}$&1802.4\\
&\multicolumn{1}{c|}{$C_{22}$}&\multicolumn{1}{r}{16.8}&\multicolumn{1}{r}{49.7}&&(2,3)&-3.0&\multicolumn{1}{c|}{-22.7($\Omega_{ccc}$)}&
&\multirow{3}*{\makecell[c]{$V^{C}(12)+V^{C}(13)+V^{C}(23)$\\$\frac{1}{2}[V^{C}(14)+V^{C}(14)+V^{C}(34)]$\\ $\frac{1}{2}[V^{C}(15)+V^{C}(23)+V^{C}(14)]$\\ $-\frac32D$ }}&\multirow{3}*{\makecell[r]{-9.0\\-79.5\\-79.5\\-1474.5}}&$\frac{m_{c}}{m_{b}+m_{c}}$$\frac98$ $m_{c\bar{b}}$&1788.4\\
&\multicolumn{1}{c|}{$C_{33}$}&\multicolumn{1}{r}{13.2}&&&(1,4)&-53.0&&&&\\
\Xcline{1-5}{0.5pt}
\multicolumn{2}{c}{Quark Mass}&\multicolumn{1}{r}{16440.0}&\multicolumn{1}{r}{16440.0}&0.0&(2,4)&-53.0&&&&\\
\Xcline{10-13}{0.5pt}
\multicolumn{2}{c}{\multirow{2}{*}{\makecell[c]{Confinement\\ Potential}}}&\multicolumn{1}{r}{\multirow{2}{*}{-2748.7}}&\multicolumn{1}{r}{\multirow{2}{*}{-3322.2}}&\multicolumn{1}{c|}{\multirow{2}{*}{573.5}}&(3,4)&-53.0&&&Subtotal&4805.0&&4780.1\\
\Xcline{9-13}{0.5pt}
\multicolumn{2}{c}{\multirow{5}{*}{\makecell[c]{$V^{C}$ \\ Subtotal}}}&\multicolumn{1}{r}{\multirow{5}{*}{-291.2}}&\multicolumn{1}{r}{\multirow{5}{*}{-864.8}}&\multicolumn{1}{c|}{\multirow{5}{*}{573.6}}&(1,5)&-53.0&&\multirow{5}*{$b$-quark}&\multirow{4}*{\makecell[c]{$m_{b}$\\ $\frac{1}{2}\frac{\textbf{p}^{2}_{x_{3}}}{2m'_{3}}$+$\frac{1}{2}\frac{3m_{\bar{c}}}{3m_{c}+2m_{\bar{b}}}\frac{\textbf{p}^{2}_{x_{4}}}{2m'_{4}}$\\ $\frac{1}{2}[V^{C}(14)+V^{C}(24)+V^{C}(34)]$\\ $\frac{1}{2}V^{C}(45)$\\ $-\frac{1}{2}D$}}&\multirow{4}*{\makecell[r]{5343.0\\ 92.1\\+43.1\\-79.5\\17.8\\-491.5}}&$\frac{m_{b}}{m_{b}+m_{c}}$$\frac98$ $m_{cb}$&5415.7\\
\Xcline{1-5}{0.5pt}
\multicolumn{5}{c}{}&(2,5)&-53.0&&&&&$-\frac{1}{2}\frac{1}{8}m_{b\bar{b}}$&-590.3\\
\multicolumn{5}{c}{}&(3,5)&-53.0&&&&&\\
\multicolumn{5}{c}{}&(4,5)&35.5&-796.7($\Upsilon$)&&&&\\
\Xcline{6-8}{0.5pt}\Xcline{10-13}{0.5pt}
\multicolumn{5}{c|}{}&\multicolumn{3}{c|}{Relative Lengths (fm)}&&Subtotal&4925.0&&4825.4\\
\Xcline{1-13}{0.5pt}
\multicolumn{2}{c}{\multirow{3}*{\makecell[c]{Kinetic \\ Energy}}}&\multicolumn{1}{r}{\multirow{3}{*}{963.8}}&\multicolumn{1}{r}{\multirow{3}{*}{1107.2}}&\multicolumn{1}{c|}{\multirow{3}{*}{-143.4}}&(1,2)&0.380&$0.370(\Omega_{ccc})$&\multirow{5}*{$\bar{b}$-quark}&\multirow{4}*{\makecell[c]{$m_{b}$\\ $\frac{1}{2}\frac{\textbf{p}^{2}_{x_{3}}}{2m'_{3}}$+$\frac{1}{2}\frac{3m_{\bar{c}}}{3m_{c}+2m_{\bar{b}}}\frac{\textbf{p}^{2}_{x_{4}}}{2m'_{4}}$\\ $\frac{1}{2}[V^{C}(15)+V^{C}(25)+V^{C}(35)]$\\ $\frac{1}{2}V^{C}(45)$\\ $-\frac{1}{2}D$}}&\multirow{4}*{\makecell[r]{5343.0\\ 92.1\\+43.1\\-79.5\\17.8\\-491.5}}&$\frac{m_{b}}{m_{b}+m_{c}}$$\frac98$ $m_{c\bar{b}}$&5326.0\\
\multicolumn{5}{c|}{}&(1,3)&0.380&$0.370(\Omega_{ccc})$&&&&$-\frac{1}{2}\frac{1}{8}m_{b\bar{b}}$&-590.3\\
\multicolumn{5}{c|}{}&(2,3)&0.380&$0.370(\Omega_{ccc})$&&&&\\
\multicolumn{5}{c|}{}&(1,4)&0.327&&&&\\
\Xcline{10-13}{0.5pt}
\multicolumn{5}{c|}{}&(2,4)&0.327&&&Subtotal&4925.0&&4735.7\\
\Xcline{9-13}{0.5pt}
\multicolumn{2}{c}{\multirow{2}*{\makecell[c]{CS \\ Interaction}}}&\multicolumn{1}{r}{\multirow{2}{*}{31.8}}&\multicolumn{1}{r}{\multirow{2}{*}{45.3}}&\multicolumn{1}{c|}{\multirow{2}{*}{-13.5}}&(3,4)&0.327&&\multirow{4}*{\makecell[c]{CS\\  Interaction}}&\multirow{3}*{\makecell[c]{$\frac{5}{8}[V^{S}(12)+V^{S}(13)+V^{S}(23)]$
\\\\$+\frac{3}{8}V^{S}(45)$}}&\multirow{3}*{\makecell[c]{31.5\\\\3.1}}
&\multirow{3}*{\makecell[c]{$\frac{5}{8}v_{cc}$\\\\$\frac{1}{8}v_{b\bar{b}}$}}
&\multirow{3}*{\makecell[c]{35.5\\\\5.7}}\\
\multicolumn{5}{c|}{}&(1,5)&0.327&&&&\\
\multicolumn{2}{c}{\multirow{3}*{\makecell[c]{Total \\ Contribution}}}&\multicolumn{1}{c}{\multirow{3}*{\makecell[c]{704.4}}}&\multicolumn{1}{c}{\multirow{3}*{\makecell[c]{287.7}}}&\multicolumn{1}{c|}{\multirow{3}*{\makecell[c]{416.6}}}&(2,5)&0.327&&&&\\
\Xcline{10-13}{0.5pt}
\multicolumn{5}{c|}{}&(3,5)&0.327&&&Subtotal&34.6&&41.2\\
\Xcline{9-13}{0.5pt}
\multicolumn{5}{c|}{}&(4,5)&0.274&$0.160(\Upsilon)$&Total&&14676.3&&14357.9\\
\toprule[0.5pt]
\toprule[1.0pt]
\end{tabular}
\end{lrbox}\scalebox{0.9005}{\usebox{\tablebox}}
\end{table*}

\begin{table*}
\caption{ The masses, variational parameters, the contribution from each term in the Hamiltonian, and the relative lengths between quarks for $bbbc\bar{c}$ system and their baryon-meson thresholds.
The notations are same as those of Table. \ref{nr1}.
}\label{nr8}
\begin{lrbox}{\tablebox}
\renewcommand\arraystretch{1.55}
\renewcommand\tabcolsep{2.185pt}
\begin{tabular}{c|c|ccc|ccc|c|cr|ccc}
\midrule[1.5pt]
\toprule[0.5pt]
\multicolumn{1}{c}{$bbbc\bar{c}$}&\multicolumn{5}{c}{The contribution from each term}&\multicolumn{2}{c|}{$V^{C}$}&\multirow{2}*{Overall}&\multicolumn{2}{c}{Present Work}&\multicolumn{2}{c}{CMI Model}\\
\Xcline{1-8}{0.5pt}\Xcline{10-13}{0.5pt}
\multicolumn{1}{r}{$J^{P}=\frac32^{-}$}&&Value&\multicolumn{1}{r}{$\Omega_{bbb}\eta_{c}$}&\multicolumn{1}{r|}{Difference}&$(i,j)$&\multicolumn{1}{r}{Vaule}
&\multicolumn{1}{c|}{$\Omega_{bbb}\eta_{c}$}&&Contribution&Value&Contribution&Value\\ \Xcline{1-13}{0.5pt}
\multicolumn{2}{c}{Mass}&\multicolumn{1}{r}{17891.2}&17420.1&471.1&(1,2)&-41.6&\multicolumn{1}{c|}{-287.8($\Omega_{bbb}$)}&\multirow{6}*{$b$-quark}&3$m_{b}$&16029.0&$\frac38$ $m_{bb}$&3573.6\\
\Xcline{1-5}{0.5pt}
\multirow{3}*{\makecell[c]{Variational\\ Parameters\\ (fm$^{-2}$)}}&$C_{11}$&\multicolumn{1}{r}{18.9}&\multicolumn{1}{r}{32.5}&&(1,3)&-41.6&
\multicolumn{1}{c|}{-287.8($\Omega_{bbb}$)}&&$[\frac{\textbf{p}^{2}_{x_{1}}}{2m'_{1}}+\frac{\textbf{p}^{2}_{x_{2}}}{2m'_{2}}]$+
$[\frac{2m_{c}}{3m_{b}+2m_{c}}\frac{\textbf{p}^{2}_{x_{4}}}{2m'_{4}}]$&\multirow{1}*{\makecell[r]{$414.1$\\ $+58.1$}}&$\frac{m_{b}}{m_{b}+m_{c}}$$\frac98$ $m_{cb}$&5415.7\\
&\multicolumn{1}{c|}{$C_{22}$}&\multicolumn{1}{r}{9.2}&\multicolumn{1}{r}{15.0}&&(2,3)&-41.6&\multicolumn{1}{c|}{-287.8($\Omega_{bbb}$)}&
&\multirow{3}*{\makecell[c]{$V^{C}(12)+V^{C}(13)+V^{C}(23)$\\$\frac{1}{2}[V^{C}(14)+V^{C}(24)+V^{C}(34)]$\\ $\frac{1}{2}[V^{C}(15)+V^{C}(25)+V^{C}(35)]$\\ $-\frac32D$ }}&\multirow{3}*{\makecell[r]{-124.9\\-103.0\\-103.0\\-1474.5}}&$\frac{m_{b}}{m_{b}+m_{\bar{c}}}$$\frac98$ $m_{b\bar{c}}$&5336.5\\
&\multicolumn{1}{c|}{$C_{33}$}&\multicolumn{1}{r}{13.3}&&&(1,4)&-68.4&&&&\\
\Xcline{1-5}{0.5pt}
\multicolumn{2}{c}{Quark Mass}&\multicolumn{1}{r}{19865.0}&\multicolumn{1}{r}{19865.0}&0.0&(2,4)&-68.4&&&&\\
\Xcline{10-13}{0.5pt}
\multicolumn{2}{c}{\multirow{2}{*}{\makecell[c]{Confinement\\ Potential}}}&\multicolumn{1}{r}{\multirow{2}{*}{-2987.5}}&\multicolumn{1}{r}{\multirow{2}{*}{-3558.2}}&\multicolumn{1}{c|}{\multirow{2}{*}{570.7}}&(3,4)&-68.4&&&Subtotal&14695.8&&14325.8\\
\Xcline{9-13}{0.5pt}
\multicolumn{2}{c}{\multirow{5}{*}{\makecell[c]{$V^{C}$ \\ Subtotal}}}&\multicolumn{1}{r}{\multirow{5}{*}{-530.0}}&\multicolumn{1}{r}{\multirow{5}{*}{-1100.7}}&\multicolumn{1}{c|}{\multirow{5}{*}{570.7}}&(1,5)&-68.4&&\multirow{5}*{$c$-quark}&\multirow{4}*{\makecell[c]{$m_{c}$\\ $\frac{1}{2}\frac{\textbf{p}^{2}_{x_{3}}}{2m'_{3}}$+$\frac{1}{2}\frac{3m_{b}}{3m_{b}+2m_{c}}\frac{\textbf{p}^{2}_{x_{4}}}{2m'_{4}}$\\ $\frac{1}{2}[V^{C}(14)+V^{C}(24)+V^{C}(34)]$\\ $\frac{1}{2}V^{C}(45)$\\ $-\frac{1}{2}D$}}&\multirow{4}*{\makecell[r]{1918.0\\ 139.5\\+121.3\\-103.0\\2.6\\-491.5}}&$\frac{m_{c}}{m_{b}+m_{c}}$$\frac98$ $m_{cb}$&1802.4\\
\Xcline{1-5}{0.5pt}
\multicolumn{5}{c}{}&(2,5)&-68.4&&&&&$-\frac{1}{2}\frac{1}{8}m_{c\bar{c}}$&-191.8\\
\multicolumn{5}{c}{}&(3,5)&-68.4&&&&&\\
\multicolumn{5}{c}{}&(4,5)&5.2&-237.2($\eta_{c}$)&&&&\\
\Xcline{6-8}{0.5pt}\Xcline{10-13}{0.5pt}
\multicolumn{5}{c|}{}&\multicolumn{3}{c|}{Relative Lengths (fm)}&&Subtotal&1586.8&&1610.6\\
\Xcline{1-13}{0.5pt}
\multicolumn{2}{c}{\multirow{3}*{\makecell[c]{Kinetic \\ Energy}}}&\multicolumn{1}{r}{\multirow{3}{*}{993.7}}&\multicolumn{1}{r}{\multirow{3}{*}{1167.1}}&\multicolumn{1}{c|}{\multirow{3}{*}{-173.4}}&(1,2)&0.258&$0.197(\Omega_{bbb})$&\multirow{5}*{$\bar{b}$-quark}&\multirow{4}*{\makecell[c]{$m_{b}$\\ $\frac{1}{2}\frac{\textbf{p}^{2}_{x_{3}}}{2m'_{3}}$+$\frac{1}{2}\frac{3m_{\bar{c}}}{3m_{c}+2m_{\bar{b}}}\frac{\textbf{p}^{2}_{x_{4}}}{2m'_{4}}$\\ $\frac{1}{2}[V^{C}(15)+V^{C}(25)+V^{C}(35)]$\\ $\frac{1}{2}V^{C}(45)$\\ $-\frac{1}{2}D$}}&\multirow{4}*{\makecell[r]{1918.0\\ 139.5\\+121.3\\-103.0\\2.6\\-491.5}}&$\frac{m_{\bar{c}}}{m_{\bar{c}}+m_{b}}$$\frac98$ $m_{b\bar{c}}$&1776.0\\
\multicolumn{5}{c|}{}&(1,3)&0.258&$0.197(\Omega_{bbb})$&&&&$-\frac{1}{2}\frac{1}{8}m_{c\bar{c}}$&-191.8\\
\multicolumn{5}{c|}{}&(2,3)&0.258&$0.197(\Omega_{bbb})$&&&&\\
\multicolumn{5}{c|}{}&(1,4)&0.311&&&&\\
\Xcline{10-13}{0.5pt}
\multicolumn{5}{c|}{}&(2,4)&0.311&&&Subtotal&1586.8&&1584.2\\
\Xcline{9-13}{0.5pt}
\multicolumn{2}{c}{\multirow{2}*{\makecell[c]{CS \\ Interaction}}}&\multicolumn{1}{r}{\multirow{2}{*}{14.9}}&\multicolumn{1}{r}{\multirow{2}{*}{-53.8}}&\multicolumn{1}{c|}{\multirow{2}{*}{68.7}}&(3,4)&0.311&&\multirow{4}*{\makecell[c]{CS\\  Interaction}}&\multirow{3}*{\makecell[c]{$\frac{5}{8}[V^{S}(12)+V^{S}(13)+V^{S}(23)]$\\$-\frac{5}{24}[V^{S}(14)+V^{S}(24)+V^{S}(34)]$
\\$\frac{5}{24}[V^{S}(15)+V^{S}(25)+V^{S}(35)]$\\$-\frac{1}{8}V^{S}(45)$}}&\multirow{3}*{\makecell[c]{17.1\\6.8\\-6.8\\-2.2}}
&\multirow{3}*{\makecell[c]{$\frac{5}{8}v_{bb}$\\$\frac{5}{24}v_{cb}$\\$-\frac{5}{24}v_{b\bar{c}}$\\$-\frac{1}{24}v_{c\bar{c}}$}}
&\multirow{3}*{\makecell[c]{19.2\\6.6\\-9.8\\-3.5}}\\
\multicolumn{5}{c|}{}&(1,5)&0.311&&&&\\
\multicolumn{2}{c}{\multirow{3}*{\makecell[c]{Total \\ Contribution}}}&\multicolumn{1}{c}{\multirow{3}*{\makecell[c]{478.6}}}&\multicolumn{1}{c}{\multirow{3}*{\makecell[c]{12.6}}}&\multicolumn{1}{c|}{\multirow{3}*{\makecell[c]{466.0}}}&(2,5)&0.311&&&&\\
\Xcline{10-13}{0.5pt}
\multicolumn{5}{c|}{}&(3,5)&0.311&&&Subtotal&14.9&&12.3\\
\Xcline{9-13}{0.5pt}
\multicolumn{5}{c|}{}&(4,5)&0.372&$0.290(\eta_{c})$&Total&&17884.3&&17532.9\\
\toprule[1.2pt]
\multicolumn{1}{r}{$J^{P}=\frac12^{-}$}&&Value&\multicolumn{1}{r}{$\Omega_{bbb}J/\psi$}&\multicolumn{1}{r|}{Difference}&$(i,j)$&\multicolumn{1}{r}{Vaule}
&\multicolumn{1}{c|}{$\Omega_{bbb}J/\psi$}&&Contribution&Value&Contribution&Value\\ \Xcline{1-13}{0.5pt}
\multicolumn{2}{c}{Mass}&\multicolumn{1}{r}{17883.8}&17513.8&370.0&(1,2)&-42.6&\multicolumn{1}{c|}{-287.8($\Omega_{bbb}$)}&\multirow{6}*{$b$-quark}&3$m_{b}$&16029.0&$\frac38$ $m_{bb}$&3573.6\\
\Xcline{1-5}{0.5pt}
\multirow{3}*{\makecell[c]{Variational\\ Parameters\\ (fm$^{-2}$)}}&$C_{11}$&\multicolumn{1}{r}{19.3}&\multicolumn{1}{r}{32.5}&&(1,3)&-42.6&
\multicolumn{1}{c|}{-287.8($\Omega_{bbb}$)}&&$[\frac{\textbf{p}^{2}_{x_{1}}}{2m'_{1}}+\frac{\textbf{p}^{2}_{x_{2}}}{2m'_{2}}]$+
$[\frac{2m_{c}}{3m_{b}+2m_{c}}\frac{\textbf{p}^{2}_{x_{4}}}{2m'_{4}}]$&\multirow{1}*{\makecell[r]{$421.5$\\ $+55.1$}}&$\frac{m_{b}}{m_{b}+m_{c}}$$\frac98$ $m_{cb}$&5415.7\\
&\multicolumn{1}{c|}{$C_{22}$}&\multicolumn{1}{r}{9.6}&\multicolumn{1}{r}{12.5}&&(2,3)&-42.6&\multicolumn{1}{c|}{-287.8($\Omega_{bbb}$)}&
&\multirow{3}*{\makecell[c]{$V^{C}(12)+V^{C}(13)+V^{C}(23)$\\$\frac{1}{2}[V^{C}(14)+V^{C}(24)+V^{C}(34)]$\\ $\frac{1}{2}[V^{C}(15)+V^{C}(25)+V^{C}(35)]$\\ $-\frac32D$ }}&\multirow{3}*{\makecell[r]{-127.8\\-102.0\\-102.0\\-1474.5}}&$\frac{m_{b}}{m_{b}+m_{\bar{c}}}$$\frac98$ $m_{b\bar{c}}$&5336.5\\
&\multicolumn{1}{c|}{$C_{33}$}&\multicolumn{1}{r}{12.6}&&&(1,4)&-68.0&&&&\\
\Xcline{1-5}{0.5pt}
\multicolumn{2}{c}{Quark Mass}&\multicolumn{1}{r}{19865.0}&\multicolumn{1}{r}{19865.0}&0.0&(2,4)&-68.0&&&&\\
\Xcline{10-13}{0.5pt}
\multicolumn{2}{c}{\multirow{2}{*}{\makecell[c]{Confinement\\ Potential}}}&\multicolumn{1}{r}{\multirow{2}{*}{-2985.8}}&\multicolumn{1}{r}{\multirow{2}{*}{-3485.2}}&\multicolumn{1}{c|}{\multirow{2}{*}{499.4}}&(3,4)&-68.0&&&Subtotal&14699.3&&14325.8\\
\Xcline{9-13}{0.5pt}
\multicolumn{2}{c}{\multirow{5}{*}{\makecell[c]{$V^{C}$ \\ Subtotal}}}&\multicolumn{1}{r}{\multirow{5}{*}{-528.3}}&\multicolumn{1}{r}{\multirow{5}{*}{-1027.7}}&\multicolumn{1}{c|}{\multirow{5}{*}{499.4}}&(1,5)&-68.0&&\multirow{5}*{$c$-quark}&\multirow{4}*{\makecell[c]{$m_{c}$\\ $\frac{1}{2}\frac{\textbf{p}^{2}_{x_{3}}}{2m'_{3}}$+$\frac{1}{2}\frac{3m_{b}}{3m_{b}+2m_{c}}\frac{\textbf{p}^{2}_{x_{4}}}{2m'_{4}}$\\ $\frac{1}{2}[V^{C}(14)+V^{C}(24)+V^{C}(34)]$\\ $\frac{1}{2}V^{C}(45)$\\ $-\frac{1}{2}D$}}&\multirow{4}*{\makecell[r]{1918.0\\ 145.5\\+115.2\\-102.0\\3.6\\-491.5}}&$\frac{m_{c}}{m_{b}+m_{c}}$$\frac98$ $m_{cb}$&1802.4\\
\Xcline{1-5}{0.5pt}
\multicolumn{5}{c}{}&(2,5)&-68.0&&&&&$-\frac{1}{2}\frac{1}{8}m_{c\bar{c}}$&-191.8\\
\multicolumn{5}{c}{}&(3,5)&-68.0&&&&&\\
\multicolumn{5}{c}{}&(4,5)&7.2&-164.2($J/\psi$)&&&&\\
\Xcline{6-8}{0.5pt}\Xcline{10-13}{0.5pt}
\multicolumn{5}{c|}{}&\multicolumn{3}{c|}{Relative Lengths (fm)}&&Subtotal&1588.8&&1610.6\\
\Xcline{1-13}{0.5pt}
\multicolumn{2}{c}{\multirow{3}*{\makecell[c]{Kinetic \\ Energy}}}&\multicolumn{1}{r}{\multirow{3}{*}{998.2}}&\multicolumn{1}{r}{\multirow{3}{*}{1091.1}}&\multicolumn{1}{c|}{\multirow{3}{*}{-92.9}}&(1,2)&0.256&$0.197(\Omega_{bbb})$&\multirow{5}*{$\bar{b}$-quark}&\multirow{4}*{\makecell[c]{$m_{b}$\\ $\frac{1}{2}\frac{\textbf{p}^{2}_{x_{3}}}{2m'_{3}}$+$\frac{1}{2}\frac{3m_{\bar{c}}}{3m_{c}+2m_{\bar{b}}}\frac{\textbf{p}^{2}_{x_{4}}}{2m'_{4}}$\\ $\frac{1}{2}[V^{C}(15)+V^{C}(25)+V^{C}(35)]$\\ $\frac{1}{2}V^{C}(45)$\\ $-\frac{1}{2}D$}}&\multirow{4}*{\makecell[r]{1918.0\\ 145.5\\+115.2\\-102.0\\3.6\\-491.5}}&$\frac{m_{\bar{c}}}{m_{\bar{c}}+m_{b}}$$\frac98$ $m_{b\bar{c}}$&1776.0\\
\multicolumn{5}{c|}{}&(1,3)&0.256&$0.197(\Omega_{bbb})$&&&&$-\frac{1}{2}\frac{1}{8}m_{c\bar{c}}$&-191.8\\
\multicolumn{5}{c|}{}&(2,3)&0.256&$0.197(\Omega_{bbb})$&&&&\\
\multicolumn{5}{c|}{}&(1,4)&0.311&&&&\\
\Xcline{10-13}{0.5pt}
\multicolumn{5}{c|}{}&(2,4)&0.311&&&Subtotal&1588.8&&1584.2\\
\Xcline{9-13}{0.5pt}
\multicolumn{2}{c}{\multirow{2}*{\makecell[c]{CS \\ Interaction}}}&\multicolumn{1}{r}{\multirow{2}{*}{24.1}}&\multicolumn{1}{r}{\multirow{2}{*}{43.0}}&\multicolumn{1}{c|}{\multirow{2}{*}{-18.9}}&(3,4)&0.311&&\multirow{4}*{\makecell[c]{CS\\  Interaction}}&\multirow{3}*{\makecell[c]{$\frac{5}{8}[V^{S}(12)+V^{S}(13)+V^{S}(23)]$
\\\\$+\frac{3}{8}V^{S}(45)$}}&\multirow{3}*{\makecell[c]{17.3\\\\6.7}}
&\multirow{3}*{\makecell[c]{$\frac{5}{8}v_{bb}$\\\\$\frac{1}{8}v_{c\bar{c}}$}}
&\multirow{3}*{\makecell[c]{19.2\\\\10.6}}\\
\multicolumn{5}{c|}{}&(1,5)&0.311&&&&\\
\multicolumn{2}{c}{\multirow{3}*{\makecell[c]{Total \\ Contribution}}}&\multicolumn{1}{r}{\multirow{3}*{\makecell[r]{494.0}}}&\multicolumn{1}{r}{\multirow{3}*{\makecell[r]{106.3}}}&\multicolumn{1}{c|}{\multirow{3}*{\makecell[c]{387.7}}}&(2,5)&0.311&&&&\\
\Xcline{10-13}{0.5pt}
\multicolumn{5}{c|}{}&(3,5)&0.311&&&Subtotal&24.1&&29.8\\
\Xcline{9-13}{0.5pt}
\multicolumn{5}{c|}{}&(4,5)&0.364&$0.318(J/\psi)$&Total&&17901.0&&17550.4\\
\toprule[0.5pt]
\toprule[1.0pt]
\end{tabular}
\end{lrbox}\scalebox{0.9005}{\usebox{\tablebox}}
\end{table*}

The $cccb\bar{c}$, $bbbc\bar{b}$, $cccb\bar{b}$, and $bbbc\bar{c}$ systems need to satisfy the \{123\}45 exchange antisymmetry.
There is one $J^{P}=5/2^{-}$ state, three $J^{P}=3/2^{-}$ states, and three $J^{P}=1/2^{-}$ states in every system.

Here we take the $cccb\bar{c}$ system as an example.
For a $J^{P}=5/2^{-}$ state, its mass is 11151.9 MeV, which is very close to the sum of the masses of the $\Omega_{ccc}$ and $B^{*}_{c}$.
Moreover, its variational parameters are $C_{11}=9.3~{\rm fm}^{-2}$, $C_{22}=20.2~{\rm fm}^{-2}$, and $C_{33}\sim 0 ~{\rm fm}^{-2}$, respectively.
The first and the second parameters are relevant to the size of the baryon and meson clusters, respectively,
while the last parameter reflects that the distance between the the baryon and meson clusters approaches infinity.
Thus we regard these states as a scattering state of $\Omega_{ccc}$ and $B^{*}_{c}$.
Similarly, the lowest two $J^{P}=3/2^{-}$ states and the lowest $J^{P}=1/2^{-}$ state have similar situations, in which the variational parameters $C_{33}$ all trend to be 0.
In conclusion, only the highest $J^{P}=3/2^{-}$ state and two higher $J^{P}=1/2^{-}$ states are genuine pentaquark states in these four systems.

Here, we show the mass, corresponding variational parameters, the internal contribution from each term, and the relative lengths between quarks in Tables \ref{nr5}-\ref{nr8} for lowest genuine states, respectively.

According to Tables \ref{nr5}-\ref{nr8}, we find that among the four systems, the $J^{P}=1/2^{-}$ $bbbc\bar{c}$ state is most
likely to be stable against the strong decay.
However, even for this state, its binding energy $B_{T}=+370.0$ MeV.
Thus all pentaquarks are considered as unstable states in these four systems.


For the kinetic energy part, the lowest $J^{P}=1/2^{-}$ $bbbc\bar{c}$ state obtains 998.2 MeV, which is smaller than that of the baryon-meson threshold  $\Omega^{*}_{ccb}\eta_{c}$.
The potential parts of pentaquark state are far smaller than those of the baryon-meson threshold.

For the potential part, we notice that the $V^{C}$ for most of pentaquarks is attractive according to Tables \ref{nr5}-\ref{nr8}.
Because the internal distances of pentaquark states are bigger than the lowest corresponding baryon-meson thresholds,
the $V^{C}$ contributions in the pentaquarks are much smaller.
For example, in Table \ref{nr8}
the quark distance of the (1,2) pair is 0.256 fm in the pentaquark state while it is 0.197 fm in $\Omega_{bbb}$.

The $V^{C}$ value of (4,5) is repulsive in the $J^{P}=3/2^{-}$ $bbbc\bar{c}$ state; thus this state seems to decay to $\Omega_{bbc}B^{*}_{c}$ easily, and the $\Omega_{bbb}\eta_{c}$ decay process may be suppressed.

There is a slight difference between the binding energy $B_{T}$ and the difference of the total contributions in Tables \ref{nr5}-\ref{nr10}. This is because the eigenstate $|\psi^{\rm eigen}\rangle$ of the Hamiltonian is the superposition of the color-spin states with special exchange symmetry, $|\psi^{\rm eigen}\rangle=c_1|\psi^{\rm cs;sym}_1\rangle+c_2|\psi^{\rm cs;sym}_2\rangle+...$ and we approximately use $|\psi^{\rm cs;sym}_1\rangle$ to calculate the matrix elements of the interaction since $|c_1|>90 \%$ mostly in this work.

In Tables \ref{nr5}-\ref{nr8}, we also give the comparisons for its mass according to the constituent quark model and the CMI model.
Here, we take $bbbc\bar{c}$ system as an example,
and we also absorb the quark mass term, the color potential term, and kinetic energy term of constituent quark model into the effective quark masses $b$, $c$, and $\bar{c}$ in Table \ref{nr8}. Here, we notice that the effective quark mass $c$, $\bar{c}$ and color-spin interaction term have less differences.
The main differences come from the effective $b$ quark mass, which leads to the pentaquark masses in the constituent quark model being about 300 MeV larger than those in the CMI model directly.

\subsection{$ccbb\bar{c}$ and $bbcc\bar{b}$ systems}

\begin{table*}
\caption{ The masses, variational parameters, the contribution from each term in the Hamiltonian, and the relative lengths between quarks for $ccbb\bar{b}$ system and their baryon-meson thresholds.
The notations are same as those of Table. \ref{nr1}.
}\label{nr9}
\begin{lrbox}{\tablebox}
\renewcommand\arraystretch{1.55}
\renewcommand\tabcolsep{2.2pt}
\begin{tabular}{c|c|ccc|ccc|c|cr|crr}
\midrule[1.5pt]
\toprule[0.5pt]
\multicolumn{1}{c}{$ccbb\bar{b}$}&\multicolumn{5}{c}{The contribution from each term}&\multicolumn{2}{c|}{$V^{C}$}&\multirow{2}*{Overall}&\multicolumn{2}{c}{Present Work}&\multicolumn{2}{c}{CMI Model}\\
\Xcline{1-8}{0.5pt}\Xcline{10-13}{0.5pt}
\multicolumn{1}{r}{$J^{P}=\frac32^{-}$}&&Value&\multicolumn{1}{r}{$\Omega^{*}_{ccb}\eta_{b}$}&\multicolumn{1}{r|}{Difference}&$(i,j)$&\multicolumn{1}{r}{Vaule}
&\multicolumn{1}{c|}{$\Omega^{*}_{ccb}\eta_{b}$}&&Contribution&Value&Contribution&Value\\ \Xcline{1-13}{0.5pt}
\multicolumn{2}{c}{Mass}&\multicolumn{1}{r}{17784.9}&17452.9&332.0&(1,2)&76.2&&\multirow{6}*{$b$-quark}&\multirow{5}*{\makecell[c]{$2m_{b}$\\
$[\frac{\textbf{p}^{2}_{x_{1}}}{2m'_{1}}]+[\frac{1}{3}\frac{\textbf{p}^{2}_{x_{2}}}{2m'_{2}}]+[\frac{1}{3}\frac{2m_{c}}{3m_{b}+2m_{c}}\frac{\textbf{p}^{2}_{x_{4}}}{2m'_{4}}]$\\
$[V^{C}(12)]$+$\frac12[V^{C}(15)+V^{C}(25)]$\\$\frac12[V^{C}(13)+V^{C}(23)]$\\$+V^{C}(14)+V^{C}(24)]$\\-D}}&\multirow{5}*{\makecell[r]{10686.0\\193.5\\118.2\\+13.1\\76.2\\-304.0\\-77.6\\-983.0}}
&$\frac{1}{2}m_{bb}$&4763.8\\
\Xcline{1-5}{0.5pt}
\multirow{3}*{\makecell[c]{Variational\\ Parameters\\ (fm$^{-2}$)}}&\multirow{3}*{\makecell[c]{$C_{11}$\\$C_{22}$\\$C_{33}$\\$C_{44}$}}&\multirow{3}*{\makecell[r]{$17.7$\\$32.4$\\$9.8$\\$9.0$}}&
\multirow{3}*{\makecell[r]{$10.4$\\$15.1$\\$57.4$\\$$}}&&(1,3)&-38.8&\multicolumn{1}{c|}{-98.4($\Omega^{*}_{ccb}$)}&&&&$\frac{m_{b}}{m_{c}+m_{b}}\frac{5}{4}m_{cb}$&6017.1\\
\multicolumn{1}{c|}{}&\multicolumn{1}{c|}{}&\multicolumn{3}{c|}{}&(2,3)&-38.8&&&&&$\frac{1}{2}\times-\frac{1}{4}m_{b\bar{b}}$&-1180.6\\
\multicolumn{1}{c|}{}&\multicolumn{1}{c|}{}&\multicolumn{3}{c|}{}&(1,4)&-38.8&\multicolumn{1}{c|}{-98.4($\Omega^{*}_{ccb}$)}&&&\\
\Xcline{1-5}{0.5pt}
\multicolumn{2}{c}{Quark Mass}&\multicolumn{1}{r}{19865.0}&\multicolumn{1}{r}{19865.0}&\multicolumn{1}{c|}{0.0}&(2,4)&-38.8&&&&\\
\Xcline{10-13}{0.5pt}
\multicolumn{2}{c}{\multirow{2}{*}{\makecell[c]{Confinement\\ Potential}}}&\multicolumn{1}{r}{\multirow{2}{*}{-3140.7}}
&\multicolumn{1}{r}{\multirow{2}{*}{-3578.7}}&\multicolumn{1}{c|}{\multirow{2}{*}{438.0}}&(3,4)&-34.7&\multicolumn{1}{c|}{-45.4($\Omega^{*}_{ccb}$)}&&Subtotal&9722.4&&9600.3\\
\Xcline{9-13}{0.5pt}
\multicolumn{2}{c}{\multirow{5}{*}{\makecell[c]{$V^{C}$ \\ Subtotal}}}&\multicolumn{1}{r}{\multirow{5}{*}{-683.2}}&
\multicolumn{1}{r}{\multirow{5}{*}{-1121.2}}&\multicolumn{1}{c|}{\multirow{5}{*}{438.0}}&(1,5)&-304.0&&\multirow{6}*{$c$-quark}&\multirow{5}*{\makecell[c]{$2m_{c}$\\
$[\frac{\textbf{p}^{2}_{x_{3}}}{2m'_{3}}]+[\frac{3m_{b}}{3m_{b}+2m_{c}}\frac{\textbf{p}^{2}_{x_{4}}}{2m'_{4}}]$\\
$[V^{C}(34)]$+$\frac12[V^{C}(35)+V^{C}(45)]$\\$\frac12[V^{C}(13)+V^{C}(23)]$\\$+V^{C}(14)+V^{C}(24)]$\\-D}}&3836.0& $-\frac{1}{4}m_{cc}$&-792.9\\
\Xcline{1-5}{0.5pt}
\multicolumn{5}{c}{}&(2,5)&-304.0&-879.1($\eta_{b}$)&&&\multirow{1}*{\makecell[r]{299.7\\+164.5}}&$\frac{m_{c}}{m_{c}+m_{b}}\frac{5}{4}m_{cb}$&2003.0\\
\multicolumn{5}{c}{}&(3,5)&19.2&&&&\multirow{3}*{\makecell[r]{-34.7\\19.2\\-77.6\\-983.0}}&$\frac{m_{c}}{m_{c}+m_{b}}\frac{5}{4}m_{c\bar{b}}$&1973.7\\
\multicolumn{5}{c}{}&(4,5)&19.2&&&&\\
\Xcline{6-8}{0.5pt}
\multicolumn{5}{c|}{}&\multicolumn{3}{c|}{Relative Lengths (fm)}&&&\\
\Xcline{1-8}{0.5pt}\Xcline{10-13}{0.5pt}
\multicolumn{2}{c}{\multirow{3}*{\makecell[c]{Kinetic \\ Energy}}}&\multicolumn{1}{r}{\multirow{3}{*}{1051.7}}&\multicolumn{1}{r}{\multirow{3}{*}{1208.6}}&\multicolumn{1}{c|}{\multirow{3}{*}{-156.9}}&(1,2)&0.267&&&Subtotal&3224.1&&3183.8\\
\Xcline{9-13}{0.5pt}
\multicolumn{5}{c|}{}&(1,3)&0.334&\multicolumn{1}{c|}{0.305($\Omega^{*}_{ccb}$)}&\multirow{4}*{$\bar{b}$-quark}&\multirow{3}*{\makecell[c]{$m_{b}$\\
$[\frac{2}{3}\frac{\textbf{p}^{2}_{x_{2}}}{2m'_{2}}]+[\frac{2}{3}\frac{2m_{c}}{3m_{b}+2m_{c}}\frac{\textbf{p}^{2}_{x_{4}}}{2m'_{4}}]$
\\$\frac12[V^{C}(15)+V^{C}(25)]$\\$\frac{1}{2}[V^{C}(35)+V^{C}(45)]$\\-$\frac12$D}}&\multirow{3}*{\makecell[c]{5343.0\\236.4\\+26.2\\-304.0\\19.2\\-491.5}}
&$\frac{m_{b}}{m_{c}+m_{b}}\frac{5}{4}m_{c\bar{b}}$&5929.2\\
\multicolumn{5}{c|}{}&(2,3)&0.334&&&&&$\frac{1}{2}\times-\frac{1}{4}m_{b\bar{b}}$&-1180.6\\
\multicolumn{5}{c|}{}&(1,4)&0.334&\multicolumn{1}{c|}{0.305($\Omega^{*}_{ccb}$)}&&&&\\
\multicolumn{5}{c|}{}&(2,4)&0.334&&&&\\
\Xcline{10-13}{0.5pt}
\multicolumn{2}{c}{\multirow{2}*{\makecell[c]{CS \\ Interaction}}}&\multicolumn{1}{r}{\multirow{2}{*}{13.1}}&
\multicolumn{1}{r}{\multirow{2}{*}{-42.0}}&\multicolumn{1}{c|}{\multirow{2}{*}{55.1}}&(3,4)&0.359&\multicolumn{1}{c|}{0.349($\Omega^{*}_{ccb}$)}&&Subtotal&4829.3&&4748.6\\
\Xcline{9-13}{0.5pt}
\multicolumn{5}{c|}{}&(1,5)&0.217&&\multirow{3}*{\makecell[c]{CS\\  Interaction}}&$\frac{3}{4}[V^{S}(12)]+\frac12[V^{S}(34)]$&\multirow{1}*{\makecell[r]{6.4\\+9.2}}&$\frac14v_{bb}+\frac{1}{6}v_{cc}$&\multirow{1}*{\makecell[r]{7.7\\+9.5}}\\
\multicolumn{5}{c|}{}&(2,5)&0.217&0.148($\eta_{b}$)&&$-\frac{1}{8}[V^{S}(35)+V^{S}(45)]$&-2.6&$-\frac{1}{12}v_{c\bar{b}}$&-3.9\\
\Xcline{10-13}{0.5pt}
\multicolumn{2}{c}{\multirow{3}*{\makecell[c]{Total \\ Contribution}}}&\multicolumn{1}{r}{\multirow{3}*{\makecell[c]{17789.1}}}&\multicolumn{1}{r}{\multirow{3}*{\makecell[c]{17452.9}}}&\multicolumn{1}{c |}{\multirow{3}*{\makecell[c]{336.2}}}&(3,5)&0.322&&&Subtotal&13.1&&13.2\\
\Xcline{9-13}{0.5pt}
\multicolumn{5}{c|}{}&(4,5)&0.322&&Total&&17788.9&&17554.2\\
\toprule[1.0pt]
\multicolumn{1}{r}{$J^{P}=\frac12^{-}$}&&Value&\multicolumn{1}{r}{$\Omega_{ccb}\eta_{b}$}&\multicolumn{1}{r|}{Difference}&$(i,j)$&\multicolumn{1}{r}{Vaule}
&\multicolumn{1}{c|}{$\Omega_{ccb}\eta_{b}$}&&Contribution&Value&Contribution&Value\\ \Xcline{1-13}{0.5pt}
\multicolumn{2}{c}{Mass}&\multicolumn{1}{r}{17784.5}&17418.5&366.0&(1,2)&79.7&&\multirow{6}*{$b$-quark}&\multirow{5}*{\makecell[c]{$2m_{b}$\\
$[\frac{\textbf{p}^{2}_{x_{1}}}{2m'_{1}}]+[\frac{1}{3}\frac{\textbf{p}^{2}_{x_{2}}}{2m'_{2}}]+[\frac{1}{3}\frac{2m_{c}}{3m_{b}+2m_{c}}\frac{\textbf{p}^{2}_{x_{4}}}{2m'_{4}}]$\\
$[V^{C}(12)]$+$\frac12[V^{C}(15)+V^{C}(25)]$\\$\frac12[V^{C}(13)+V^{C}(23)]$\\$+V^{C}(14)+V^{C}(24)]$\\-D}}&\multirow{5}*{\makecell[r]{10686.0\\200.1\\116.5\\+13.3\\79.7\\-305.0\\-80.6\\-983.0}}
&$\frac{1}{2}m_{bb}$&4763.8\\
\Xcline{1-5}{0.5pt}
\multirow{3}*{\makecell[c]{Variational\\ Parameters\\ (fm$^{-2}$)}}&\multirow{3}*{\makecell[c]{$C_{11}$\\$C_{22}$\\$C_{33}$\\$C_{44}$}}&\multirow{3}*{\makecell[r]{$18.3$\\$32.0$\\$9.9$\\$9.1$}}&
\multirow{3}*{\makecell[r]{$10.8$\\$16.1$\\$57.4$\\$$}}&&(1,3)&-40.3&\multicolumn{1}{c|}{-109.4($\Omega_{ccb}$)}&&&&$\frac{m_{b}}{m_{c}+m_{b}}\frac{5}{4}m_{cb}$&6017.1\\
\multicolumn{1}{c|}{}&\multicolumn{1}{c|}{}&\multicolumn{3}{c|}{}&(2,3)&-40.3&&&&&$\frac{1}{2}\times-\frac{1}{4}m_{b\bar{b}}$&-1180.6\\
\multicolumn{1}{c|}{}&\multicolumn{1}{c|}{}&\multicolumn{3}{c|}{}&(1,4)&-40.3&\multicolumn{1}{c|}{-109.4($\Omega_{ccb}$)}&&&\\
\Xcline{1-5}{0.5pt}
\multicolumn{2}{c}{Quark Mass}&\multicolumn{1}{r}{19865.0}&\multicolumn{1}{r}{19865.0}&\multicolumn{1}{c|}{0.0}&(2,4)&-40.3&&&&\\
\Xcline{10-13}{0.5pt}
\multicolumn{2}{c}{\multirow{2}{*}{\makecell[c]{Confinement\\ Potential}}}&\multicolumn{1}{r}{\multirow{2}{*}{-3145.7}}
&\multicolumn{1}{r}{\multirow{2}{*}{-3608.2}}&\multicolumn{1}{c|}{\multirow{2}{*}{462.5}}&(3,4)&-36.0&\multicolumn{1}{c|}{-52.8($\Omega_{ccb}$)}&&Subtotal&9727.0&&9600.3\\
\Xcline{9-13}{0.5pt}
\multicolumn{2}{c}{\multirow{5}{*}{\makecell[c]{$V^{C}$ \\ Subtotal}}}&\multicolumn{1}{r}{\multirow{5}{*}{-688.2}}&
\multicolumn{1}{r}{\multirow{5}{*}{-1150.7}}&\multicolumn{1}{c|}{\multirow{5}{*}{462.5}}&(1,5)&-305.0&&\multirow{6}*{$c$-quark}&\multirow{5}*{\makecell[c]{$2m_{c}$\\
$[\frac{\textbf{p}^{2}_{x_{3}}}{2m'_{3}}]+[\frac{3m_{b}}{3m_{b}+2m_{c}}\frac{\textbf{p}^{2}_{x_{4}}}{2m'_{4}}]$\\
$[V^{C}(34)]$+$\frac12[V^{C}(35)+V^{C}(45)]$\\$\frac12[V^{C}(13)+V^{C}(23)]$\\$+V^{C}(14)+V^{C}(24)]$\\-D}}&3836.0& $-\frac{1}{4}m_{cc}$&-792.9\\
\Xcline{1-5}{0.5pt}
\multicolumn{5}{c}{}&(2,5)&-305.0&-879.1($\eta_{b}$)&&&\multirow{1}*{\makecell[r]{301.7\\+166.2}}&$\frac{m_{c}}{m_{c}+m_{b}}\frac{5}{4}m_{cb}$&2003.0\\
\multicolumn{5}{c}{}&(3,5)&19.5&&&&\multirow{3}*{\makecell[r]{-36.0\\19.5\\-80.6\\-983.0}}&$\frac{m_{c}}{m_{c}+m_{b}}\frac{5}{4}m_{c\bar{b}}$&1973.7\\
\multicolumn{5}{c}{}&(4,5)&19.5&&&&\\
\Xcline{6-8}{0.5pt}
\multicolumn{5}{c|}{}&\multicolumn{3}{c|}{Relative Lengths (fm)}&&&\\
\Xcline{1-8}{0.5pt}\Xcline{10-13}{0.5pt}
\multicolumn{2}{c}{\multirow{3}*{\makecell[c]{Kinetic \\ Energy}}}&\multicolumn{1}{r}{\multirow{3}{*}{1057.2}}&\multicolumn{1}{r}{\multirow{3}{*}{1238.0}}&\multicolumn{1}{c|}{\multirow{3}{*}{-181.0}}&(1,2)&0.263&&&Subtotal&3223.8&&3183.8\\
\Xcline{9-13}{0.5pt}
\multicolumn{5}{c|}{}&(1,3)&0.334&\multicolumn{1}{c|}{0.305($\Omega^{*}_{ccb}$)}&\multirow{4}*{$\bar{b}$-quark}&\multirow{3}*{\makecell[c]{$m_{b}$\\
$[\frac{2}{3}\frac{\textbf{p}^{2}_{x_{2}}}{2m'_{2}}]+[\frac{2}{3}\frac{2m_{c}}{3m_{b}+2m_{c}}\frac{\textbf{p}^{2}_{x_{4}}}{2m'_{4}}]$
\\$\frac12[V^{C}(15)+V^{C}(25)]$\\$\frac{1}{2}[V^{C}(35)+V^{C}(45)]$\\-$\frac12$D}}&\multirow{3}*{\makecell[r]{5343.0\\232.9\\+26.5\\-305.0\\19.5\\-491.5}}
&$\frac{m_{b}}{m_{c}+m_{b}}\frac{5}{4}m_{c\bar{b}}$&5929.2\\
\multicolumn{5}{c|}{}&(2,3)&0.332&&&&&$\frac{1}{2}\times-\frac{1}{4}m_{b\bar{b}}$&-1180.6\\
\multicolumn{5}{c|}{}&(1,4)&0.332&\multicolumn{1}{c|}{0.305($\Omega^{*}_{ccb}$)}&&&&\\
\multicolumn{5}{c|}{}&(2,4)&0.332&&&&\\
\Xcline{10-13}{0.5pt}
\multicolumn{2}{c}{\multirow{2}*{\makecell[c]{CS \\ Interaction}}}&\multicolumn{1}{r}{\multirow{2}{*}{21.1}}&
\multicolumn{1}{r}{\multirow{2}{*}{-76.3}}&\multicolumn{1}{c|}{\multirow{2}{*}{97.4}}&(3,4)&0.357&\multicolumn{1}{c|}{0.349($\Omega^{*}_{ccb}$)}&&Subtotal&4825.4&&4748.6\\
\Xcline{9-13}{0.5pt}
\multicolumn{5}{c|}{}&(1,5)&0.217&&\multirow{3}*{\makecell[c]{CS\\  Interaction}}&$\frac{3}{4}[V^{S}(12)]+\frac12[V^{S}(34)]$&\multirow{1}*{\makecell[r]{6.6\\+9.2}}&$\frac14v_{bb}+\frac{1}{6}v_{cc}$&\multirow{1}*{\makecell[r]{7.7\\+9.5}}\\
\multicolumn{5}{c|}{}&(2,5)&0.217&0.148($\eta_{b}$)&&$\frac{1}{4}[V^{S}(35)+V^{S}(45)]$&5.2&$\frac{1}{6}v_{c\bar{b}}$&7.6\\
\Xcline{10-13}{0.5pt}
\multicolumn{2}{c}{\multirow{3}*{\makecell[c]{Total \\ Contribution}}}&\multicolumn{1}{r}{\multirow{3}*{\makecell[c]{17797.5}}}&\multicolumn{1}{r}{\multirow{3}*{\makecell[c]{17418.5}}}&\multicolumn{1}{c|}{\multirow{3}*{\makecell[c]{378.9}}}&(3,5)&0.321&&&Subtotal&21.1&&26.9\\
\Xcline{9-13}{0.5pt}
\multicolumn{5}{c|}{}&(4,5)&0.321&&Total&&17797.3&&17554.2\\
\toprule[0.5pt]
\toprule[1.0pt]
\end{tabular}
\end{lrbox}\scalebox{0.90}{\usebox{\tablebox}}
\end{table*}

\begin{table*}
\caption{ The masses, variational parameters, the contribution from each term in the Hamiltonian, and the relative lengths between quarks for $ccbb\bar{c}$ system and their baryon-meson thresholds.
The notations are same as those of Table. \ref{nr1}.
}\label{nr10}
\begin{lrbox}{\tablebox}
\renewcommand\arraystretch{1.55}
\renewcommand\tabcolsep{2.2pt}
\begin{tabular}{c|c|ccc|ccc|c|cr|crr}
\midrule[1.5pt]
\toprule[0.5pt]
\multicolumn{1}{c}{$ccbb\bar{c}$}&\multicolumn{5}{c}{The contribution from each term}&\multicolumn{2}{c|}{$V^{C}$}&\multirow{2}*{Overall}&\multicolumn{2}{c}{Present Work}&\multicolumn{2}{c}{CMI Model}\\
\Xcline{1-8}{0.5pt}\Xcline{10-13}{0.5pt}
\multicolumn{1}{r}{$J^{P}=\frac32^{-}$}&&Value&\multicolumn{1}{r}{$\Omega^{*}_{bbc}\eta_{c}$}&\multicolumn{1}{r|}{Difference}&$(i,j)$&\multicolumn{1}{r}{Vaule}
&\multicolumn{1}{c|}{$\Omega^{*}_{bbc}\eta_{c}$}&&Contribution&Value&Contribution&Value\\ \Xcline{1-13}{0.5pt}
\multicolumn{2}{c}{Mass}&\multicolumn{1}{r}{14579.4}&14271.7&307.7&(1,2)&2.0&&\multirow{6}*{$c$-quark}&\multirow{5}*{\makecell[c]{$2m_{c}$\\
$[\frac{\textbf{p}^{2}_{x_{1}}}{2m'_{1}}]+[\frac{1}{3}\frac{\textbf{p}^{2}_{x_{2}}}{2m'_{2}}]+[\frac{1}{3}\frac{2m_{b}}{3m_{c}+2m_{b}}\frac{\textbf{p}^{2}_{x_{4}}}{2m'_{4}}]$\\
$[V^{C}(12)]$+$\frac12[V^{C}(15)+V^{C}(25)]$\\$\frac12[V^{C}(13)+V^{C}(23)]$\\$+V^{C}(14)+V^{C}(24)]$\\-D}}&\multirow{5}*{\makecell[r]{3836.0\\256.0\\+106.1\\+40.5\\2.0\\-43.7\\-73.6\\-983.0}}
&$\frac{1}{2}m_{cc}$&1585.8\\
\Xcline{1-5}{0.5pt}
\multirow{3}*{\makecell[c]{Variational\\ Parameters\\ (fm$^{-2}$)}}&\multirow{3}*{\makecell[c]{$C_{11}$\\$C_{22}$\\$C_{33}$\\$C_{44}$}}&\multirow{3}*{\makecell[r]{$8.4$\\$10.5$\\$24.1$\\$10.0$}}&
\multirow{3}*{\makecell[r]{$26.0$\\$8.5$\\$15.0$\\$$}}&&(1,3)&-36.8&\multicolumn{1}{c|}{-131.2($\Omega^{*}_{bbc}$)}&&&&$\frac{m_{c}}{m_{c}+m_{b}}\frac{5}{4}m_{cb}$&2003.0\\
\multicolumn{1}{c|}{}&\multicolumn{1}{c|}{}&\multicolumn{3}{c|}{}&(2,3)&-36.8&&&&&$\frac{1}{2}\times-\frac{1}{4}m_{c\bar{c}}$&-383.6\\
\multicolumn{1}{c|}{}&\multicolumn{1}{c|}{}&\multicolumn{3}{c|}{}&(1,4)&-36.8&\multicolumn{1}{c|}{-131.2($\Omega^{*}_{bbc}$)}&&&\\
\Xcline{1-5}{0.5pt}
\multicolumn{2}{c}{Quark Mass}&\multicolumn{1}{r}{16440.0}&\multicolumn{1}{r}{16440.0}&\multicolumn{1}{c|}{0.0}&(2,4)&-36.8&&&&\\
\Xcline{10-13}{0.5pt}
\multicolumn{2}{c}{\multirow{2}{*}{\makecell[c]{Confinement\\ Potential}}}&\multicolumn{1}{r}{\multirow{2}{*}{-2872.6}}
&\multicolumn{1}{r}{\multirow{2}{*}{-3193.0}}&\multicolumn{1}{c|}{\multirow{2}{*}{320.4}}&(3,4)&-218.8&\multicolumn{1}{c|}{-235.9($\Omega^{*}_{bbc}$)}&&Subtotal&3140.4&&3205.2\\
\Xcline{9-13}{0.5pt}
\multicolumn{2}{c}{\multirow{5}{*}{\makecell[c]{$V^{C}$ \\ Subtotal}}}&\multicolumn{1}{r}{\multirow{5}{*}{-415.1}}&
\multicolumn{1}{r}{\multirow{5}{*}{-735.5}}&\multicolumn{1}{c|}{\multirow{5}{*}{320.4}}&(1,5)&-43.7&&\multirow{6}*{$b$-quark}&\multirow{5}*{\makecell[c]{$2m_{b}$\\
$[\frac{\textbf{p}^{2}_{x_{3}}}{2m'_{3}}]+[\frac{3m_{c}}{3m_{c}+2m_{b}}\frac{\textbf{p}^{2}_{x_{4}}}{2m'_{4}}]$\\
$[V^{C}(34)]$+$\frac12[V^{C}(35)+V^{C}(45)]$\\$\frac12[V^{C}(13)+V^{C}(23)]$\\$+V^{C}(14)+V^{C}(24)]$\\-D}}&10686.0& $-\frac{1}{4}m_{bb}$&-2381.9\\
\Xcline{1-5}{0.5pt}
\multicolumn{5}{c}{}&(2,5)&-43.7&-237.2($\eta_{c}$)&&&\multirow{1}*{\makecell[r]{263.4\\+65.4}}&$\frac{m_{b}}{m_{c}+m_{b}}\frac{5}{4}m_{cb}$&6017.1\\
\multicolumn{5}{c}{}&(3,5)&18.0&&&&\multirow{3}*{\makecell[r]{-218.8\\+18.0\\-73.6\\-983.0}}&$\frac{m_{b}}{m_{\bar{c}}+m_{b}}\frac{5}{4}m_{b\bar{c}}$&5929.2\\
\multicolumn{5}{c}{}&(4,5)&18.0&&&&\\
\Xcline{6-8}{0.5pt}
\multicolumn{5}{c|}{}&\multicolumn{3}{c|}{Relative Lengths (fm)}&&&\\
\Xcline{1-8}{0.5pt}\Xcline{10-13}{0.5pt}
\multicolumn{2}{c}{\multirow{3}*{\makecell[c]{Kinetic \\ Energy}}}&\multicolumn{1}{r}{\multirow{3}{*}{1024.9}}&\multicolumn{1}{r}{\multirow{3}{*}{1080.9}}&\multicolumn{1}{c|}{\multirow{3}{*}{-56.0}}&(1,2)&0.388&&&Subtotal&9757.4&&9564.4\\
\Xcline{9-13}{0.5pt}
\multicolumn{5}{c|}{}&(1,3)&0.337&\multicolumn{1}{c|}{0.281($\Omega^{*}_{bbc}$)}&\multirow{4}*{$\bar{c}$-quark}&\multirow{3}*{\makecell[c]{$m_{c}$\\
$[\frac{2}{3}\frac{\textbf{p}^{2}_{x_{2}}}{2m'_{2}}]+[\frac{2}{3}\frac{2m_{b}}{3m_{c}+2m_{b}}\frac{\textbf{p}^{2}_{x_{4}}}{2m'_{4}}]$
\\$\frac12[V^{C}(15)+V^{C}(25)]$\\$\frac{1}{2}[V^{C}(35)+V^{C}(45)]$\\-$\frac12$D}}&\multirow{3}*{\makecell[c]{1918.0\\212.2\\+81.1\\-43.7\\18.0\\-491.5}}
&$\frac{m_{\bar{c}}}{m_{\bar{c}}+m_{b}}\frac{5}{4}m_{b\bar{c}}$&1973.7\\
\multicolumn{5}{c|}{}&(2,3)&0.337&&&&&$\frac{1}{2}\times-\frac{1}{4}m_{c\bar{c}}$&-383.6\\
\multicolumn{5}{c|}{}&(1,4)&0.337&\multicolumn{1}{c|}{0.281($\Omega^{*}_{bbc}$)}&&&&\\
\multicolumn{5}{c|}{}&(2,4)&0.337&&&&\\
\Xcline{10-13}{0.5pt}
\multicolumn{2}{c}{\multirow{2}*{\makecell[c]{CS \\ Interaction}}}&\multicolumn{1}{r}{\multirow{2}{*}{15.3}}&
\multicolumn{1}{r}{\multirow{2}{*}{-56.3}}&\multicolumn{1}{c|}{\multirow{2}{*}{71.6}}&(3,4)&0.229&\multicolumn{1}{c|}{0.221($\Omega^{*}_{bbc}$)}&&Subtotal&1694.1&&1590.1\\
\Xcline{9-13}{0.5pt}
\multicolumn{5}{c|}{}&(1,5)&0.358&&\multirow{3}*{\makecell[c]{CS\\  Interaction}}&$\frac{3}{4}[V^{S}(12)]+\frac12[V^{S}(34)]$&\multirow{1}*{\makecell[r]{12.2\\+5.6}}&$\frac14v_{cc}+\frac{1}{6}v_{bb}$&\multirow{1}*{\makecell[r]{14.2\\+5.1}}\\
\multicolumn{5}{c|}{}&(2,5)&0.358&0.290($\eta_{c}$)&&$-\frac{1}{8}[V^{S}(35)+V^{S}(45)]$&-2.5&$-\frac{1}{12}v_{b\bar{c}}$&-3.9\\
\Xcline{10-13}{0.5pt}
\multicolumn{2}{c}{\multirow{3}*{\makecell[c]{Total \\ Contribution}}}&\multicolumn{1}{r}{\multirow{3}*{\makecell[c]{14607.5}}}&\multicolumn{1}{r}{\multirow{3}*{\makecell[c]{14271.6}}}&\multicolumn{1}{c |}{\multirow{3}*{\makecell[c]{336.0}}}&(3,5)&0.326&&&Subtotal&15.3&&15.4\\
\Xcline{9-13}{0.5pt}
\multicolumn{5}{c|}{}&(4,5)&0.326&&Total&&14607.1&&14369.8\\
\toprule[1.0pt]
\multicolumn{1}{r}{$J^{P}=\frac12^{-}$}&&Value&\multicolumn{1}{r}{$\Omega_{bbc}\eta_{c}$}&\multicolumn{1}{r|}{Difference}&$(i,j)$&\multicolumn{1}{r}{Vaule}
&\multicolumn{1}{c|}{$\Omega_{bbc}\eta_{c}$}&&Contribution&Value&Contribution&Value\\ \Xcline{1-13}{0.5pt}
\multicolumn{2}{c}{Mass}&\multicolumn{1}{r}{14566.0}&14232.7&333.3&(1,2)&6.6&&\multirow{6}*{$c$-quark}&\multirow{5}*{\makecell[c]{$2m_{c}$\\
$[\frac{\textbf{p}^{2}_{x_{1}}}{2m'_{1}}]+[\frac{1}{3}\frac{\textbf{p}^{2}_{x_{2}}}{2m'_{2}}]+[\frac{1}{3}\frac{2m_{b}}{3m_{c}+2m_{b}}\frac{\textbf{p}^{2}_{x_{4}}}{2m'_{4}}]$\\
$[V^{C}(12)]$+$\frac12[V^{C}(15)+V^{C}(25)]$\\$\frac12[V^{C}(13)+V^{C}(23)]$\\$+V^{C}(14)+V^{C}(24)]$\\-D}}&\multirow{5}*{\makecell[r]{3836.0\\268.3\\+104.2\\+41.3\\6.6\\-43.8\\-79.4\\-983.0}}
&$\frac{1}{2}m_{cc}$&1585.8\\
\Xcline{1-5}{0.5pt}
\multirow{3}*{\makecell[c]{Variational\\ Parameters\\ (fm$^{-2}$)}}&\multirow{3}*{\makecell[c]{$C_{11}$\\$C_{22}$\\$C_{33}$\\$C_{44}$}}&\multirow{3}*{\makecell[r]{$8.8$\\$10.3$\\$24.4$\\$10.2$}}&
\multirow{3}*{\makecell[r]{$26.8$\\$15.2$\\$15.0$\\$$}}&&(1,3)&-39.7&\multicolumn{1}{c|}{-145.0($\Omega_{bbc}$)}&&&&$\frac{m_{c}}{m_{c}+m_{b}}\frac{5}{4}m_{cb}$&2003.0\\
\multicolumn{1}{c|}{}&\multicolumn{1}{c|}{}&\multicolumn{3}{c|}{}&(2,3)&-39.7&&&&&$\frac{1}{2}\times-\frac{1}{4}m_{c\bar{c}}$&-383.6\\
\multicolumn{1}{c|}{}&\multicolumn{1}{c|}{}&\multicolumn{3}{c|}{}&(1,4)&-39.7&\multicolumn{1}{c|}{-237.2($\Omega_{bbc}$)}&&&\\
\Xcline{1-5}{0.5pt}
\multicolumn{2}{c}{Quark Mass}&\multicolumn{1}{r}{16440.0}&\multicolumn{1}{r}{16440.0}&\multicolumn{1}{c|}{0.0}&(2,4)&-39.7&&&&\\
\Xcline{10-13}{0.5pt}
\multicolumn{2}{c}{\multirow{2}{*}{\makecell[c]{Confinement\\ Potential}}}&\multicolumn{1}{r}{\multirow{2}{*}{-2882.5}}
&\multicolumn{1}{r}{\multirow{2}{*}{-3227.8}}&\multicolumn{1}{c|}{\multirow{2}{*}{438.0}}&(3,4)&-221.7&\multicolumn{1}{c|}{-243.2($\Omega_{bbc}$)}&&Subtotal&3150.2&&3205.2\\
\Xcline{9-13}{0.5pt}
\multicolumn{2}{c}{\multirow{5}{*}{\makecell[c]{$V^{C}$ \\ Subtotal}}}&\multicolumn{1}{r}{\multirow{5}{*}{-425.0}}&
\multicolumn{1}{r}{\multirow{5}{*}{-770.4}}&\multicolumn{1}{c|}{\multirow{5}{*}{438.0}}&(1,5)&-43.8&&\multirow{6}*{$b$-quark}&\multirow{5}*{\makecell[c]{$2m_{b}$\\
$[\frac{\textbf{p}^{2}_{x_{3}}}{2m'_{3}}]+[\frac{3m_{c}}{3m_{c}+2m_{b}}\frac{\textbf{p}^{2}_{x_{4}}}{2m'_{4}}]$\\
$[V^{C}(34)]$+$\frac12[V^{C}(35)+V^{C}(45)]$\\$\frac12[V^{C}(13)+V^{C}(23)]$\\$+V^{C}(14)+V^{C}(24)]$\\-D}}&10686.0& $-\frac{1}{4}m_{bb}$&-2381.9\\
\Xcline{1-5}{0.5pt}
\multicolumn{5}{c}{}&(2,5)&-43.8&-237.2($\eta_{c}$)&&&\multirow{1}*{\makecell[r]{266.8\\+66.7}}&$\frac{m_{b}}{m_{c}+m_{b}}\frac{5}{4}m_{cb}$&6017.1\\
\multicolumn{5}{c}{}&(3,5)&18.2&&&&\multirow{3}*{\makecell[r]{-221.7\\+18.2\\-79.4\\-983.0}}&$\frac{m_{b}}{m_{\bar{c}}+m_{b}}\frac{5}{4}m_{b\bar{c}}$&5929.2\\
\multicolumn{5}{c}{}&(4,5)&18.2&&&&\\
\Xcline{6-8}{0.5pt}
\multicolumn{5}{c|}{}&\multicolumn{3}{c|}{Relative Lengths (fm)}&&&\\
\Xcline{1-8}{0.5pt}\Xcline{10-13}{0.5pt}
\multicolumn{2}{c}{\multirow{3}*{\makecell[c]{Kinetic \\ Energy}}}&\multicolumn{1}{r}{\multirow{3}{*}{1038.4}}&\multicolumn{1}{r}{\multirow{3}{*}{1116.0}}&\multicolumn{1}{c|}{\multirow{3}{*}{-156.9}}&(1,2)&0.379&&&Subtotal&9753.6&&9564.4\\
\Xcline{9-13}{0.5pt}
\multicolumn{5}{c|}{}&(1,3)&0.333&\multicolumn{1}{c|}{0.272($\Omega_{bbc}$)}&\multirow{4}*{$\bar{c}$-quark}&\multirow{3}*{\makecell[c]{$m_{c}$\\
$[\frac{2}{3}\frac{\textbf{p}^{2}_{x_{2}}}{2m'_{2}}]+[\frac{2}{3}\frac{2m_{b}}{3m_{c}+2m_{b}}\frac{\textbf{p}^{2}_{x_{4}}}{2m'_{4}}]$
\\$\frac12[V^{C}(15)+V^{C}(25)]$\\$\frac{1}{2}[V^{C}(35)+V^{C}(45)]$\\-$\frac12$D}}&\multirow{3}*{\makecell[c]{1918.0\\208.4\\+82.6\\-43.8\\18.2\\-491.5}}
&$\frac{m_{\bar{c}}}{m_{\bar{c}}+m_{b}}\frac{5}{4}m_{b\bar{c}}$&1973.7\\
\multicolumn{5}{c|}{}&(2,3)&0.333&&&&&$\frac{1}{2}\times-\frac{1}{4}m_{c\bar{c}}$&-383.6\\
\multicolumn{5}{c|}{}&(1,4)&0.333&\multicolumn{1}{c|}{0.272($\Omega_{bbc}$)}&&&&\\
\multicolumn{5}{c|}{}&(2,4)&0.333&&&&\\
\Xcline{10-13}{0.5pt}
\multicolumn{2}{c}{\multirow{2}*{\makecell[c]{CS \\ Interaction}}}&\multicolumn{1}{r}{\multirow{2}{*}{23.4}}&
\multicolumn{1}{r}{\multirow{2}{*}{-95.4}}&\multicolumn{1}{c|}{\multirow{2}{*}{55.1}}&(3,4)&0.228&\multicolumn{1}{c|}{0.217($\Omega_{bbc}$)}&&Subtotal&1691.9&&1590.1\\
\Xcline{9-13}{0.5pt}
\multicolumn{5}{c|}{}&(1,5)&0.358&&\multirow{3}*{\makecell[c]{CS\\  Interaction}}&$\frac{3}{4}[V^{S}(12)]+\frac12[V^{S}(34)]$&\multirow{1}*{\makecell[r]{12.7\\+5.6}}&$\frac14v_{cc}+\frac{1}{6}v_{bb}$&\multirow{1}*{\makecell[r]{14.2\\+5.1}}\\
\multicolumn{5}{c|}{}&(2,5)&0.358&0.290($\eta_{c}$)&&$\frac{1}{4}[V^{S}(35)+V^{S}(45)]$&5.1&$\frac{1}{6}v_{b\bar{c}}$&7.9\\
\Xcline{10-13}{0.5pt}
\multicolumn{2}{c}{\multirow{3}*{\makecell[c]{Total \\ Contribution}}}&\multicolumn{1}{r}{\multirow{3}*{\makecell[c]{17789.1}}}&\multicolumn{1}{r}{\multirow{3}*{\makecell[c]{17452.9}}}&\multicolumn{1}{c |}{\multirow{3}*{\makecell[c]{336.2}}}&(3,5)&0.325&&&Subtotal&23.4&&27.1\\
\Xcline{9-13}{0.5pt}
\multicolumn{5}{c|}{}&(4,5)&0.325&&Total&&14619.1&&14391.1\\
\toprule[0.5pt]
\toprule[1.0pt]
\end{tabular}
\end{lrbox}\scalebox{0.90}{\usebox{\tablebox}}
\end{table*}

The $ccbb\bar{c}$ and $ccbb\bar{b}$ systems need to satisfy the \{12\}\{34\}5 symmetry.
There is one $J^{P}=5/2^{-}$ state, four $J^{P}=3/2^{-}$ states, and four $J^{P}=1/2^{-}$ states in these two systems.
Meanwhile, we think all of these states are genuine pentaquark states.

For $J^{P}=5/2^{-}$ $ccbb\bar{c}$ and $ccbb\bar{b}$ states, their masses are 14637.5 MeV and 17851.7 MeV, respectively.
Accordingly, their blind energies $B_{T}$ are $+272.1$ MeV and $+319.0$ MeV, respectively.
Relative to other lowest states, we find that the $J^{P}=5/2^{-}$ $ccbb\bar{c}$ state is most likely to be stable against the strong decay.
However, even this state can still decay into a baryon and a meson through strong interaction.

Here, we show the masses, corresponding variational parameters, the internal contribution from each term, and the relative lengths between quarks for the $J^{P}=3/2^{-}$ and $J^{P}=1/2^{-}$ $ccbb\bar{c}$ ($ccbb\bar{b}$) states in Table \ref{nr9} (\ref{nr10}). Based on Tables \ref{nr9} and \ref{nr10},
we find the $ccbb\bar{c}$ and $ccbb\bar{b}$ systems have similar situations as previously discussed systems.
One notes that the $V^{C}$ of $ccbb\bar{c}$ system is much more attractive than that of $bbcc\bar{b}$ system.

In  Tables \ref{nr9} and \ref{nr10}, we also give the comparisons for the masses according to the constituent quark model and the CMI model.
According to Table \ref{nr10}, we notice that the effective $c$ quark mass of the constituent quark model is slightly larger than that from the CMI model.
For color spin interaction term, the differences between each other are negligible.
The main difference between the constituent quark model and the CMI model comes from the effective quark masses $b$ and $\bar{b}$,
which lead to the $ccbb\bar b$ masses in the constituent quark model being about 250 MeV larger than those in the CMI model directly.
This seems to suggest that the effective quark mass increases as the number of hadronic quarks increases.

As for the color spin interaction term, we find that the $J^{P}=3/2^{-}$ $ccbb\bar{c}$ ($ccbb\bar{b}$) state has a similar values while the $J^{P}=1/2^{-}$ $ccbb\bar{c}$ ($ccbb\bar{b}$) state has small difference between the constituent quark model and the CMI model.
However, the small differences are still negligible. In summary, the color spin interaction of quark and antiquark results in the mass gaps of corresponding mesons, and thus the mass gaps in the two quark models are consistent.

\section{summary}\label{sec5}

The discovery of fully charmed tetraquark state give us strong confidence to find the fully heavy pentaquark state.
Furthermore, all of the fully heavy pentaquarks are flavor exotic.
In this work,
we use the variational method with the spatial wave function in the a simple Gaussian form to systematically investigate the masses of fully heavy pentaquark states within the constituent quark model.
Moreover, we also give the corresponding internal contributions, relative lengths, and the comparisons with the CMI model.

We repeat to calculate the masses of traditional hadrons including the $\Xi_{cc}$ with the variational method and the same set of parameters in order to check the reliability.
We construct the spatial wave functions in a simple Gaussian form and the wave functions in the color $\otimes$ spin space based on the permutation group property.
Based on these wave functions, we obtain the masses for the lowest states with different $J^{P}$ quantum numbers.
Then we also give the contributions from the quark mass term, kinetic energy part, confinement potential part, and color spin interaction part.
Meanwhile, we also calculate the length between quarks to explain the magnitude of confinement potential part.
Correspondingly, we also provide the numerical results for lowest baryon-meson threshold.

There is only a $J^{P}=3/2^{-}$ and a $J^{P}=1/2^{-}$ state in each of the $cccc\bar{c}$, $bbbb\bar{b}$, $cccc\bar{b}$, and $bbbb\bar{c}$ systems due to the \{1234\}5 symmetry,
and the $V^{C}$ of two $bbbb\bar{b}$ states seems to be more attractive relative to other systems.
For the $cccb\bar{c}$, $bbbc\bar{b}$, $cccb\bar{b}$, and $bbbc\bar{c}$ systems, there is only one $J^{P}=3/2^{-}$ and two $J^{P}=1/2^{-}$ genuine states in every system.
The reason is that other states are considered as scattering states whose variational parameter $C_{33}\sim 0$ meaning the distance between the baryon and the meson approaches infinity.
For the $ccbb\bar{c}$ and $bbcc\bar{b}$ systems, there is one $J^{P}=5/2^{-}$, four $J^{P}=1/2^{-}$, and four $J^{P}=1/2^{-}$ genuine states in every system.

In summary, we find that all of the lowest states have a large positive binding energy $B_{T}$.
Hence, we conclude that there are no stable fully heavy pentaquark states, which means that all of them can decay into a baryon and a meson through the strong interaction.
This conclusion is same with Ref. \cite{Richard:2021jgp} in which it is pointed out that no bound multiquark state is found that contains solely heavy quarks $c$ or $b$ within standard quark models.

As for the comparison with the CMI model, we have found that the masses calculated in constituent model are generally larger than the results in the CMI model.
The main differences come from the effective quark mass.
On the contrary, the contribution from the color spin terms from two different models are similar, and thus the mass gaps in the two quark models are consistent.
All in all, we hope our work will stimulate the interests in the fully heavy pentaquark system.

\section{Acknowledgments}
This work is supported by the China National Funds for Distinguished Young Scientists under Grant No. 11825503, National Key Research and Development Program of China under Contract No. 2020YFA0406400, the 111 Project under Grant No. B20063, and the National Natural Science Foundation of China under Grant No. 12047501. This project is also supported by the National Natural Science Foundation of China under Grants No. 12175091, and 11965016, and CAS Interdisciplinary Innovation Team.


\begin{thebibliography}{300}
\bibitem{GellMann:1964nj}
M.~Gell-Mann,
A schematic model of baryons and mesons,
Phys. Lett. \textbf{8}, 214 (1964).

\bibitem{Zweig:1981pd}
G.~Zweig,
An SU(3) model for strong interaction symmetry and its breaking. Version 1,
CERN-TH-401.

\bibitem{Zweig:1964jf}
G.~Zweig,
An SU(3) model for strong interaction symmetry and its breaking. Version 2,
CERN-TH-412.


\bibitem{LHCb:2016nsl}
R.~Aaij \textit{et al.} [LHCb],
Amplitude analysis of $B^+\to J/\psi \phi K^+$ decays,
Phys. Rev. D \textbf{95} (2017) no.1, 012002.

\bibitem{BESIII:2016bnd}
M.~Ablikim \textit{et al.} [BESIII],
Precise measurement of the $e^+e^-\to \pi^+\pi^-J/\psi$ cross section at center-of-mass energies from 3.77 to 4.60 GeV,
Phys. Rev. Lett. \textbf{118} (2017) no.9, 092001.

\bibitem{Ablikim:2015vvn}
  M.~Ablikim {\it et al.} [BESIII Collaboration],
Observation of a neutral charmoniumlike state $Z_c(4025)^0$ in $e^{+} e^{-} \to (D^{*} \bar{D}^{*})^{0} \pi^0$,
  Phys.\ Rev.\ Lett.\  {\bf 115}, no. 18, 182002 (2015).

\bibitem{Ablikim:2016qzw}
M.~Ablikim \textit{et al.} [BESIII],
Precise measurement of the $e^+e^-\to \pi^+\pi^-J/\psi$ cross section at center-of-mass energies from 3.77 to 4.60 GeV,
Phys. Rev. Lett. \textbf{118}, no.9, 092001 (2017).

\bibitem{Ablikim:2017oaf}
M.~Ablikim \textit{et al.} [BESIII],
Measurement of $e^{+}e^{-}\rightarrow \pi^{+}\pi^{-}\psi(3686)$ from 4.008 to 4.600~GeV and observation of a charged structure in the $\pi^{\pm}\psi(3686)$ mass spectrum,
Phys. Rev. D \textbf{96}, no.3, 032004 (2017).



\bibitem{Belle:2011aa}
  A.~Bondar {\it et al.} [Belle Collaboration],
  Observation of two charged bottomonium-like resonances in $\Upsilon(5S)$ decays,
  Phys.\ Rev.\ Lett.\  {\bf 108}, 122001 (2012).

\bibitem{Mizuk:2008me}
  R.~Mizuk {\it et al.} [Belle Collaboration],
Observation of two resonance-like structures in the $\pi^+ \chi_{c1}$ mass distribution in exclusive $\bar{B}^0\to K^- \pi^+ \chi_{c1}$ decays,
  Phys.\ Rev.\ D {\bf 78}, 072004 (2008)

\bibitem{Choi:2003ue}
S.~K.~Choi {\it et al.} [Belle Collaboration],
Observation of a narrow charmonium - like state in exclusive $B^{+-} \rightarrow K^{+-} \pi^+ \pi^- J / \psi$ decays,''
Phys.\ Rev.\ Lett.\  {\bf 91}, 262001 (2003).
\bibitem{LHCb:2021auc}
R.~Aaij \textit{et al.} [LHCb],
Study of the doubly charmed tetraquark $T_{cc}^+$,
[arXiv:2109.01056 [hep-ex]].

\bibitem{LHCb:2021vvq}
R.~Aaij \textit{et al.} [LHCb],
Observation of an exotic narrow doubly charmed tetraquark,
[arXiv:2109.01038 [hep-ex]].





\bibitem{Aaij:2015tga}
  R.~Aaij {\it et al.} [LHCb Collaboration],
Observation of $J/\psi p$ Resonances Consistent with Pentaquark States in $\Lambda_b^0 \to J/\psi K^- p$ Decays,
  Phys.\ Rev.\ Lett.\  {\bf 115}, 072001 (2015).
\bibitem{Aaij:2016phn}
  R.~Aaij {\it et al.} [LHCb Collaboration],
  ``Model-independent evidence for $J/\psi p$ contributions to $\Lambda_b^0\to J/\psi p K^-$ decays,''
  Phys.\ Rev.\ Lett.\  {\bf 117}, no. 8, 082002 (2016).


\bibitem{Aaij:2019vzc}
  R.~Aaij {\it et al.} [LHCb Collaboration],
  Observation of a narrow pentaquark state, $P_c(4312)^+$, and of two-peak structure of the $P_c(4450)^+$,
  Phys.\ Rev.\ Lett.\  {\bf 122}, no. 22, 222001 (2019).

\bibitem{Faldt:2011zv}
G.~Faldt and C.~Wilkin,
Estimation of the ratio of the $pn\rightarrow pn\pi^{0}\pi^{0}$/$pn\rightarrow d\pi^{0}\pi^{0}$ cross sections,
Phys. Lett. B \textbf{701} (2011), 619-622.

\bibitem{Adlarson:2011bh}
P.~Adlarson \textit{et al.} [WASA-at-COSY],
ABC Effect in Basic Double-Pionic Fusion --- Observation of a new resonance?,
Phys. Rev. Lett. \textbf{106}, 242302 (2011).

\bibitem{Adlarson:2012fe}
P.~Adlarson \textit{et al.} [WASA-at-COSY],
Isospin Decomposition of the Basic Double-Pionic Fusion in the Region of the ABC Effect,
Phys. Lett. B \textbf{721} (2013), 229-236.


\bibitem{LHCb:2020bwg}
R.~Aaij \textit{et al.} [LHCb],
Observation of structure in the $J /\psi$ -pair mass spectrum,
Sci. Bull. \textbf{65} (2020) no.23, 1983-1993.

\bibitem{Becchi:2020uvq}
C.~Becchi, J.~Ferretti, A.~Giachino, L.~Maiani and E.~Santopinto,
A study of $c c\bar{c}\bar{c}$ tetraquark decays in 4 muons and in $D^{(*)} \bar{D}^{(*)}$ at LHC,
Phys. Lett. B \textbf{811} (2020), 135952.

\bibitem{Wan:2020fsk}
B.~D.~Wan and C.~F.~Qiao,
Gluonic tetracharm configuration of $X (6900)$,
Phys. Lett. B \textbf{817} (2021), 136339.

\bibitem{Guo:2020pvt}
Z.~H.~Guo and J.~A.~Oller,
Insights into the inner structures of the fully charmed tetraquark state $X(6900)$,
Phys. Rev. D \textbf{103} (2021) no.3, 034024.

\bibitem{Ke:2021iyh}
H.~W.~Ke, X.~Han, X.~H.~Liu and Y.~L.~Shi,
Tetraquark state $X(6900)$ and the interaction between diquark and antidiquark,
Eur. Phys. J. C \textbf{81} (2021) no.5, 427.

\bibitem{Deng:2020iqw}
C.~Deng, H.~Chen and J.~Ping,
Towards the understanding of fully-heavy tetraquark states from various models,
Phys. Rev. D \textbf{103} (2021) no.1, 014001.


\bibitem{Jin:2020jfc}
X.~Jin, Y.~Xue, H.~Huang and J.~Ping,
Full-heavy tetraquarks in constituent quark models,
Eur. Phys. J. C \textbf{80} (2020) no.11, 1083

\bibitem{Li:2021ygk}
Q.~Li, C.~H.~Chang, G.~L.~Wang and T.~Wang,
Mass spectra and wave functions of $T_{QQ\bar{Q}\bar{Q}}$ tetraquarks,
Phys. Rev. D \textbf{104} (2021) no.1, 014018.


\bibitem{Lu:2020cns}
Q.~F.~L\"u, D.~Y.~Chen and Y.~B.~Dong,
Masses of fully heavy tetraquarks $QQ {\bar{Q}} {\bar{Q}}$ in an extended relativized quark model,
Eur. Phys. J. C \textbf{80}, no.9, 871 (2020).

\bibitem{Albuquerque:2020hio}
R.~M.~Albuquerque, S.~Narison, A.~Rabemananjara, D.~Rabetiarivony and G.~Randriamanatrika,
Doubly-hidden scalar heavy molecules and tetraquarks states from QCD at NLO,''
Phys. Rev. D \textbf{102} (2020) no.9, 094001.


\bibitem{Wang:2020dlo}
Z.~G.~Wang,
Revisit the tetraquark candidates in the $J/\psi J/\psi$ mass spectrum,
Int. J. Mod. Phys. A \textbf{36} (2021), 2150014.


\bibitem{Zhang:2020xtb}
J.~R.~Zhang,
``$0^{+}$ fully-charmed tetraquark states,''
Phys. Rev. D \textbf{103} (2021) no.1, 014018.


\bibitem{Faustov:2020qfm}
R.~N.~Faustov, V.~O.~Galkin and E.~M.~Savchenko,
Masses of the $QQ\bar Q\bar Q$ tetraquarks in the relativistic diquark--antidiquark picture,
Phys. Rev. D \textbf{102} (2020), 114030.


\bibitem{Giron:2020wpx}
J.~F.~Giron and R.~F.~Lebed,
Simple spectrum of $c\bar c c\bar c$ states in the dynamical diquark model,
Phys. Rev. D \textbf{102} (2020) no.7, 074003.


\bibitem{Gordillo:2020sgc}
M.~C.~Gordillo, F.~De Soto and J.~Segovia,
Diffusion Monte Carlo calculations of fully-heavy multiquark bound states,
Phys. Rev. D \textbf{102} (2020) no.11, 114007.


\bibitem{Weng:2020jao}
X.~Z.~Weng, X.~L.~Chen, W.~Z.~Deng and S.~L.~Zhu,
``Systematics of fully heavy tetraquarks,''
Phys. Rev. D \textbf{103} (2021), 034001.


\bibitem{Wang:2020gmd}
X.~Y.~Wang, Q.~Y.~Lin, H.~Xu, Y.~P.~Xie, Y.~Huang and X.~Chen,
``Discovery potential for the LHCb fully-charm tetraquark $X(6900)$ state via $\bar{p}p$ annihilation reaction,''
Phys. Rev. D \textbf{102} (2020), 116014.

\bibitem{Dong:2020nwy}
X.~K.~Dong, V.~Baru, F.~K.~Guo, C.~Hanhart and A.~Nefediev,
``Coupled-channel interpretation of the LHCb double-$J/\psi$ spectrum and hints of a new state near $J/\psi J/\psi$ threshold,''
[arXiv:2009.07795 [hep-ph]].



\bibitem{Feng:2020riv}
F.~Feng, Y.~Huang, Y.~Jia, W.~L.~Sang, X.~Xiong and J.~Y.~Zhang,
``Fragmentation production of fully-charmed tetraquarks at LHC,''
[arXiv:2009.08450 [hep-ph]].

\bibitem{Ma:2020kwb}
Y.~Q.~Ma and H.~F.~Zhang,
``Exploring the Di-$J/\psi$ Resonances around 6.9 $\mathrm{GeV}$ Based on $ab$ $initio$ Perturbative QCD,''
[arXiv:2009.08376 [hep-ph]].


\bibitem{Zhu:2020xni}
R.~Zhu,
Fully-heavy tetraquark spectra and production at hadron colliders,
Nucl. Phys. B \textbf{966} (2021), 115393.

\bibitem{Karliner:2020dta}
M.~Karliner and J.~L.~Rosner,
``Interpretation of structure in the di- $J/\psi$ spectrum,''
Phys. Rev. D \textbf{102} (2020) no.11, 114039.


\bibitem{Maciula:2020wri}
R.~Maciu\l{}a, W.~Sch\"afer and A.~Szczurek,
``On the mechanism of $T_{4c}$(6900) tetraquark production,''
Phys. Lett. B \textbf{812} (2021), 136010.


\bibitem{Szczurek:2021orw}
A.~Szczurek, R.~Maciu\l{}a and W.~Sch\"afer,
What is the mechanism of the $T_{4c}(6900)$ tetraquark production?,
[arXiv:2107.13285 [hep-ph]].

\bibitem{Wang:2020wrp}
J.~Z.~Wang, D.~Y.~Chen, X.~Liu and T.~Matsuki,
Producing fully charm structures in the $J/\psi$ -pair invariant mass spectrum,
Phys. Rev. D \textbf{103} (2021) no.7, 071503.



\bibitem{Bhaduri:1981pn}
R.~K.~Bhaduri, L.~E.~Cohler and Y.~Nogami,
A Unified Potential for Mesons and Baryons,
Nuovo Cim. A \textbf{65} (1981), 376-390.

\bibitem{Brink:1998as}
D.~M.~Brink and F.~Stancu,
Tetraquarks with heavy flavors,
Phys. Rev. D \textbf{57} (1998), 6778-6787.

\bibitem{Karliner:2017qjm}
M.~Karliner and J.~L.~Rosner,
Discovery of doubly-charmed $\Xi_{cc}$ baryon implies a stable ($b b \bar{u} \bar{d}$) tetraquark,
Phys. Rev. Lett. \textbf{119} (2017) no.20, 202001.

\bibitem{Eichten:2017ffp}
E.~J.~Eichten and C.~Quigg,
Heavy-quark symmetry implies stable heavy tetraquark mesons $Q_iQ_j \bar q_k \bar q_l$,
Phys. Rev. Lett. \textbf{119} (2017) no.20, 202002.

\bibitem{Lu:2020rog}
Q.~F.~L\"u, D.~Y.~Chen and Y.~B.~Dong,
Masses of doubly heavy tetraquarks $T_{QQ^\prime}$ in a relativized quark model,
Phys. Rev. D \textbf{102} (2020) no.3, 034012.

\bibitem{Cheng:2020wxa}
J.~B.~Cheng, S.~Y.~Li, Y.~R.~Liu, Z.~G.~Si and T.~Yao,
Double-heavy tetraquark states with heavy diquark-antiquark symmetry,
Chin. Phys. C \textbf{45} (2021) no.4, 043102.

\bibitem{Luo:2017eub}
S.~Q.~Luo, K.~Chen, X.~Liu, Y.~R.~Liu and S.~L.~Zhu,
Exotic tetraquark states with the $qq\bar{Q}\bar{Q}$ configuration,
Eur. Phys. J. C \textbf{77} (2017) no.10, 709.

\bibitem{Bicudo:2017szl}
P.~Bicudo, M.~Cardoso, A.~Peters, M.~Pflaumer and M.~Wagner,
$u d \bar{b} \bar{b}$ tetraquark resonances with lattice QCD potentials and the Born-Oppenheimer approximation,
Phys. Rev. D \textbf{96} (2017) no.5, 054510.

\bibitem{Bicudo:2016ooe}
P.~Bicudo, J.~Scheunert and M.~Wagner,
Including heavy spin effects in the prediction of a $\bar{b} \bar{b} u d$ tetraquark with lattice QCD potentials,
Phys. Rev. D \textbf{95} (2017) no.3, 034502.
\bibitem{Noh:2021lqs}
S.~Noh, W.~Park and S.~H.~Lee,
The Doubly-heavy Tetraquarks ($qq'\bar{Q}\bar{Q'}$) in a Constituent Quark Model with a Complete Set of Harmonic Oscillator Bases,
Phys. Rev. D \textbf{103} (2021), 114009.

\bibitem{Park:2013fda}
W.~Park and S.~H.~Lee,
Color spin wave functions of heavy tetraquark states,
Nucl. Phys. A \textbf{925} (2014), 161-184.

\bibitem{Park:2015nha}
W.~Park, A.~Park and S.~H.~Lee,
Dibaryons in a constituent quark model,
Phys. Rev. D \textbf{92} (2015) no.1, 014037.


\bibitem{Park:2017jbn}
W.~Park, A.~Park, S.~Cho and S.~H.~Lee,
$P_c(4380)$ in a constituent quark model,
Phys. Rev. D \textbf{95} (2017) no.5, 054027.


\bibitem{Park:2018wjk}
W.~Park, S.~Noh and S.~H.~Lee,
Masses of the doubly heavy tetraquarks in a constituent quark model,
Nucl. Phys. A \textbf{983} (2019), 1-19.

\bibitem{Park:2016mez}
A.~Park, W.~Park and S.~H.~Lee,
Dibaryons with two strange quarks and one heavy flavor in a constituent quark model,
Phys. Rev. D \textbf{94} (2016) no.5, 054027.

\bibitem{Park:2016cmg}
W.~Park, A.~Park and S.~H.~Lee,
Dibaryons with two strange quarks and total spin zero in a constituent quark model,
Phys. Rev. D \textbf{93} (2016) no.7, 074007.



\bibitem{Wang:2019spc}
G.~J.~Wang, L.~Y.~Xiao, R.~Chen, X.~H.~Liu, X.~Liu and S.~L.~Zhu,
Probing hidden-charm decay properties of $P_c$ states in a molecular scenario,
Phys. Rev. D \textbf{102} (2020) no.3, 036012.

\bibitem{Ke:2019bkf}
H.~W.~Ke, M.~Li, X.~H.~Liu and X.~Q.~Li,
Study on possible molecular states composed of $\Lambda_c\bar D$ ($\Lambda_b B$) and $\Sigma_c\bar D$ ($\Sigma_b B$) within the Bethe-Salpeter framework,
Phys. Rev. D \textbf{101} (2020) no.1, 014024.

\bibitem{He:2019rva}
J.~He and D.~Y.~Chen,
Molecular states from $\Sigma^{(*)}_c\bar{D}^{(*)}-\Lambda_c\bar{D}^{(*)}$ interaction,
Eur. Phys. J. C \textbf{79} (2019) no.11, 887.

\bibitem{Lin:2019qiv}
Y.~H.~Lin and B.~S.~Zou,
Strong decays of the latest LHCb pentaquark candidates in hadronic molecule pictures,
Phys. Rev. D \textbf{100} (2019) no.5, 056005.


\bibitem{Chen:2019asm}
R.~Chen, Z.~F.~Sun, X.~Liu and S.~L.~Zhu,
Strong LHCb evidence supporting the existence of the hidden-charm molecular pentaquarks,
Phys. Rev. D \textbf{100} (2019) no.1, 011502.


\bibitem{Chen:2019bip}
H.~X.~Chen, W.~Chen and S.~L.~Zhu,
Possible interpretations of the $P_c(4312)$, $P_c(4440)$, and $P_c(4457)$,
Phys. Rev. D \textbf{100} (2019) no.5, 051501.

\bibitem{Xiao:2019mvs}
C.~J.~Xiao, Y.~Huang, Y.~B.~Dong, L.~S.~Geng and D.~Y.~Chen,
Exploring the molecular scenario of Pc(4312) , Pc(4440) , and Pc(4457),
Phys. Rev. D \textbf{100} (2019) no.1, 014022.




\bibitem{An:2020jix}
H.~T.~An, K.~Chen, Z.~W.~Liu and X.~Liu,
Fully heavy pentaquarks,
Phys. Rev. D \textbf{103} (2021) no.7, 074006.

\bibitem{Yan:2021glh}
Y.~Yan, Y.~Wu, X.~Hu, H.~Huang and J.~Ping,
Fully heavy pentaquarks in quark models,
Phys. Rev. D \textbf{105} (2022) no.1, 014027.


\bibitem{Richard:2021jgp}
J.~M.~Richard,
Fully Heavy Multiquarks,
Few Body Syst. \textbf{62} (2021) no.3, 37.


\bibitem{Richard:2018yrm}
J.~M.~Richard, A.~Valcarce and J.~Vijande,
Few-body quark dynamics for doubly heavy baryons and tetraquarks,
Phys. Rev. C \textbf{97} (2018) no.3, 035211.



\bibitem{Richard:2020zxb}
J.~M.~Richard, A.~Valcarce and J.~Vijande,
Very heavy flavored dibaryons,
Phys. Rev. Lett. \textbf{124} (2020) no.21, 212001.


\bibitem{Zhang:2020vpz}
J.~R.~Zhang,
Fully-heavy pentaquark states,
Phys. Rev. D \textbf{103} (2021) no.7, 074016.

\bibitem{Wang:2021xao}
Z.~G.~Wang,
Analysis of the fully-heavy pentaquark states via the QCD sum rules,
Nucl. Phys. B \textbf{973} (2021), 115579.





\bibitem{Stancu:1999qr}
F.~Stancu and S.~Pepin,
Isoscalar factors of the permutation group,
Few Body Syst. \textbf{26} (1999), 113-133.



\bibitem{Weng:2021hje}
X.~Z.~Weng, W.~Z.~Deng and S.~L.~Zhu,
Doubly heavy tetraquarks in an extended chromomagnetic model,
[arXiv:2108.07242].

\end{thebibliography}
\end{document}